\documentclass[twocolumn]{aastex7}
\usepackage{natbib}
\usepackage{makecell}
\usepackage{rotating}
\usepackage{booktabs}
\usepackage{longtable}
\usepackage{wrapfig}
\usepackage{amsmath}
\usepackage{graphicx} 
\usepackage{multirow}
\usepackage{hyperref}
\usepackage{tabularx}

\newcolumntype{Y}{>{\raggedright\arraybackslash}X} 

\begin{document}

\title{Planets Across Space and Time (PAST). \uppercase\expandafter{\romannumeral8}: Kinematic Characterization and Identification of Radial Velocity Variables for the LAMOST-Gaia-TESS Stars}

\author[orcid=0009-0001-7966-7021]{Di Wu}
\affiliation{School of Astronomy and Space Science, Nanjing University, Nanjing 210023, China}
\affiliation{Key Laboratory of Modern Astronomy and Astrophysics, Ministry of Education, Nanjing 210023, China}
\email[]{502023260014@smail.nju.edu.cn}

\author[orcid=0000-0003-0707-3213]{Di-Chang Chen}
\affiliation{School of Physics and Astronomy, Sun Yat-sen University, Zhuhai 519082, China}
\affiliation{School of Astronomy and Space Science, Nanjing University, Nanjing 210023, China}
\affiliation{Key Laboratory of Modern Astronomy and Astrophysics, Ministry of Education, Nanjing 210023, China}
\email[]{dcchen@nju.edu.cn}

\author[orcid= 0000-0002-6472-5348]{Ji-Wei Xie}
\affiliation{School of Astronomy and Space Science, Nanjing University, Nanjing 210023, China}
\affiliation{Key Laboratory of Modern Astronomy and Astrophysics, Ministry of Education, Nanjing 210023, China}
\email[]{jwxie@nju.edu.cn}

\author[orcid= 0000-0003-1680-2940]{Ji-Lin Zhou}
\affiliation{School of Astronomy and Space Science, Nanjing University, Nanjing 210023, China}
\affiliation{Key Laboratory of Modern Astronomy and Astrophysics, Ministry of Education, Nanjing 210023, China}
\email[]{zhoujl@nju.edu.cn}

\author[orcid= 0000-0001-8459-1036]{Hai-Feng Wang}
\affiliation{Dipartimento di Fisica e Astronomia "Galileo Galilei", Universit$\acute{a}$ degli Studi di Padova, Vicolo Osservatorio 3, I-35122, Padova, Italy}
\email[]{haifeng.wang.astro@gmail.com}

\author[orcid= 0000-0002-7660-9803]{Weikai Zong}
\affiliation{Institute for Frontiers in Astronomy and Astrophysics, Beijing Normal University, Beijing 102206, China} 
\affiliation{School of Physics and Astronomy, Beijing Normal University, Beijing 100875, China}
\email[]{weikai.zong@bnu.edu.cn}

\author[orcid= 0000-0002-1027-0990]{Subo Dong}
\affiliation{Department of Astronomy, School of Physics, Peking University, 5 Yiheyuan Road, Haidian District, Beijing 100871, China}
\affiliation{Kavli Institute of Astronomy and Astrophysics, Peking University, 5 Yiheyuan Road, Haidian District, Beijing 100871, China}
\affiliation{National Astronomical Observatories, Chinese Academy of Science, 20A Datun Road, Chaoyang District, Beijing 100101, China}
\email[]{dongsubo@pku.edu.cn}

\author[orcid= 0000-0002-5818-8769]{Maosheng Xiang}
\affiliation{National Astronomical Observatories, Chinese Academy of Science, 20A Datun Road, Chaoyang District, Beijing 100101, China}
\email[]{msxiang@nao.cas.cn}

\author[orcid= 0000-0001-7865-2648]{A-Li Luo}
\affiliation{National Astronomical Observatories, Chinese Academy of Science, 20A Datun Road, Chaoyang District, Beijing 100101, China}
\email[]{lal@nao.cas.cn}

\correspondingauthor{Di-Chang Chen; Ji-Wei Xie}
\email[show]{chendch28@mail.sysu.edu.cn; jwxie@nju.edu.cn}



\begin{abstract}
The Transiting Exoplanet Survey Satellite (TESS) has discovered over 6700 nearby exoplanets candidates using the transit method through its all-sky survey. Characterizing the kinematic properties and identifying variable stars for the TESS stellar sample is crucial for revealing the correlations between the properties of planetary systems and the properties of stars (e.g., Galactic components, age, chemistry, dynamics, radiation).
Based on data from TESS, Gaia DR3, and LAMOST DR10, we present a catalog of kinematic properties (i.e., Galactic positions, velocities, orbits, Galactic components, and kinematic age) as well as other basic stellar parameters for $\sim 660,000$ TESS stars. Our analysis of the kinematic catalog reveals that stars belonging to different Galactic components (i.e., thin disk, thick disk, halo and 12 streams in the disk) display distinctive kinematic and chemical properties.
We also find that hot planets with period less then 10 days in the TESS sample favor thin disk stars compared to thick disk stars, consistent with previous studies. 
Furthermore, using the LAMOST multiple-epoch observations, we identify 41,445 stars exhibiting significant radial velocity variations, among which 7,846 are classified as binary stars. 
By fitting the radial velocity curves, we further derive orbital parameters (e.g., mass ratio, orbital period and eccentricity) for 297 binaries. 
The catalogs constructed in this work have laid a solid foundation for future work on the formation and evolution of stellar and planetary systems in different Galactic environments.
\end{abstract}

\keywords{\uat{Catalogs}{205} --- \uat{Exoplanets}{498} --- \uat{Stellar ages}{1581} --- \uat{Binary stars}{154} --- \uat{Stellar Kinematics}{1608}}

\section{Introduction} \label{sec:intro}
Since the first discovery of exoplanet orbiting Sun-like stars \citep{1995Natur.378..355M}, nearly 6,000 exoplanets have been discovered to date using various detection methods (e.g., radial velocity, direct imaging, astrometry, microlensing, and transit photometry) and thousands of candidates remain yet to be confirmed. 
Among these methods, the transit method has proven to be one of the most effective, accounting for the majority of known exoplanets (candidates) discoveries. 
Over the past two decades, several space-based transit surveys have been designed to search for large amounts of transiting exoplanets, from CoRoT \citep{2006cosp...36.3749B} and Kepler \citep{2010Sci...327..977B,2016RPPh...79c6901B}, to the ongoing Transiting Exoplanet Survey Satellite (TESS) mission \citep{rickerTransitingExoplanetSurvey2015}.


In its first six years, TESS has identified approximately 7,000 planet candidates, with several hundred confirmed exoplanets \citep{stassunTESSInputCatalog2018,guerreroTESSObjectsInterest2021a,winnTransitingExoplanetSurvey2024}. 
TESS has revealed numerous intriguing targets orbiting bright and nearby stars that are particularly well-suited for follow-up observations. 
For example, it has contributed to the discovery of hundreds of super-Earths and mini-Neptunes \citep{2018ApJ...868L..39H,2019NatAs...3.1099G}, many of which have enabled precise mass measurements through radial velocity (RV) follow-up and detailed atmospheric characterization with the James Webb Space Telescope (JWST) \citep{2023NatAs...7.1317L,2024ApJ...961L..44Z}. 
At the same time, TESS’s all-sky coverage yields a large and diverse exoplanet sample, making it particularly valuable for statistical studies of planet populations. Ongoing efforts are exploring correlations between planetary properties and stellar characteristics, e.g., hot Jupiter occurrence across different stellar mass \citep{2022MNRAS.516...75B,2023MNRAS.521.3663B}, stellar types \citep{2023AJ....165...17G,2024AJ....167..161K}, and metallicities \citep{2025ApJS..276...47G}.
Moreover, TESS has also significantly benefited the study in stellar physics, especially the identification and analyses of brown dwarfs \citep[e.g.,][]{2023MNRAS.519.5177C} and the variability of stars, including eclipsing binaries \citep{prsaTESSEclipsingBinary2022}
, flaring stars \citep{2022ApJ...925L...9F,2024ApJS..271...57X}, oscillating stars \citep{2021ApJ...919..131H} and other types of variable stars \citep{2021ApJS..253...11P,2021ApJS..253...35T,2021MNRAS.505.1476H}.

However, the kinematics of TESS stars is yet to be well characterized and their impacts on the exoplanetary systems remain to be explored. 
Previous studies based on the Kepler field have already revealed several intriguing discoveries \citep[e.g.,][]{2019AJ....158...61B,2019MNRAS.489.2505M,2022MNRAS.510.3449B,2023AJ....165..262Z}.
For example, hot Jupiters are found to be preferentially hosted by kinematically younger stars, with their frequency declining significantly with stellar age, providing evidence for tidal decay and offering crucial insights into their origins and evolution \citep[Chen et al. 2025, submitted]{2019AJ....158..190H,2023PNAS..12004179C}.
\citet{yangPlanetsSpaceTime2023} demonstrated that Kepler planetary systems generally become dynamically hotter over time using the LAMOST-Gaia-Kepler kinematic catalogs \citep{chenPlanetsSpaceTime2021}.
However, Kepler's coverage is limited to a relative small region of the sky, raising the question of whether its findings can be generalized to the whole Milky Way. 
Moreover, plenty of Galactic substructures (e.g., associations, streams) have been identified in an all-sky region with distance $< 500$ pc to our Sun \citep{2018A&A...618A..93C,2021MNRAS.505.1476H,2025NewAR.10001713B}, offering a valuable opportunity to study the impacts of Galactic histories and stellar interactions on the planetary systems.
With its all-sky survey of stars, TESS is well-suited to reveal the global picture of exoplanets across diverse Galactic environments and locations within the Milky Way.




To address these questions, the basis is to build a well-constructed kinematic catalog of TESS targets, containing their Galactic positions, velocities, orbits, component memberships, ages, and chemical abundances. 
The study conducted by \citet{2020MNRAS.491.4365C} constitutes a pioneering endeavor in this direction, which provided the spatial positions, Galactic velocities and Galactic membership for $\sim 2$ million TESS stars by using astrometric and radial velocity data from Gaia DR2 and collected chemical properties (i.e., $\rm [Fe/H]$ and $\rm [\alpha/Fe]$) from several large spectroscopic surveys.
However, their kinematic characterization was based on the methods of \citet{bensbyElementalAbundanceTrends2003,bensbyExploringMilkyWay2014}, which were primarily calibrated for stars within 100 pc and thus cannot be directly applied to TESS stars, which are distributed across a much broader range. 
Furthermore, their catalogs did not account for the identification of nearby substructures (e.g., stellar streams), which are suggested to have (potentially significant) impacts on planetary system properties \citep{2020Natur.586..528W,2023AJ....166..219D}.
In addition, their use of heterogeneous spectroscopic datasets without cross-calibration introduces inconsistencies and selection biases toward planet-hosting stars (e.g., APOGEE).
In this work, to improve upon these limitations, we first adopt the revised kinematic methods from the PAST series \citep{chenPlanetsSpaceTime2021,yangPlanetsSpaceTime2023}, which can be applied to stars within $\sim$2,000 pc for classification of Galactic components, thus covering most of TESS stars. 
The refined age-velocity dispersion relation (AVR) can also provide age estimations for groups of stars with a typical uncertainty of $\sim$10\% - 20\%. 
Secondly, we calculate the angular momentum and orbital actions to identify nearby stellar substructures following the methodology of \citet{2019A&A...631A..47K}.
Moreover, to obtain a homogeneous catalog and ensure that the sample is unbiased toward planets, we rely on the spectroscopic data from LAMOST \citep[the Large Sky Area Multi-Object Fiber Spectroscopic Telescope, also known as Goushoujing Telescope,][]{2012RAA....12.1197C,2012RAA....12..723Z}, which provides stellar parameters, radial velocities, and elemental abundances (e.g., [Fe/H], [$\alpha$/Fe]) uniformly derived from the LAMOST stellar parameter pipeline  \citep{2014IAUS..306..340W,luoFirstDataRelease2015,xiangLAMOSTStellarParameter2015}. The LAMOST-TESS kinematic catalog enable statistic studies which can directly compare with those from LAMOST-Kepler kinematic data \citep[e.g.,][]{2022AJ....163..249C,chenEvolutionHotJupiters2023,2024AJ....167...55H}, revealing the similarities and differences in the distribution and evolution of exoplanets in different positions of the Milky way.

Furthermore, this work leverages multiple epoch spectroscopy observations of the LAMOST medium-resolution (MRS) time-domain survey \citep{liuLAMOSTMediumResolutionSpectroscopic2020,zongPhaseIILAMOSTKepler2020,yanOverviewLAMOSTSurvey2022,liLAMOSTMediumresolutionSpectroscopic2023}, enabling the detection of variable and binary stars through RV variability. Using the systemic velocities derived from RV curve fitting, we can also characterize their kinematic properties. Numerous studies have used multi-epoch data for individual stars from wide-field spectroscopic surveys such as LAMOST, GALAH and APOGEE to search for and study RV variables and binaries \citep{gaoBinarityGalacticDwarf2017,2018ApJ...854..147B,price-whelanBinaryCompanionsEvolved2018,price-whelanCloseBinaryCompanions2020,2019ApJ...875...61M,2020MNRAS.499.1607M,2022MNRAS.517.3888B,guoBinarityEarlytypeStars2022}. Variables and binary stars offer natural laboratories for testing fundamental astrophysics \citep{percyUnderstandingVariableStars2007}. 
For instance, RR Lyrae variables are essential standard candles for cosmic distance measurements \citep{2018SSRv..214..113B} and also have been used to derive the structural parameters of the inner halo and thick disk \citep[e.g.,][]{2018MNRAS.479..211M,hanStructurePopulationsKinematics2025}.
$\delta$ Scuti variables can provide insights into the scaling relations between astroseismology and stellar parameters \citep{2021MNRAS.505.1476H}. 
On the other hand, interactions within binary systems can significantly influence stellar evolution \citep[e.g., tidal decay, mergers;][]{raghavanSurveyStellarFamilies2010,sanaBinaryInteractionDominates2012,ducheneStellarMultiplicity2013,moeMindYourPs2017,chenBinaryStarsNew2024}. 
Moreover, the gravitational perturbations induced by binary companions can significantly affect planet formation and evolution \citep{1999AJ....117..621H,2015ARA&A..53..409W}. 
For example, previous studies suggest that close stellar companions can influence protoplanetary disk dynamics and suppress planet formation, depending on the binary separation \citep{2014ApJ...783....4W,2015ApJ...813..130W,2016AJ....152....8K,2016ApJ...827....8N,2019ApJ...875...61M,2019MNRAS.482L.139E}. 
Binary-induced Kozai interactions can also affect the stability of planetary systems \citep{2011PhRvL.107r1101K,2011ApJ...742...94L,2014ApJ...781L...5D,2015ApJ...799...27P,2015ApJ...805...75P,2015ARA&A..53..409W,2021ARA&A..59..291Z}, and
statistical studied have suggested that this mechanism can be crucial  in the formation of hot Jupiters \citep{2014ApJ...785..126K,2015ApJ...800..138N,2025ApJ...980L..31W}.
Therefore, the synergy combining TESS and LAMOST can provide a large dataset of variable and binary stars, together with the kinematic properties, it offers a valuable opportunity to explore stellar evolution and interaction, as well as their influence on planetary systems.
This work constitutes the eighth paper of the series of PAST (Planets Across Space and Time). Previously, we revised the kinematic method \citep[PAST I,][]{chenPlanetsSpaceTime2021}, constructed the LAMOST-Gaia-Kepler kinematic catalog \citep[PAST II,][]{chenPlanetsSpaceTime2021a}, and applied the kinematic method and catalog to reveal the formation and evolution of various types of planets, e.g., Super Earth and Sub-Neptunes \citep[PAST III-IV,][]{2022AJ....163..249C,yangPlanetsSpaceTime2023}, Hot Jupiters \citep[PAST V and VII,][]{chenEvolutionHotJupiters2023,chenOriginTidalEvolution2025} and Ultra Short Period (USP) planets \citep[PAST VI,][]{2025NatAs...9..995T}. In this paper (PAST \uppercase\expandafter{\romannumeral8}), we kinematically characterize a large, homogeneous sample of TESS stars by using data from Gaia and LAMOST. Using multiple-epoch RV measurements from LAMOST, we identify RV variables and binary stars within the sample. For the binaries, we derive orbital parameters by fitting their RV curves, enabling reliable determination of their kinematic properties. The resulting binary catalog, which combines kinematic and orbital parameters, provides a solid foundation for investigating correlations between binary characteristics and stellar age, mass, and metallicity individually. Furthermore, the method of identifying RV variables can be applied to future LAMOST datasets or other large spectroscopic surveys to study how stellar multiplicity influences planetary systems. The catalogs presented in this work will thus facilitate future exploration of the formation and evolution of both planets and binary stars in the Galactic context.

This paper is organized as follows. In Section \ref{sec: section 2. Data collection}, we describe the construction of our sample from TESS, Gaia, and LAMOST. 
Section \ref{sec: section 3. classification} outlines our methodology for kinematic characterization and stellar age estimation. In Section \ref{sec: section 4. MRS single catalog}, we present the LAMOST-Gaia-TESS stellar catalog with kinematic and chemical properties. Section \ref{sec: Section 5. catalog of RV variables} demonstrates the process of identifying RV variables and binaries using the LAMOST MRS multiple-epoch catalog. Caveats and limitations are discussed in Section \ref{sec: section 6. caveats}. Finally, we summarize our main findings in Section \ref{sec: summary}.


\section{Data Collection} \label{sec: section 2. Data collection}

This section describes how we constructed the stellar sample from TESS (Section \ref{subsubsec: section 2.1.1}), Gaia (Section \ref{subsubsec: section 2.1.2}), and LAMOST (Section \ref{subsubsec: section 2.1.3}) for further kinematic characterization. In brief, we began with the TESS catalog and cross-matched it with Gaia to obtain astrometric data, i.e. positions, proper motions, and parallaxes. Radial velocities and other stellar spectral parameters were then incorporated from LAMOST spectroscopic data. 
To ensure data quality, we then applied filters to RV and spectroscopic data (Section \ref{subsec: section 2.2 data reduction}).
Based on these datasets, we can derive the space velocities and other kinematic properties for each star (with details discussed later in Section \ref{sec: section 3. classification}).

\subsection{Cross-match}\label{subsec: section 2.1}

\subsubsection{TESS: Transiting Exoplanet Survey Satellite}\label{subsubsec: section 2.1.1}

Our sample was initialized from the TESS Candidate Target List (CTL) v8.1, which is a curated subset of the TESS Input Catalog (TIC) v8.2 \citep{stassunRevisedTESSInput2019b,2021arXiv210804778P}. The CTL consists of 9,488,282 targets selected as promising candidates for TESS transit detection, either with surface gravity values ($\log g$) indicative of dwarf stars or satisfying the criteria of TESS magnitudes $T \le 13$ and stellar radii $\le 5 R_\odot$ \citep{stassunRevisedTESSInput2019b}. 


For the planetary sample, we incorporated data from the TESS Objects of Interest (TOI) catalog retrieved on November, 2024, which includes a total of 6,767 planet candidates after excluding false positives (FPs). Out of these candidates, 347 planets were first confirmed by TESS, while an additional 529 were classified as previously known planets.

\subsubsection{Astrometric Parameters from Gaia DR3}\label{subsubsec: section 2.1.2}
We then cross-matched the CTL sample with Gaia DR3 to obtain astrometric data. Gaia DR3 provides sky positions ($\alpha$, $\delta$), parallaxes, and proper motions ($\mu_{\alpha}$, $\mu_{\delta}$) for more than 1.5 billion stars, spanning Gaia magnitudes from $G = 3$ to $G = 21$ \citep{lindegrenGaiaEarlyData2021,gaiacollaborationGaiaDataRelease2023}. 
Radial velocities (RVs) for $\sim$33.81 million stars were derived by combining the low-resolution BP/RP spectra with the medium-resolution RVS spectra (R $\sim$ 11,500) \citep{katzGaiaDataRelease2023,2023A&A...674A..29R}. The uncertainties in RV depend on the brightness of the stars, increasing from $\sim$1.3 $\mathrm{km\ s^{-1}}$ at $G \approx 12$ to $\sim$6.4 $\mathrm{km\ s^{-1}}$ at $G \approx 14$. For a subset of 5.5 million bright stars, Gaia DR3 further provided astrophysical parameters including effective temperature ($T_{\text{eff}}$), surface gravity ($\log g$), and metallicity ([M/H]), with median uncertainties of 120 K, 0.2 dex, and 0.14 dex, respectively. Additionally, the reliability of these parameters may be affected by systematic errors \citep{fouesneauGaiaDataRelease2023,gaiacollaborationGaiaDataRelease2023,2023A&A...674A..29R}, and previous studies indicate the accuracy of some derived parameters can be affected by input features such as extinction ($A_G$)  \citep{zongPhaseIILAMOSTKepler2020}.
The cross-match between the CTL sample and Gaia DR3 was performed using the X-match service provided by CDS (\url{http://cdsxmatch.u-strasbg.fr}). A separation limit of 1.5 arcseconds was adopted for the cross-matching. 
To ensure consistency in brightness, we applied a magnitude cut by inspecting the distribution of magnitude differences, setting the $\mathit{G}$ magnitude difference to be less than 2. For cases where multiple matches satisfied these criteria for the same star, we kept the one with the smallest angular separation. Alternative methods were also tested, and yielded consistent results. We also removed potential binaries by eliminating stars with Gaia renormalized unit weight error (RUWE) $> 1.2$ following the recommendations of \cite{lindegrenGaiaEarlyData2021} as binary orbits could affect the results of kinematic characterization. This criterion may also remove sources in crowded regions, intrinsically variable or bright stars, and active stars, whose astrometric solutions can deviate from the single-star model \citep{lindegrenGaiaEarlyData2021,2021A&A...649A...3R}. The high-RUWE stars will be included in Section \ref{sec: subsec 5.1 crossmatch with LAMOST multiple epoch}, where we provide a detailed analysis of binaries in our sample.

\subsubsection{RV and Stellar Properties from LAMOST}\label{subsubsec: section 2.1.3}




We supplemented our catalog with spectra data from LAMOST DR10, which provides RVs, metallicity, and other stellar parameter measurements. LAMOST DR10 includes both low- and medium-resolution spectroscopic surveys (hereinafter LRS and MRS respectively). 
The LRS observational mode has a resolution of R $\sim$ 1800 and covers wavelengths from 3700 – 9000\r{A}, providing over 11.4 million spectra. 
Stellar parameters derived from LRS are available for over 7.4 million A, F, G, and K-type stars, uncertainties of 120 K for $T_{\text{eff}}$, 0.2 dex for $\log g$, 0.1 dex for [Fe/H] and $\sim$ 3 - 5 $\mathrm{km\ s^{-1}}$ for RV, depending on the signal-to-noise ratio (SNR) \citep{luoFirstDataRelease2015}.



The MRS mode offers higher precision with R $\sim$ 7500 and two wavelength ranges: 4950 – 5350\r{A} and 6300 – 6800\r{A}, covering 2,148,470 stars. Compared to LRS, MRS improves parameter accuracy, with typical uncertainties for $T_{\text{eff}}$, $\log g$, [Fe/H] and RV reduced to $\sim$100 K, $\sim$0.1 dex, 0.06 – 0.12 dex and $\sim$1 $\mathrm{km\ s^{-1}}$, respectively \citep{liuLAMOSTMediumResolutionSpectroscopic2020}. Moreover, the higher resolution of MRS provided measurements of detailed chemical abundances for 12 individual elements, including C, N, O, Mg, and Ti \citep{yanOverviewLAMOSTSurvey2022}.  

\begin{figure}[t]
	\centering
	\includegraphics[width=\linewidth]{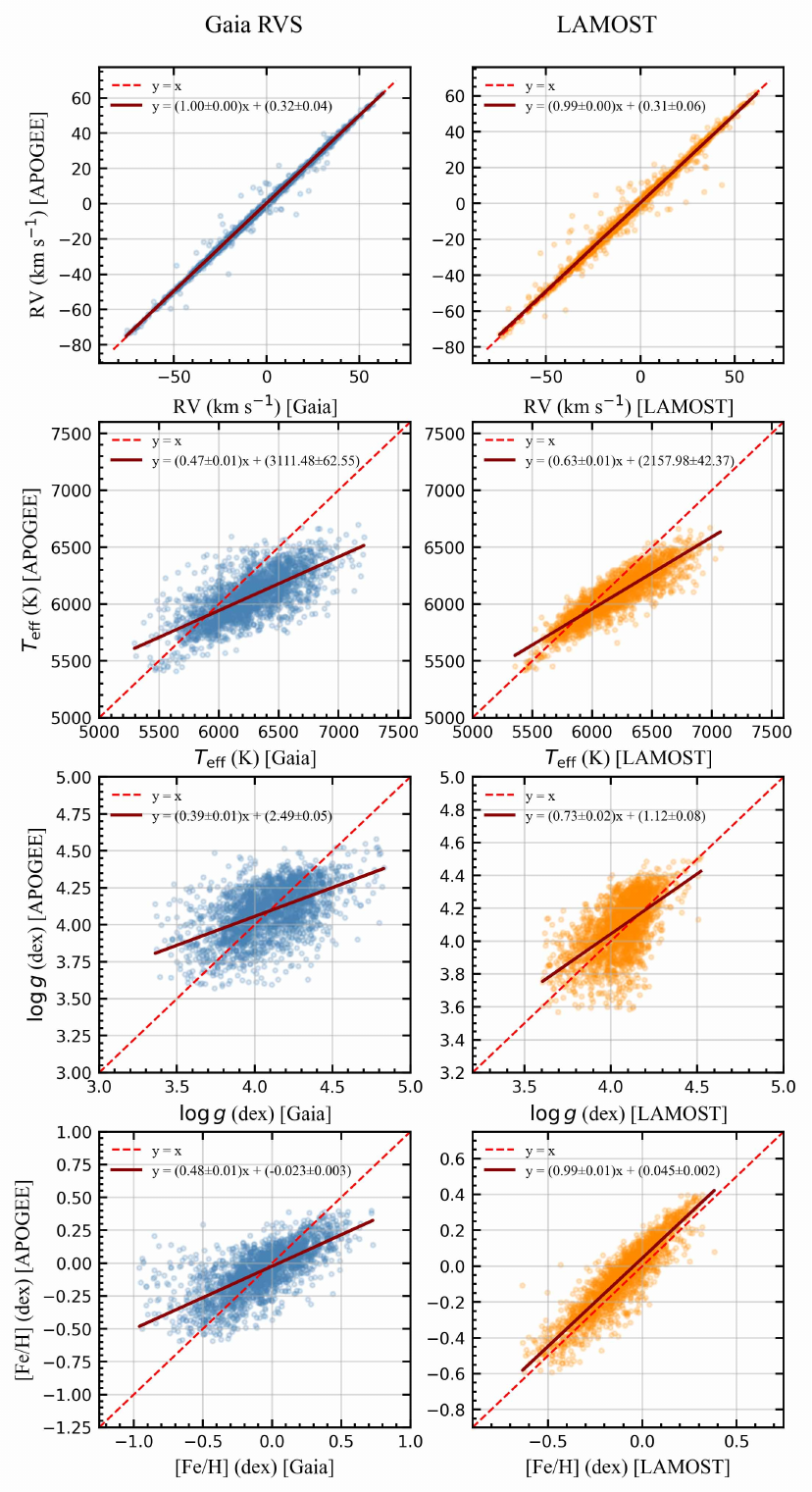}
	\caption{Radial velocity and stellar parameters obtained from Gaia RVS (left panel) and LAMOST (right panel) in comparison with APOGEE. Dark red fitted regression lines illustrate the correlations between the respective datasets, with slope and intercept values indicated. The line \(y = x\) is plotted as a red dashed line for reference.}
	\label{Fig: fig 1. apogee comparison}
\end{figure}

To evaluate the reliability of the LAMOST MRS and Gaia RVS data, we compared RV, $T_\text{eff}$, $\log g$, and [Fe/H] measurements from both LAMOST and Gaia with APOGEE as a benchmark. As shown in Figure \ref{Fig: fig 1. apogee comparison}, both LAMOST and Gaia RVS show good agreement with APOGEE for RV, exhibiting a linear correlation with the slope close to unity. However, in the case of other parameters, LAMOST demonstrates better consistency, while Gaia exhibits more significant deviations. For $T_{\text{eff}}$, LAMOST has a slope of 0.63, with deviations of $\sim$250 K for $T_{\text{eff}}$ ranging from 6000K to 6500 K. In contrast, Gaia’s $T_{\text{eff}}$ scale has a shallower slope (0.47). For stars with $T_{\text{eff}} >$  6000 K, Gaia can underestimate temperatures by $>500$ K compared to APOGEE. In the case of $\log g$, LAMOST maintains a slope of 0.73, while Gaia shows a slope of 0.39, resulting in underestimates of $\sim$0.5 dex at $\log g$ = 4.2. The difference is most prominent for [Fe/H], where LAMOST measurements are nearly identical to APOGEE (slope of $0.99 \pm 0.01$), whereas Gaia shows a significant deviation (slope of $0.48 \pm 0.01$), suggesting systematic errors in Gaia’s measurements. Given the consistency of LAMOST MRS with APOGEE and the systematic biases in Gaia’s spectroscopic parameters, we adopt LAMOST MRS measurements for characterizing the kinematic and chemical properties of the CTL-Gaia sample.




The CTL-Gaia sample was cross-matched with both the LAMOST DR10 LRS and MRS stellar parameter catalogs using the same procedure described in Section \ref{subsubsec: section 2.1.2}, with a separation limit of 1.5 arcseconds. Additionally, we applied a quality cut of SNR $>$ 10.
After the cross-matching and selections, we constructed two samples consisting of astrometric data from Gaia, along with RVs, atmospheric parameters, and chemical compositions from LAMOST. The MRS sample included 386,867 stars and 357 planets, while the LRS sample included 950,503 stars and 656 planets.




\subsection{Data Reduction}
\label{subsec: section 2.2 data reduction}
To ensure data quality, we further applied the following filters: 
\begin{enumerate}
    
    \item Consistency Check Between LRS and MRS: Stars with significant differences in $T_{\text{eff}}$, RV, [Fe/H], and [$\alpha$/Fe] measurements between DR10 LRS and MRS were removed. To address systematic differences and offsets in the measurements between the DR10 LRS and MRS, we performed a fitting analysis. First, outliers beyond a 3$\sigma$ threshold were removed from the common sample. A linear relation of the form $X_{\text{LRS}} = k \cdot X_{\text{MRS}} + b$ was then fitted using least-squares regression, where $X$ represents the measured parameters. Residuals were calculated as $\Delta X = X_{\text{LRS}} - k \cdot X_{\text{MRS}} - b$ and measurements with residuals exceeding twice their typical uncertainties ($\left| \Delta X \right| > 2\sigma$) were considered as inconsistent and excluded from the sample.

    \item Comparison Between LAMOST DR10 and DR9:
    Following the similar procedure, stars with significant differences in $T_{\text{eff}}$, RV, [Fe/H], and [$\alpha$/Fe] measurements between DR10 and DR9 were removed.
    

    \item Comparison with TESS and Gaia:
    Stars with significant differences in $T_{\text{eff}}$ measurements between TESS, Gaia, and LAMOST were removed with similar procedure as before. 
    
    
\end{enumerate}

After completing the above cross-match and data reduction process, we only kept stars with a distance less than 2 kpc to the sun, which is the applicable limit of the revised kinematic characteristics and age-velocity relation (AVR) in PAST series \citep{chenPlanetsSpaceTime2021,yangPlanetsSpaceTime2023}.  
The final MRS sample consists of 207,690 stars and 223 planet candidates. We found that 117,632 stars in the LRS sample also have corresponding MRS spectra, as they were observed by both programs. To ensure the LRS and MRS samples are mutually exclusive, for the LRS sample, we retained only those stars without MRS coverage, resulting in a final LRS sample of 452,803 stars and 321 planet candidates.
The composition of the sample after each processing step is summarized in Table \ref{tab: Table 1. construction}.


Additionally, LAMOST MRS time-domain survey provided repeated observations and RV measurements for 547,726 stars in the LAMOST DR10 MRS dataset. These multiple-epoch observations are utilized in Section \ref{sec: Section 5. catalog of RV variables} to construct a catalog of RV variable stars for subsequent binary analysis, without applying RUWE or RV consistency filters.





\begin{table}[!ht]
\caption{Construction of the LAMOST-Gaia-TESS star sample}
\label{tab: Table 1. construction}

\begin{minipage}[t]{0.42\textwidth}
\centering
\textbf{(a) LRS}
\begin{tabular}{lcc}
\hline
\textbf{Selection Criteria} & \textbf{$N_{\text{s}}$} & \textbf{$N_{\text{p}}$} \\
\hline

            CTL & 9{,}488{,}272 & 6{,}767 \\
            Crossmatch with LAMOST\_LRS DR10 & 970{,}373 & 659 \\
            SNR $>$ 10 & 950{,}503 & 656 \\
            Crossmatch with Gaia DR3 & 927{,}801 & 629 \\
            No binary & 666{,}717 & 470 \\ 
            Reliability with LAMOST DR9 & 638{,}673 & 457 \\ 
            Reliability  with LAMOST\_MRS DR10 & 620{,}292 & 438 \\ 
            $T_{\text{eff}}$ reliability & 571{,}530 & 413 \\ 
            Distance $<$ 2 kpc & 570{,}435 & 413 \\  
            Without MRS spectra &452{,}803  &321    \\  
\hline
\end{tabular}
\end{minipage}
\hfill
\begin{minipage}[t]{0.42\textwidth}
\centering
\textbf{(b) MRS}
\begin{tabular}{lcc}
\hline
\textbf{Selection Criteria} & \textbf{$N_{\text{s}}$} & \textbf{$N_{\text{p}}$} \\
\hline
            CTL & 9{,}488{,}272 & 6{,}767 \\
            Crossmatch with LAMOST\_MRS DR10 & 388{,}110 & 659 \\
            SNR $>$ 10 & 386{,}867 & 357 \\
            Crossmatch with Gaia DR3 & 381{,}290 & 349 \\
            No binary & 270{,}973 & 270 \\
            Reliability with LAMOST DR9 & 235{,}986 & 251 \\ 
            Reliability with LAMOST\_LRS DR10 & 222{,}132 & 232 \\ 
            $T_{\text{eff}}$ reliability & 207{,}776 & 223\\ 
            Distance $<$ 2 kpc & 207{,}690 &223 \\
\hline
\end{tabular}

\end{minipage}

\noindent
\tablecomments{$N_{\text{s}}$ and $N_{\text{p}}$ are the numbers of stars and planets (TOIs), respectively, during the process of sample selection in Section \ref{subsec: section 2.2 data reduction}. $T_{\text{eff}}$ reliability includes both comparison with Gaia and TESS.}
\end{table}

\section{Methods for Classification Of Galactic Components and Estimation of Ages}
\label{sec: section 3. classification}
In this Section, we describe how we calculate the stellar kinematics (i.e., space position and Galactic orbits; Section \ref{subsec: section 3.1 space velocity}), classify star into different Galactic components (Section \ref{subsec: section 3.2 Classification of Galactic Components}) and estimate stellar ages (Section \ref{subsec: section 3.3 Calculating Kinematic Age and uncertainty}).
\subsection{Space Position and Galactic Orbits} 
\label{subsec: section 3.1 space velocity}


We determined the 3D Galactocentric cylindrical coordinates ($R$, $\theta$, $Z$) using the Solar position of $R_\odot$= 8.18 kpc and $Z_\odot$ = 27 pc \citep{thegravitycollaborationGeometricDistanceMeasurement2019,gaiacollaborationGaiaDataRelease2023}. Galactic velocities ($U$, $V$, $W$) in the Cartesian Galactic system were calculated from the LAMOST RV and Gaia astrometric data, following the equations in \citet{johnsonCalculatingGalacticSpace1987}. 
We adopt a right-handed Galactic coordinate system, where $U$ points toward the Galactic center, $V$ follows the direction of rotation, and $W$ is directed toward the North Galactic Pole (NGP). 
To derive the Galactic rectangular velocities with respect to the local standard of rest (LSR), we adopt the Solar peculiar motion $[U, V, W]$ = [9.58, 10.52, 7.01] $\rm km\ s^{-1}$, as reported by \citet{tianSTELLARKINEMATICSSOLAR2015}. Uncertainties in the velocity components $U$, $V$, and $W$ were propagated from measurement errors in distances, proper motions, and RVs using the method introduced by \citet{johnsonCalculatingGalacticSpace1987}.
The cylindrical velocities $V_R$, $V_\theta$, and $V_Z$ are considered  positive with the increases in $R$, $\theta$, and $Z$ respectively, with $V_Z$ increasing towards the North Galactic Pole.
Furthermore, we calculate the orbital angular momentum($L_X$, $L_Y$, $L_Z$) and orbital actions ($J_R$, $J_Z$) using an axisymmetric potential model of the Milky Way implemented in the Python package $\it gala$ \citep{2017JOSS....2..388P} with the potential model including a spherical galactic nucleus and bulge, a three-component sum of Miyamoto–Nagai galactic disks and a spherical Navarro–Frenk–White dark matter halo model \citep{1996ApJ...462..563N}. 
The galactic rotation curve and the vertical
structure of the Galactic disks are set by the velocity curve from \cite{2019ApJ...871..120E} and  the shape of the phase-space spiral from \cite{2023ApJ...955...74D}, respectively.
    

\subsection{Classification of Galactic Components}
\label{subsec: section 3.2 Classification of Galactic Components}
We first identified stars belonging to stellar streams (e.g., Arcturus, Sirius, and Pleiades/Pleiades/Hyades) by adopting the velocity, angular momentum, and action planes in \cite{2019A&A...631A..47K}.
For the remaining stars, with the refined kinematic method \citep[Section 2.2 of][]{chenPlanetsSpaceTime2021a} and updated typical characteristics based on Gaia DR3 data \citep[Table 2 of][]{yangPlanetsSpaceTime2023}, we calculate the probabilities of stars belonging to each Galactic component,
i.e., thin disk (D), thick disk (TD), halo (H), and the Hercules stream (Herc). Stars were then classified into these Galactic components following the same criteria in \cite{bensbyExploringMilkyWay2014}.

\subsection{Kinematic Age}\label{subsec: section 3.3 Calculating Kinematic Age and uncertainty}

\subsubsection{Average Kinematic Age from Velocity Dispersion}
As described in Section 3.6 of PAST \uppercase\expandafter{\romannumeral1} \citep{chenPlanetsSpaceTime2021a}, for a group of stars, the average kinematic age can be derived using the age-velocity dispersion relation (AVR), expressed as:
\begin{equation}
{\rm Age_{kin}} = \left(\frac{\sigma}{k\rm\, km \ s^{-1}}\right)^{\frac{1}{\beta}}\, \rm Gyr,
\label{eqkineage}
\end{equation}
where $\sigma$ is the velocity dispersion, which is defined as the root mean square of stellar Galactic velocity.
The parameters $k$, $\beta$ are the fitting coefficients of AVR, adopted from Table 4 of \cite{yangPlanetsSpaceTime2023} based on Gaia DR3 and LAMOST MSTO-SG catalog.
The uncertainty of $Age_{\rm kin}$ is propagated from the errors in $k$, $\beta$ and $\sigma$, yielding a typical uncertainty of $\sim 10\% - 20\%$. 

\subsubsection{Individual Age from $J_Z$}
\cite{2024ApJ...977...49S} proposed a flexible model and a software package {\it zoomies}$\footnote{\url{www.github.com/ssagear/zoomies}}$ to predict stellar ages $\rm Age_{zoomies}$ for individual stars based on their vertical action $J_Z$.
We also used the \textit{zoomies} package to estimate ages for our LAMOST-Gaia-TESS sample.
Nevertheless, the typical uncertainties in the derived $\rm Age_{zoomies}$ are quite large as $\sim 3 - 4$ Gyr \citep[see Figure. 15 of][]{2024ApJ...977...49S}, limiting their applicability to providing only a rough estimate of the age range.

\section{The Catalog of LAMOST-Gaia-TESS Stars With Kinematic Characteristics} \label{sec: section 4. MRS single catalog}

\begin{figure}[htb]
    \centering
    \includegraphics[width=\linewidth,height=\linewidth,keepaspectratio=false]{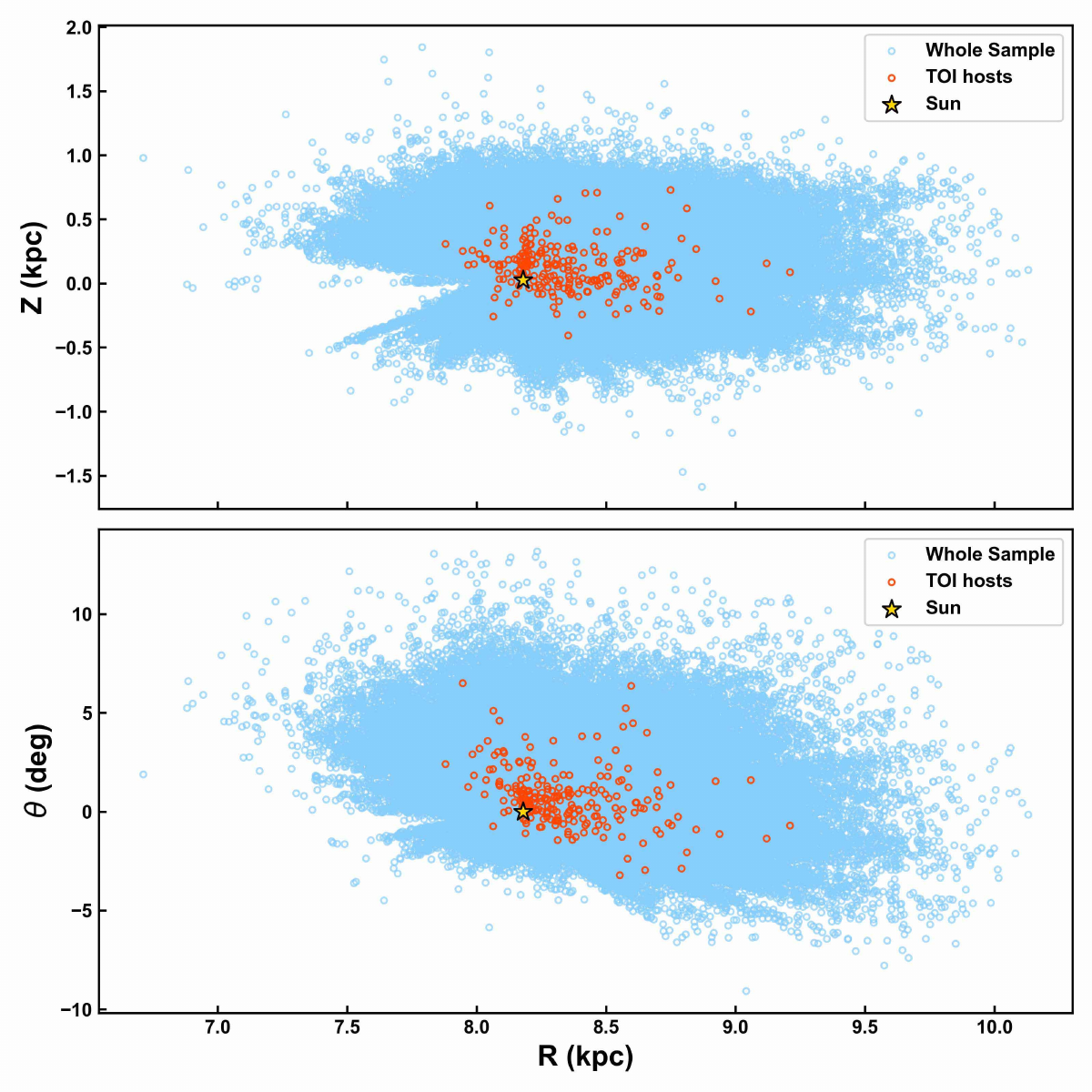}
    \caption{Top panel: Galactocentric radius (R) vs. height (Z) for the LAMOST\_MRS-Gaia-TESS sample. A yellow star marks the location of the Sun (8.18 kpc, 0.027 kpc). Bottom panel: Galactocentric radius (R) vs. angle ($\theta$) for the MRS sample, with (8.18 kpc, $0^{\circ}$) marking the position of Sun. Red points represent the TOI host stars.}
    \label{fig: fig 4. distribution of mrs sample}
\end{figure}
Applying the methods described in Section \ref{sec: section 3. classification} to the LAMOST-Gaia-TESS sample, we characterize the kinematic properties including space position, Galactic orbits, the relative membership probabilities between different Galactic components ($TD/D$, $TD/H$, $Herc/D$ and $Herc/TD$), and kinematic age for 207,690 and 452,803 TESS stars with LAMOST MRS and LRS spectral data, respectively (Table \ref{tab: MRS single catalog}).
For clarity and conciseness, we present results primarily for the MRS sample in this section, while those for the LRS sample are provided in Figures \ref{figappendix: lrs distribution}–\ref{figappendix: lrs streams} in the Appendix.
Figure \ref{fig: fig 4. distribution of mrs sample} presents the spatial distribution of stars in the MRS sample. As can be seen, the majority of stars are located within the region $|Z| < 1.0$ kpc and $7.5 < R < 9.5$ kpc.

\startlongtable
\begin{deluxetable*}{ccccl}
\tablecaption{The catalog of the LAMOST\_MRS-Gaia-TESS star sample \label{tab: MRS single catalog}}
\tablehead{
\colhead{Column} & \colhead{Name} & \colhead{Format} & \colhead{Units} & \colhead{Description}
}
\tabletypesize{\small}
\tablecolumns{5}

\startdata
\multicolumn{5}{c}{Parameters obtained from Gaia, LAMOST and TESS} \\
\hline
1 & Gaia DR3 ID & Long & & Unique Gaia source identifier \\
2 & LAMOST ID & String & & LAMOST unique spectral ID \\
3 & TIC & Integer & & TESS Input Catalog (TIC) ID \\
4 & Gaia RA & Double & deg & Barycentric right ascension \\
5 & Gaia Dec & Double & deg & Barycentric Declination \\
6 & Gaia parallax & Double & mas & Absolute stellar parallax \\
7 & Gaia $e\_\text{parallax}$ & Double & mas & Standard error of parallax \\
8 & Gaia pmra & Double & mas yr$^{-1}$ & Proper motion in right ascension direction \\
9 & Gaia $e\_\text{pmra}$ & Double & mas yr$^{-1}$ & Standard error of proper motion in right ascension direction \\
10 & Gaia pmdec & Double & mas yr$^{-1}$ & Proper motion in declination direction\\
11 & Gaia $e\_\text{pmdec}$ & Double & mas yr$^{-1}$ & Standard error of proper motion in declination direction \\
12 & Gaia $\mathit{G}$ mag & Double & & Gaia G band apparent magnitude \\
13 & TESS mag & Double & & TESS apparent magnitude \\
14 & Radius & Double & $R_\odot$ & Stellar radius from TESS \\
15 & $e\_radius$ & Double & $R_\odot$ & Uncertainty of stellar radius \\
16 & Mass & Double & $M_\odot$ & Stellar mass from TESS \\
17 & $e\_mass$ & Double & $M_\odot$ & Uncertainty of stellar mass \\
18 & $T_{\text{eff}}$ & Float & K & Effective temperature from LAMOST \\
19 & $e\_{T_{\text{eff}}}$ & Float & K & Error of effective temperature \\
20 & $\log g$ & Float & & Surface gravity from LAMOST \\
21 & $e\_{\log g}$ & Float & & Error of surface gravity \\
22 & $[\mathrm{Fe}/\mathrm{H}]$ & Float & dex & Metallicity from LAMOST \\
23 & $e\_[\mathrm{Fe}/\mathrm{H}]$ & Float & dex & Error of Fe element abundance \\
24 & $[\alpha/\mathrm{Fe}]$ & Float & dex & Alpha abundance from LAMOST \\
25 & $e\_{[\alpha/\mathrm{Fe}]}$ & Float & dex & Error of alpha elements abundance\\
26 & $[\mathrm{C}/\mathrm{Fe}]$ & Float & dex & Element abundance of C from LAMOST \\
    27 & $[\mathrm{N}/\mathrm{Fe}]$ & Float & dex & Element abundance of N from LAMOST \\
    28 & $[\mathrm{O}/\mathrm{Fe}]$ & Float & dex & Element abundance of O from LAMOST \\
    29 & $[\mathrm{Mg}/\mathrm{Fe}]$ & Float & dex & Element abundance of Mg from LAMOST \\
    30 & $[\mathrm{Al}/\mathrm{Fe}]$ & Float & dex & Element abundance of Al from LAMOST \\
    31 & $[\mathrm{Si}/\mathrm{Fe}]$ & Float & dex & Element abundance of Si from LAMOST \\
    32 & $[\mathrm{S}/\mathrm{Fe}]$ & Float & dex & Element abundance of S from LAMOST \\
    33 & $[\mathrm{Ca}/\mathrm{Fe}]$ & Float & dex & Element abundance of Ca from LAMOST \\
    34 & $[\mathrm{Ti}/\mathrm{Fe}]$ & Float & dex & Element abundance of Ti from LAMOST \\
    35 & $[\mathrm{Cr}/\mathrm{Fe}]$ & Float & dex & Element abundance of Cr from LAMOST \\
    36 & $[\mathrm{Ni}/\mathrm{Fe}]$ & Float & dex & Element abundance of Ni from LAMOST \\
    37 & $[\mathrm{Cu}/\mathrm{Fe}]$ & Float & dex & Element abundance of Cu from LAMOST \\
    38 & RV source flag & Float &  & RV source described in Section \ref{sec: section 6. caveats} \\
    39 & $v_{\mathrm{r}}$ & Double & km s$^{-1}$ & Radial velocity from LAMOST \\
    40 & $e\_{v_{\mathrm{r}}}$ & Double & km s$^{-1}$ & Error of radial velocity \\
    41 & TOI Flag & Boolean & & True if the star hosts TOI  \\
    42 & $N_{\mathrm{p}}$ & Integer & & Planet (candidate) multiplicity \\
    43 & Full TOI ID & String & & Unique TESS Object of Interest identifier \\
    44 & TESS Disposition & String & & Planet disposition from TESS\\
    45 & $M_{\mathrm{p}}$ & String & $M_\oplus$ & Planet mass \\
    46 & Period & String & days & Planet orbital period \\
    47 & $e\_{\text{period}}$ & String & days & Error of orbital period \\
    48 & $R_{\text{p}}$ & Double & $R_\oplus$ & Planet radius \\
    49 & $e\_{R_{\text{p}}}$ & Double & $R_\oplus$ & Error of planet radius \\
    50 & Temp & Double & K & Planet equilibrium temperature \\
\hline
\multicolumn{5}{c}{Parameters derived} \\
\hline
51 & $R$ & Double & kpc & Galactocentric Cylindrical radial distance \\
    52 & $\theta$ & Double & deg & Galactocentric Cylindrical azimuth angle \\
    53 & $Z$ & Double & kpc & Galactocentric Cylindrical vertical height \\
    54 & $U_{\mathrm{LSR}}$ & Double & km s$^{-1}$ & Cartesian Galactocentric x-velocity to the LSR \\
    55 & $e\_{U_{\mathrm{LSR}}}$ & Double & km s$^{-1}$ & Error of Cartesian Galactocentric x-velocity to the LSR \\
    56 & $V_{\mathrm{LSR}}$ & Double & km s$^{-1}$ & Cartesian Galactocentric y-velocity to the LSR \\
    57 & $e\_{V_{\mathrm{LSR}}}$ & Double & km s$^{-1}$ & Error of Cartesian Galactocentric y-velocity to the LSR \\
    58 & $W_{\mathrm{LSR}}$ & Double & km s$^{-1}$ & Cartesian Galactocentric z-velocity to the LSR \\
    59 & $e\_{W_{\mathrm{LSR}}}$ & Double & km s$^{-1}$ & Error of Cartesian Galactocentric z-velocity to the LSR \\
    60 & $L_X$ & Double & kpc km s$^{-1}$ &  Galactocentric x-angular momentum \\
    61 & $L_Y$ & Double & kpc km s$^{-1}$ & Galactocentric y-angular momentum \\
    62 & $L_Z$ & Double & kpc km s$^{-1}$ & Galactocentric z-angular momentum \\
    63 & $J_R$ & Double & kpc km s$^{-1}$ & Radial orbital action \\
    64 & $J_Z$ & Double & kpc km s$^{-1}$ & Vertical  orbital action \\
    65 & $t_{J_Z}$ & Double & Gyr & kinematic age derived from $J_Z$ using \textit{zoomies} \\
    66 & $t^{Up}_{J_Z}$ & Double & Gyr & 1-$\sigma$ upper limit of kinematic age derived from $J_Z$ using \textit{zoomies} \\
    67 & $t^{Lower}_{J_Z}$ & Double & Gyr & 1-$\sigma$ lower limit of kinematic age derived from $J_Z$ using \textit{zoomies} \\
    68 & $TD / D$ & Double & & Thick disc to thin disc membership probability \\
    69 & $TD / H$ & Double & & Thick disc to halo membership probability \\
    70 & $Herc / D$ & Double & & Hercules stream to thin disc membership probability \\
    71 & $Herc / TD$ & Double & & Hercules stream to thick disc membership probability \\
    72 & Component & String & & Classification of Galactic components \\ \hline
\enddata
\tablecomments{Table \ref{tab: MRS single catalog} is published in its entirety in the machine-readable format.
      A column description is shown here for guidance regarding its form and content.}
\end{deluxetable*}

\begin{figure}[ht]
    \centering
    \includegraphics[width=\linewidth]{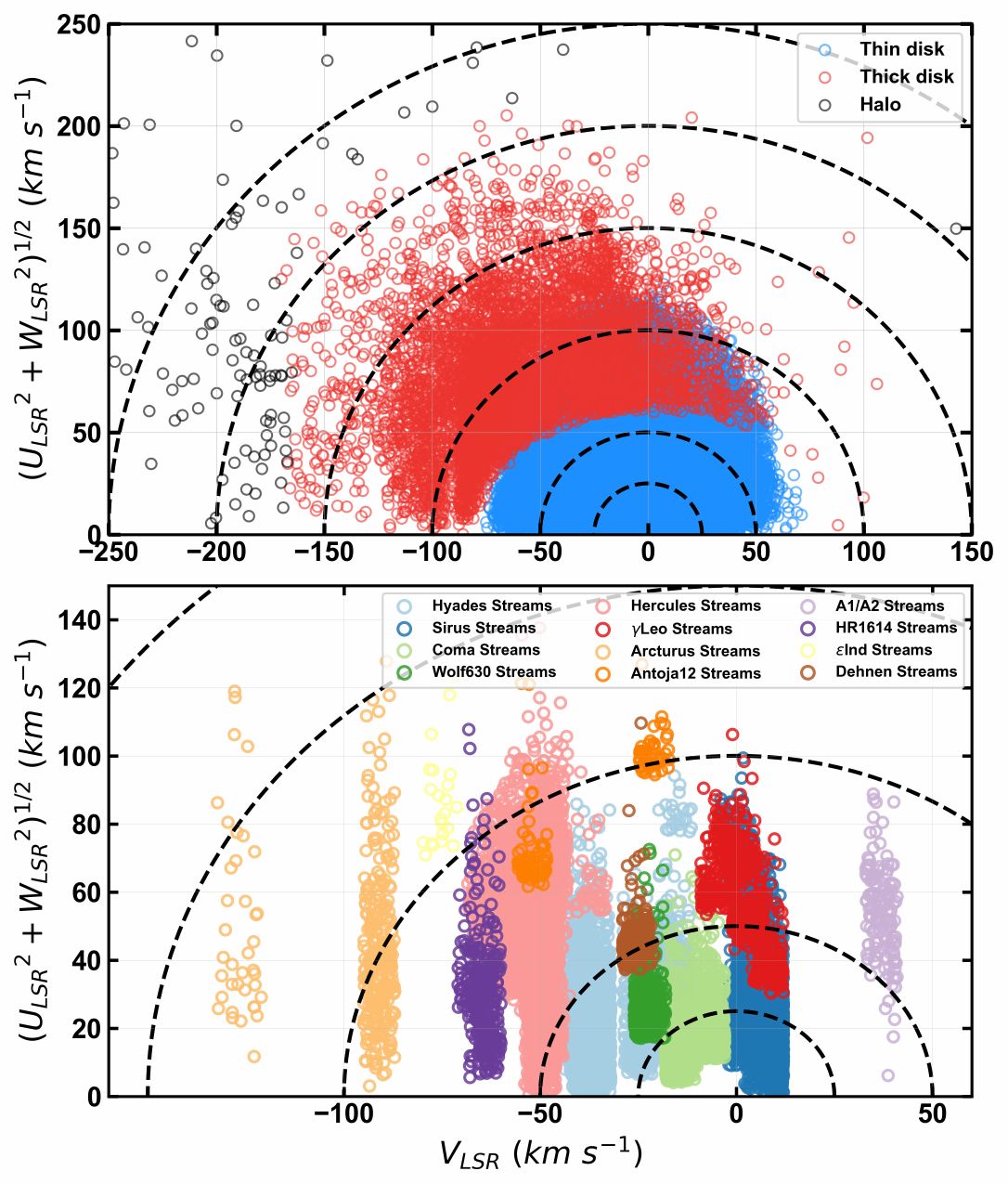}
    \caption{The Toomre diagram of LAMOST\_MRS-Gaia-TESS stars for different Galactic components (top panel) and stellar streams (bottom panel). Different color represents different memberships. Dotted lines indicate constant values of the total Galactic velocity $V_{tot}$ in stpdf of 25 $\rm km\ s^{-1}$ and 50 $\rm km\ s^{-1}$.}
    \label{fig: MRS toomre}
\end{figure}

\begin{figure}[t]
  \centering
  \includegraphics[width=\linewidth]{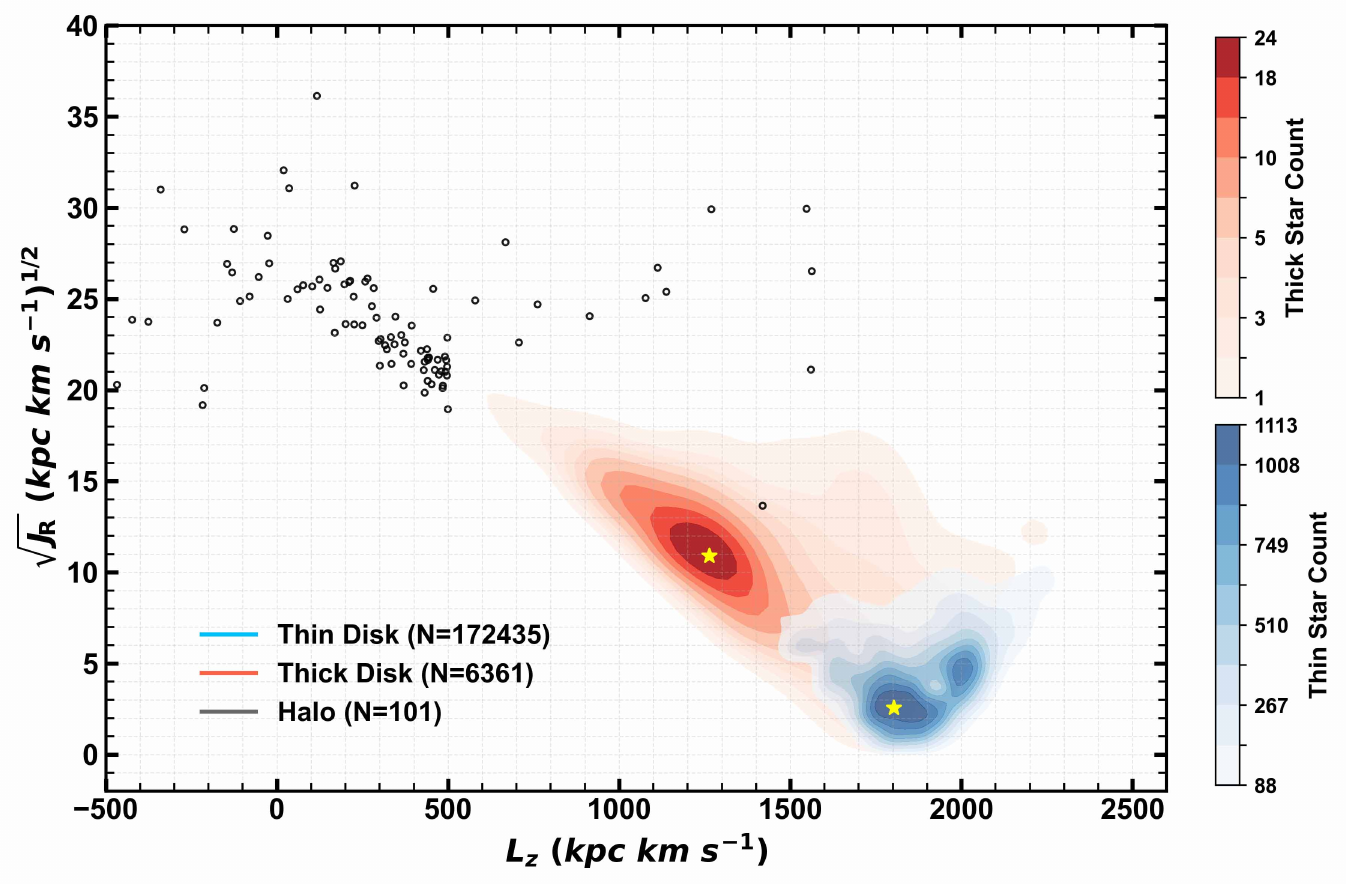}
  \caption{The distribution of disk stars in the vertical angular momentum \(L_z\) vs. radial action \(\sqrt{J_\text{R}}\) plane for the LAMOST\_MRS-Gaia-TESS sample, with the thin disk (\(N=172,435\)), thick disk (\(N=6,361\)) and halo (\(N=101\)) populations shown in blue, red, and black, respectively. Contour density maps illustrate the stellar distributions, with darker regions indicating higher star densities, the peak density location for each component is marked with a yellow star. Colorbars for each component are scaled separately since their sample sizes differ significantly, the ticks show an estimated number of sample in each grid. As shown, different components occupy distinct regions in the $L_z$-$J_R$ plane, reflecting their kinematic differences.}
  \label{fig: Lz disks MRS}
\end{figure}

\begin{figure}[t]
  \centering
  \includegraphics[width=\linewidth]{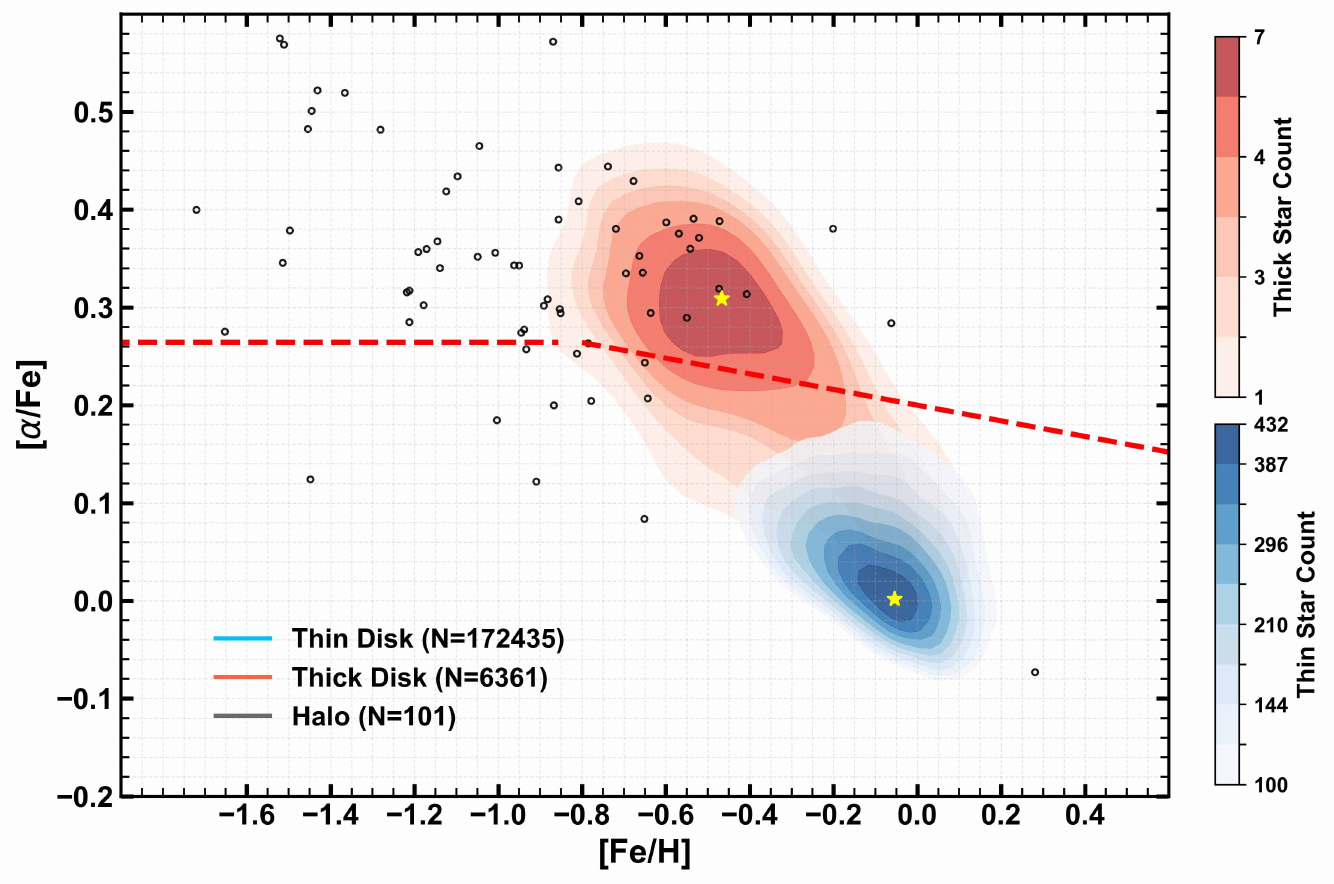}
  \caption{Same as Figure \ref{fig: Lz disks MRS} but for the distribution  in the \([\alpha/\text{Fe}]\) vs. \([\text{Fe}/\text{H}]\) plane for the LAMOST\_MRS-Gaia-TESS sample. The red dashed line represents the division between the thick and thin disks from \cite{leeFORMATIONEVOLUTIONDISK2011}, the distribution reveals distinct chemical abundance trends among these populations.}
  \label{fig: Fe/H vs alpha/Fe disks MRS}
\end{figure}

\begin{figure}
    \centering
    \includegraphics[width=\linewidth]{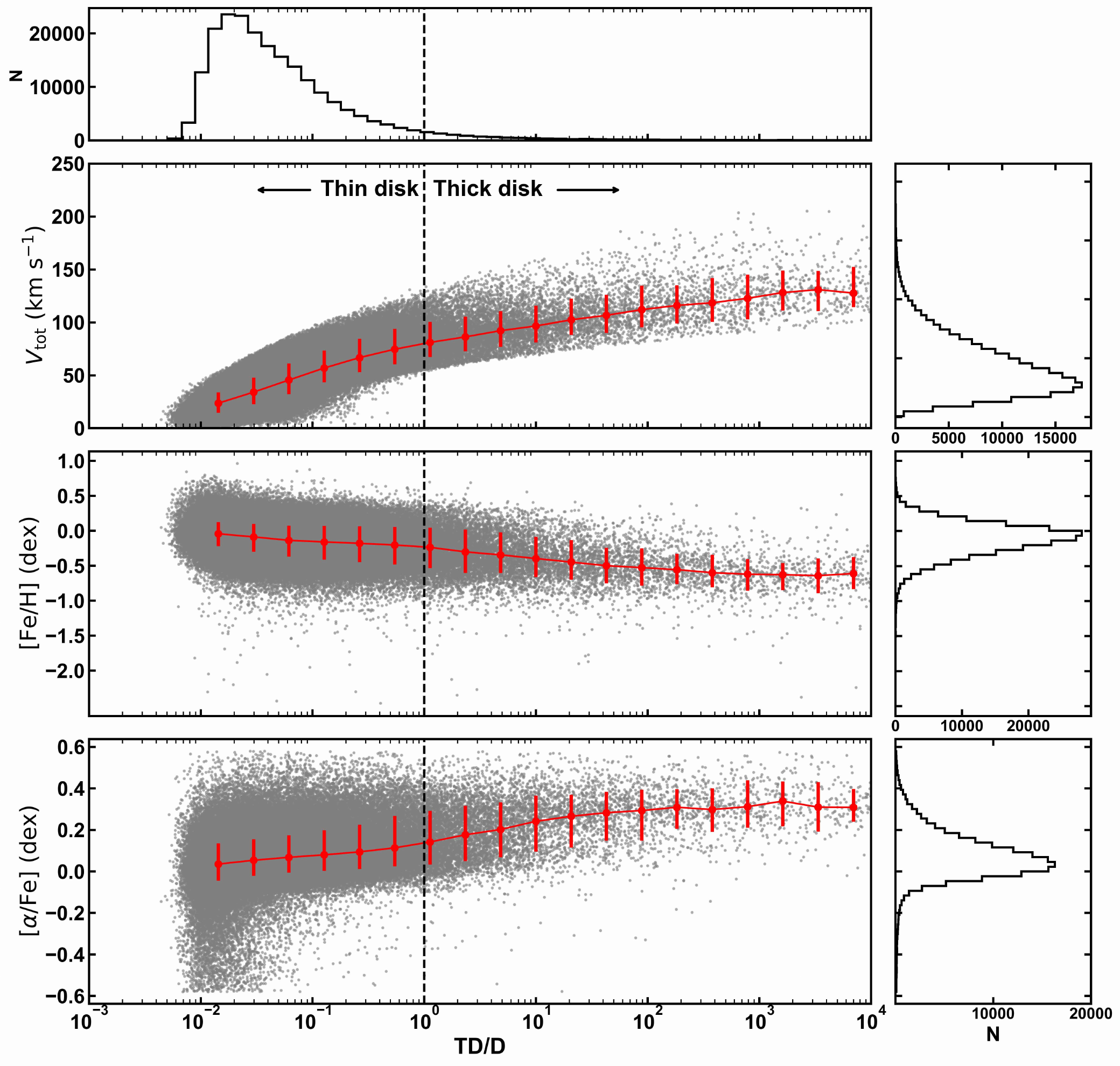}
    \caption{Kinematic and chemical properties (from top to bottom: total velocity $V_{\rm tot}$, [Fe/H] and [$\alpha$/Fe]) as functions of the relative probability between thick disk to thin disk (TD/D) for the LAMOST\_MRS-Gaia-TESS stars. The vertical dashed lines indicate where TD/D = 1. Medians and 1$\sigma$ dispersion for different TD/D groups are marked in the plots. Histograms of TD/D and kinematic/chemical properties are displayed in the topmost and right panels, respectively.}
    \label{fig: MRS chemical_TDD}
\end{figure}

Based on their velocity, angular momentum, and action, we identify stars associated with nearby stellar streams by adopting the $U-V$, $V-\sqrt{U^2+2V^2}$, $L_Z-\sqrt{L_X^2+L_Y^2}$ and $L_Z-J_R$ planes characteristics according to \cite{2019A&A...631A..47K}. The remaining stars are then classified into four Galactic components, i.e., thin disk, thick disk, Hercules stream and halo based on the calculated relative membership probabilities between different Galactic components, following the previously mentioned criteria. For stars not assigned to any of these components, we place them in a category termed “in between,” following the classification in \cite{bensbyExploringMilkyWay2014}.

Table \ref{tab: properties of LAMOST-Gaia-TESS star sample.} summarizes the number of stars and planets, as well as the kinematic properties and chemical abundances of different Galactic components for the MRS samples (See Table \ref{tab: properties of LAMOST_LRS-Gaia-TESS star sample.} in the Appendix \ref{appedndix:LRS} for the LRS sample).
As can be seen, for the MRS sample, approximately 83.0\% (172,435/207,690) of the stars belong to the thin disk, while approximately 3.1\% (6,361/207,690) are in the thick disk. The fraction of stars belonging to the halo is about 0.05\% (101/207,690). Additionally, $\sim 8.0\%$ (16,551/207,690) of the stars are classified into 12 distinct stellar streams.

To illustrate the kinematic properties, we plot the Toomre diagram and the vertical angular momentum ($L_Z$)-radial action ($\sqrt{J_\text{R}}$) distributions of the LAMOST\_MRS-Gaia-TESS stars in Figure \ref{fig: MRS toomre} and Figure \ref{fig: Lz disks MRS}, respectively. As shown, most stars in the thin disk have low velocities $\left(V_{\text{tot}} \lesssim 70 \mathrm{~km}~\mathrm{s}^{-1}\right)$, high vertical angular momentum ($L_Z \sim 1500-2500 \rm ~kpc\ km\ s^{-1}$) and low radial action ($\sqrt{J_\text{R}} \lesssim 5 \rm ~kpc^{1/2}\ km^{1/2}\ s^{-1/2}$)  while thick disk stars mainly exhibit moderate velocities $\left(V_{\text{tot}} \sim 70-180 \mathrm{~km}~\mathrm{s}^{-1}\right)$, vertical angular momentum in the range $L_Z \sim 900-2000 \rm ~kpc\ km\ s^{-1}$ and elevated radial action ($\sqrt{J_\text{R}} \sim 5-20 \rm ~kpc^{1/2}\ km^{1/2}\ s^{-1/2}$). Halo stars typically have velocities exceeding $200 \mathrm{~km}~\mathrm{s}^{-1}$, along with the lowest vertical angular momentum of $L_Z \lesssim 500 \rm ~kpc \ km\ s^{-1}$ and highest radial action of $\sqrt{J_\text{R}} > 20 \rm ~kpc^{1/2}\ km^{1/2}\ s^{-1/2} $.

\begin{figure*}[ht]
  \centering
  \includegraphics[width=\textwidth]{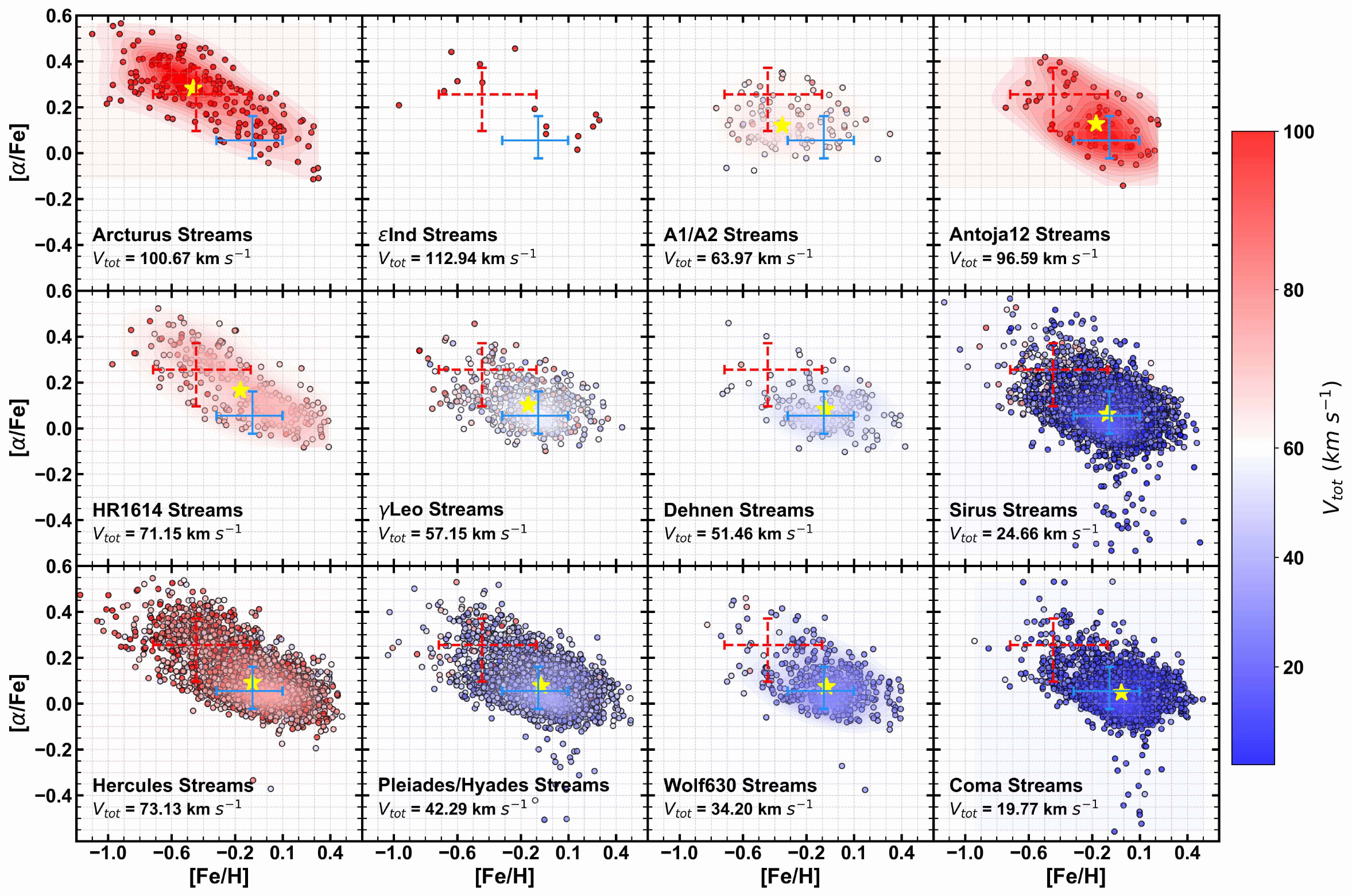}
  \caption{Distribution and number density of stellar streams in the [Fe/H] vs. [$\alpha$/Fe] plane for LAMOST\_MRS-Gaia-TESS sample. Each subplot represents a unique stellar stream, with data points color-coded by $V_{\rm tot}$ and arranged from top-left to bottom-right according to [Fe/H]. Contour lines indicate stellar density, where darker shades represent higher densities, while a yellow star denotes the typical value of [Fe/H] for each stream. The two error bars represent the values for thin disk stars (blue solid line) and thick disk stars (red dashed line) for reference.}
  \label{fig: Streams MRS}
\end{figure*}

The distribution of chemical abundances of the disk stars is presented in Figure \ref{fig: Fe/H vs alpha/Fe disks MRS}, along with their number density. As shown, thick disk stars are more metal-poor ($\rm [Fe/H] \sim-0.45$) and alpha-enhanced ($\rm [\alpha/Fe] \sim0.26$) compared to thin disk stars. These two populations are clearly distinguishable in the [Fe/H]-[$\alpha$/Fe] plane, as previously suggested by \cite{leeFORMATIONEVOLUTIONDISK2011} using a G-dwarf sample from SEGUE.
For halo stars, as expected, they exhibit the highest $[\alpha/Fe]$ values, reaching up to $\sim$ 0.4 dex (see Table \ref{tab: properties of LAMOST-Gaia-TESS star sample.}). Their metallicities span a wide range, from [Fe/H] of –2 dex to –0.2 dex, and the population can be broadly divided into metal-poor ([Fe/H]$<$–1) and metal-rich ([Fe/H]$>$–1) subgroups, which is consistent with previous studies that combined Gaia data with Apogee, RAVE, and LAMOST surveys \citep{2017ApJ...845..101B,2018A&A...615A..70P,2019MNRAS.487.1462F,2020MNRAS.494.3880B,2020ApJ...903..131Y,2021ApJ...915....9Z}. As discussed in earlier studies, metal-rich halo stars may have formed in-situ within the Galactic disk and were later dynamically heated during merger events, altering their orbits to exhibit halo-like motion. In contrast, the canonical metal-poor halo stars are generally believed to have been largely accreted from satellite galaxies \citep{2009ApJ...696..620C,2020ApJ...903..131Y}. In Figure \ref{fig: MRS chemical_TDD}, we plot the total velocity $V_{\rm tot}$, $\rm [Fe/H]$, and $\rm [\alpha/Fe]$ as a function of $TD/D$. As shown, with the increasing of $TD/D$, $V_{\rm tot}$ and $\rm [\alpha/Fe]$ increase, while $\rm [Fe/H]$ decreases.

Moreover, Figure \ref{fig: Streams MRS} displays the $\rm [Fe/H]-\rm [\alpha/Fe]$ distributions of stars in the 12 streams, color-coded by their $V_{\rm tot}$. 
Stars in a given stream exhibit similar total velocities ($V_{\rm tot}$), vertical angular momenta ($L_Z$), and radial orbital actions ($J_R$) (see Table \ref{tab: properties of LAMOST-Gaia-TESS star sample.}), suggesting that they share a common origin. 
As seen in Figure \ref{fig: Streams MRS} and Table \ref{tab: properties of LAMOST-Gaia-TESS star sample.}, from the chemo-kinematics, we find that the Coma stream is metal-richer and alpha-poorer, with lower velocities ($V_{\rm tot}$) and orbital actions ($J_R$, $J_Z$) compared to the thin disk.
The chemical and kinematic properties of the Sirius and Wolf630 streams are similar to those of the thin disk. For the Hercules, Pleiades/Hyades, Dehnen, and HR 1614 streams, their velocities, vertical angular momentums ($L_Z$), orbital actions and chemical abundances all lie between those of the thin and thick disks, which is consistent with previous studies \citep[e.g.,][]{bensbyExploringMilkyWay2014,2020A&A...638A.154K} and indicates that they are likely mixtures of the thin and thick disks. The Arcturus and $\epsilon$Ind streams in our sample have $V_{\rm tot} \sim 100 \ \rm km\ s^{-1}$, low angular momentum, and high orbital actions, with chemical compositions similar to the thick disk. 
We also notice that stars in the A1/A2 and $\gamma$ Leo streams have the highest $L_Z$ (similar to very thin disk stars), but their other properties do not match those of thin disk stars.

\begin{figure}
    \centering
    \includegraphics[width=1\linewidth]{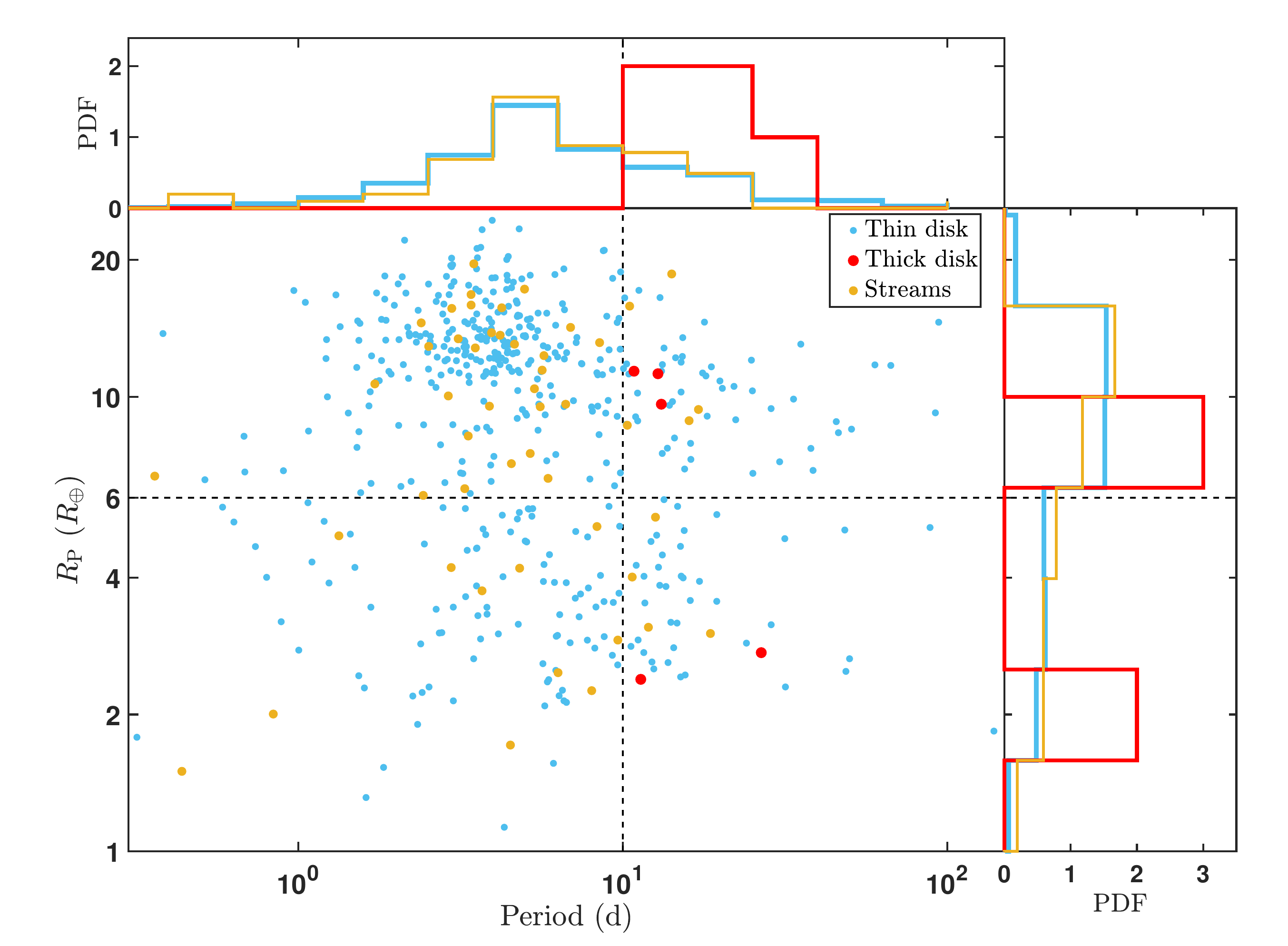}
    \caption{Planetary radius vs. orbital period for planets hosted by stars in different Galactic components. Planet candidates from both the LRS and MRS samples are shown. Different colors denote the Galactic components of the host stars, with planets in the thin disk, thick disk, and streams represented in blue, red, and orange, respectively. Two dashed lines highlight the classification criteria for different planet types: an orbital period of 10 days and a planetary radius of 6 $R_\oplus$. Histograms on the sides display the probability density functions (PDF) of planets in different components.}
    \label{fig: planet P-R}
\end{figure}

For the planetary sample, there are 544 TESS planet candidates in the LAMOST-Gaia-TESS catalog, with 321 in the LRS and 223 in the MRS. As shown in Figure \ref{fig: fig 4. distribution of mrs sample}, most planet candidates in our sample are located within the region of $R \sim 8.0-8.5$ kpc and $|Z|<0.5$ kpc. Figure \ref{fig: planet P-R} illustrates the period-radius distributions of these planet candidates from both the LRS and MRS samples across different Galactic components. As can be seen, the majority (84.7\%, 461/544) of TESS planet candidates are hosted by thin disk stars. Additionally, 0.9\% (5/544) and 9.6\%(52/544) of TESS planet candidates are found orbiting stars in the thick disk and stellar streams, respectively, while no planets are detected around halo stars. Without considering observational biases, the number ratio of planets over stars in the thin disk (0.27\%, 461/172,435) appears to be higher by a factor of three than that in the thick disk (0.08\%, 5/6,361). Most planets in the thin disk exhibit short orbital periods ($P<10$ days), with the majority being hot Jupiters, which is consistent with observational selection efficiency. In contrast, for the thick disk stars, 5 warm planets are found, but no hot Jupiters are detected. This indicates that hot Jupiters are unlikely to form or survive around thick disk stars, which is consistent with previous studies \citep{2023PNAS..12004179C}: Thick disk stars are metal-poor with a low frequency of hot Jupiters \citep{2010PASP..122..905J,2017ApJ...838...25G} and are old enough that some hot Jupiters may have undergone tidal orbital decay and disruption \citep{2009ApJ...698.1357J,2009ApJ...692L...9L}.

\begin{table*}[ht!]
    \caption{Kinematic and chemical properties of different Galactic components for the LAMOST-Gaia-TESS sample.}
    \centering
    \begin{tabular}{c|cccccccccc}
    \hline Components & $N_\mathrm{s}$ & $N_\mathrm{p}$ & $V_\mathrm{tot}$ & $[\mathrm{Fe}/\mathrm{H}]$ & $[\alpha/\mathrm{Fe}]$ & $L_Z$ & $\sqrt{J_R}$ & $\ln(J_Z)$ & {$\rm Age_{\rm kin}$} & $\rm Age_{\rm zoomies}$ \\
    \hline
    \multicolumn{11}{c}{MRS} \\ \hline
    Thin disk & 172,435 & 189  & $35.0_{-15.6}^{+22.4}$ & $-0.09_{-0.22}^{+0.18}$ & $0.05_{-0.07}^{+0.11}$ & $1858^{+162}_{-160}$ & $4.3^{+3.0}_{-2.2}$ & $1.1^{+1.2}_{-1.3}$ &  $2.60_{-0.15}^{+0.16}$ & $1.41^{+2.37}_{-1.41}$\\
    Thick disk & 6,361 & 0 & $101.2_{-18.8}^{+29.2}$ & $-0.42_{-0.28}^{+0.34}$ & $0.26_{-0.16}^{+0.11}$ & $1346^{+307}_{-208}$ & $11.3^{+3.6}_{-4.3}$  & $3.3^{+0.8}_{-1.8}$ & $13.04_{-1.56}^{+1.84}$ & $9.73^{+4.27}_{-8.13}$\\
    Halo & 101 & 0 & $238.54_{-26.99}^{+63.90}$ & $-0.94_{-0.73}^{+0.38}$ & $0.34_{-0.07}^{+0.11}$ & $320^{+189}_{-343}$ & $23.9^{+4.5}_{-2.7}$ & $3.8^{+1.4}_{-1.7}$ & \text{NA} & $10.25^{+3.75}_{-7.80}$ \\
    Sirius & 5,569 & 5 & $24.8_{-9.3}^{+11.1}$ & $-0.10_{-0.20}^{+0.17}$ & $0.06_{-0.07}^{+0.11}$ & $1927^{+27}_{-34}$ & $3.7^{+0.7}_{-0.5}$ & $1.4^{+1.2}_{-1.3}$ & NA & $2.20^{+4.09}_{-2.20}$ \\
    Hercules & 3,532 & 7 & $73.0_{-17.9}^{+11.4}$ & $-0.10_{-0.29}^{+0.25}$ & $0.09_{-0.09}^{+0.17}$ & $1466^{+45}_{-61}$ & $8.7^{+1.8}_{-2.0}$ & $1.8^{+0.9}_{-1.1}$ & NA & $2.79^{+3.93}_{-2.79}$ \\
    Pleiades/Hyades & 3,162 & 1 & $42.2_{-5.9}^{+8.3}$ & $-0.08_{-0.25}^{+0.21}$ & $0.08_{-0.09}^{+0.12}$ & $1568^{+89}_{-37}$ & $4.8^{+1.5}_{-1.0}$ & $1.6^{+1.1}_{-1.2}$ & NA & $2.76^{+4.15}_{-2.76}$ \\
    Coma & 2,129 & 8 & $19.8_{-3.4}^{+7.3}$ & $-0.02_{-0.21}^{+0.18}$ & $0.05_{-0.08}^{+0.11}$ & $1793^{+46}_{-54}$ & $1.7^{+0.7}_{-0.5}$ & $0.8^{+1.4}_{-1.2}$ & NA & $1.27^{+1.70}_{-1.27}$\\
    $\gamma$ Leo & 591 & 0 & $57.7_{-17.1}^{+16.0}$ & $-0.15_{-0.26}^{+0.22}$ & $0.10_{-0.09}^{+0.11}$ & $1900^{+50}_{-47}$ & $7.1^{+2.0}_{-1.3}$ & $2.0^{+1.0}_{-1.3}$ & NA & $4.29^{+5.56}_{-4.29}$ \\
    Wolf630 & 517 & 2 & $34.2_{-3.2}^{+5.4}$ & $-0.06_{-0.24}^{+0.20}$ & $0.07_{-0.08}^{+0.11}$ & $1696^{+27}_{-41}$ & $3.2^{+0.5}_{-0.5}$ & $1.2^{+1.2}_{-1.2}$ & NA & $1.66^{+2.43}_{-1.66}$ \\
    HR1614 & 318 & 2 & $71.3_{-6.3}^{+7.1}$ & $-0.18_{-0.33}^{+0.34}$ & $0.18_{-0.15}^{+0.16}$ & $1347^{+27}_{-32}$ & $8.8^{+0.7}_{-0.6} $ & $2.1^{+1.1}_{-1.3}$ & NA & $4.72^{+9.28}_{-4.72}$ \\
    Arcturus & 268 & 0 & $100.7_{-6.6}^{+24.7}$ & $-0.46_{-0.30}^{+0.38}$ & $0.28_{-0.15}^{+0.12}$ & $1118^{+27}_{-39}$ & $12.5^{+1.2}_{-0.8}$ & $2.5^{+1.4}_{-1.3}$  & NA & $6.31^{+7.69}_{-6.31}$ \\
    Dehnen & 212 & 0 & $51.35_{-2.9}^{+4.2}$ & $-0.11_{-0.21}^{+0.24}$ & $0.08_{-0.08}^{+0.10}$ & $1676^{+32}_{-28}$ & $5.5^{+0.6}_{-0.3}$ & $2.6^{+1.3}_{-1.0}$ & NA & $2.20^{+3.76}_{-2.20}$ \\
    A1/A2 & 139 & 0 & $64.9_{-9.3}^{+10.8}$ & $-0.35_{-0.18}^{+0.31}$ & $0.12_{-0.07}^{+0.10}$ & $2185^{+22}_{-41}$ & $10.9^{+1.3}_{-0.5}$ & $2.1^{+0.9}_{-1.3}$ & NA & $4.36^{+6.53}_{-4.36} $ \\
    Antoja12 & 94 & 0 & $96.5_{-12.2}^{+8.3}$ & $-0.18_{-0.28}^{+0.22}$ & $0.13_{-0.10}^{+0.13}$ & $1488^{+233}_{-51}$ & $9.9^{+1.6}_{-1.1}$ & $2.1^{+0.9}_{-1.2}$ & NA & $4.35^{+6.43}_{-4.35}$ \\
    $\epsilon$Ind & 20 & 0 & $111.2_{-4.4}^{+12.6}$ & $-0.40_{-0.27}^{+0.56}$ & $0.21_{-0.12}^{+0.19}$ & $1261^{+25}_{-31}$ & $12.7^{+0.3}_{-0.9}$ & $2.9^{+1.4}_{-1.4}$ & NA & $7.97^{+6.03}_{-7.97}$ \\
    \hline
    \end{tabular}
    \label{tab: properties of LAMOST-Gaia-TESS star sample.}
    \noindent
    {\raggedright \tablecomments{See Table \ref{tab: MRS single catalog} for the units of each column.}\par}
    
\end{table*}

\section{Catalogs of RV variables From LAMOST Multiple Epoch Observation}\label{sec: Section 5. catalog of RV variables}



Binary systems and variable stars can be identified using various methods. For example, by analyzing light curves, previous studies have identified eclipsing binaries by analyzing light curves based on Kepler and TESS data \citep{prsaKeplerEclipsingBinary2011,prsaTESSEclipsingBinary2022}. Through astrometric measurements, Gaia provides numerous wide binaries \citep{2021MNRAS.506.2269E}. 
In some cases, binary systems can also be detected directly from spectral features, such as double-lined spectroscopic binaries (SB2s), which exhibit two distinct sets of spectral lines \citep{2018MNRAS.476..528E,2020A&A...638A.145T}. On the other hand, analyzing time-domain RV data from spectroscopic surveys are particularly effective in identifying close binaries \citep{price-whelanBinaryCompanionsEvolved2018,price-whelanCloseBinaryCompanions2020,chenBinaryStarsNew2024}, especially for the systems remain unresolved in photometric or spectroscopic observations.



\begin{figure*}[ht]
  \centering
  \includegraphics[width=0.55\linewidth]{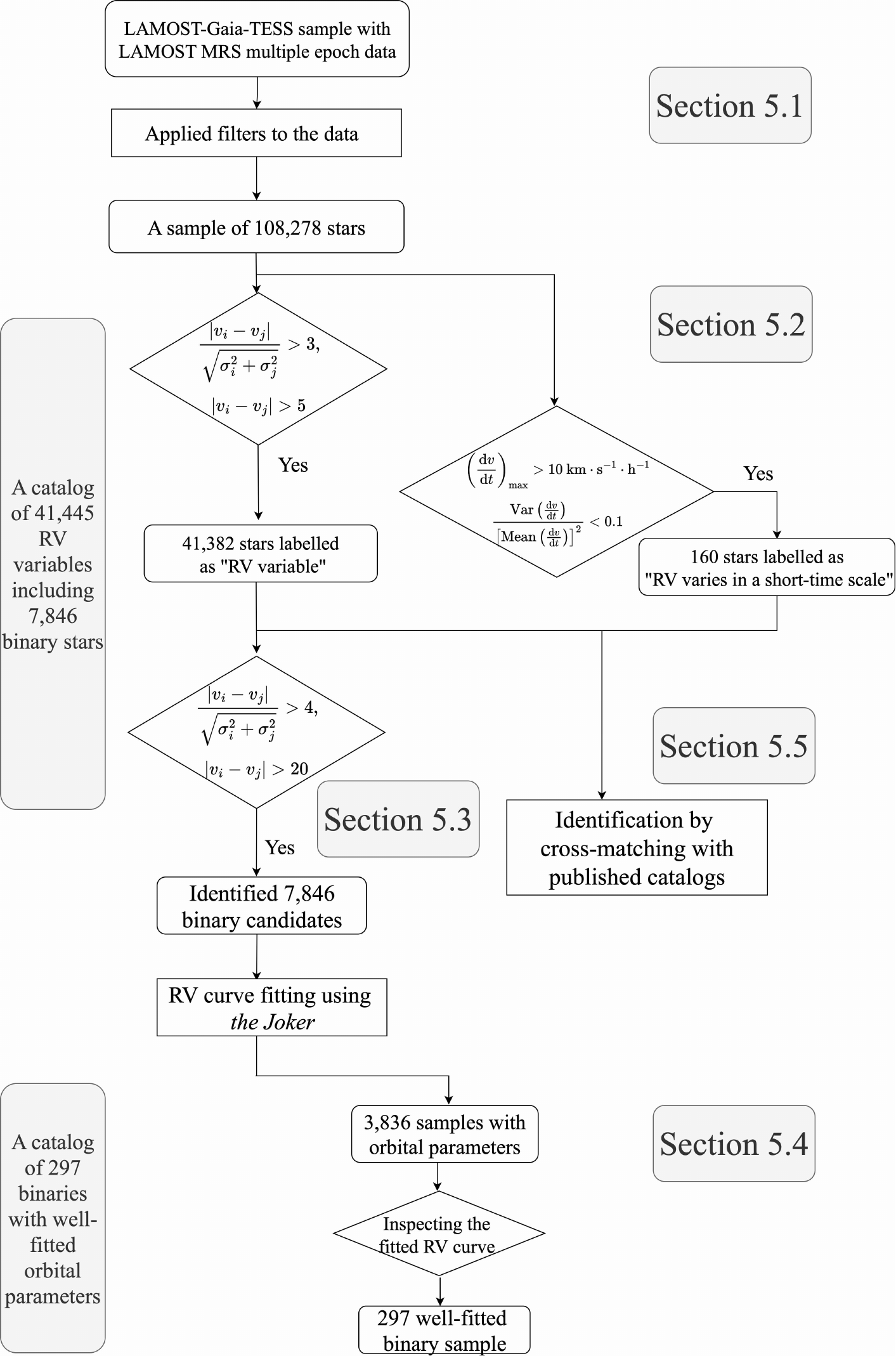}
  \caption{A summary flow chart of Section \ref{sec: Section 5. catalog of RV variables}, with the corresponding subsection shown to the right of each process.}
  \label{fig: flow chart}
\end{figure*}

In this Section, we search for RV variables within the LAMOST-Gaia-TESS catalog. Utilizing the time-domain RV data from the LAMOST MRS multiple epoch observations, we identify RV variable stars and binary stars, as presented in Table \ref{tab: sub catalog}.  For a subset of binary candidates, we further derive orbital parameters by fitting their radial velocity curves (see Table \ref{tab: sub catalog} and \ref{tab: 299 gold sample}). The flow chart of the whole procedure is illustrated in Figure \ref{fig: flow chart}.

\subsection{Crossmatch with LAMOST Multiple Epoch Data}\label{sec: subsec 5.1 crossmatch with LAMOST multiple epoch}


The LAMOST MRS multiple epoch catalog\footnote{\url{http://www.lamost.org/dr10/v1.0/catalogue}}, released as part of LAMOST MRS DR10, contains spectroscopic observations of 547,726 targets obtained over multiple observation nights. Each target in the catalog was typically observed with approximately 20 exposures spanning 5 observation nights. For each exposure of a given target, the catalog provides local modified Julian minute (lmjm), RV with associated errors, SNR and derived physical parameters (e.g., $T_{\text{eff}}$, $\log g$). 





We first applied the following filtering procedures to the LAMOST MRS multiple epoch catalog: (1) We removed exposures with invalid RV measurements and errors. (2) We excluded exposures with SNR $<$ 10, as the typical RV uncertainty for such measurements in the sample is approximately twice that of exposures with higher SNR, as indicated by \citet{xiangLAMOSTStellarParameter2015}. (3) Following our filtering process, some stars were left with fewer than two exposures across all observation nights.
For each target, we calculated the total number of observation nights ($N_{night}$) and the number of exposures per night (\(n_i\)), where $i$ denotes each observation night. To ensure reliability, we excluded observation nights with only one single exposure from the calculation of $N_{night}$, as such nights do not provide reliable information to assess RV variability. Stars with $N_{night}=0$ were excluded from the sample. (4) Following a similar procedure described in Section \ref{sec: section 2. Data collection}, we applied filters on $T_{\text{eff}}$, [Fe/H], and [$\alpha$/Fe] to ensure data quality and consistency. After applying these filters to the multiple epoch data, we cross-matched the remaining sample with Gaia and TESS catalogs, imposing a separation limit $<1.5$ arcseconds and a magnitude difference $<2$ in $\mathit{G}$-band magnitudes. Notably, we did not impose the RUWE threshold or RV consistency filter when constructing this sample, in order to retain potential RV variables and binary candidates. After applying these criteria, we obtained a final sample of 108,278 stars, including 114 TOI host stars.
Each star in the dataset was observed between 2 and 150 times, with a typical value of 13 exposures. These observations are distributed over a time span of 2 to 37 days, typically covering around 3.5 nights.



\subsection{Identifying the RV Variables}\label{sec: subsec 5.2 Identifying the RV variables}

The identification of binary stars in our sample is based on detecting significant RV variations across multiple epochs. RV variability has been demonstrated to be a reliable diagnostic for identifying RV variables and estimating binary fractions in plenty of previous studies \citep[][ additional references below]{2012ApJ...749L..11B,2012ApJ...751..143M,price-whelanBinaryCompanionsEvolved2018,price-whelanCloseBinaryCompanions2020,2022A&A...657A...4M}. One common approach classifies a star as a binary if its RV shows statistically significant variability, being assessed using $\chi^{2}$ probability that the RV is constant \citep{2002AJ....124.1144L,2003AJ....125..293C,2016A&A...586A.160H,2016A&A...588A...3H}, or assessed through comparing the maximum RV shift to measurement uncertainties and/or applying a minimum threshold \citep{sanaVLTFLAMESTarantulaSurvey2013,sanaBinaryInteractionDominates2012,2018ApJ...854..147B,2020MNRAS.499.1607M}. These methods can also derive the intrinsic binary fraction and has been adopted in large-scale analyses of various stellar populations \citep{dunstallVLTFLAMESTarantulaSurvey2015,2019ApJ...875...61M,luoBinaryFractionBtype2021,guoBinarityEarlytypeStars2022,2022A&A...667A..44G,2022MNRAS.512.2051D,2022ApJ...933L..18B}. Instead of identifying individual systems, another approach focuses on population-level inference of the overall binary fraction. It infers binary fraction statistically from the dispersion in RVs across multi-epoch observations \citep[e.g.,][]{2014ApJ...788L..37G,2017MNRAS.469L..68G,2018RAA....18...52T}. More recent work of \citet{2024A&A...692A.247B,2024MNRAS.535..949B,2025MNRAS.541.2008B} modeled the RV variation distribution as a mixture of single and binary populations, enabling binary fraction estimates without resolving each binary’s orbit. This approach enable efficient, homogeneous estimation of binary fractions across large samples.

In this work, we aim to construct a stellar catalog and characterize the kinematic properties and binarity of the sample, with the additional goal of deriving orbital parameters for selected binary systems. To this end, we focus on identifying binary stars on a star-by-star basis using the first approach described above. To identify RV variables and binary candidates, we adopted the method described by \citet{sanaBinaryInteractionDominates2012,sanaVLTFLAMESTarantulaSurvey2013}, which evaluates the significance of RV variations through Equation \ref{eq: rv criteria}. We conducted a Monte-Carlo simulation accounting for LAMOST’s observational cadence and measurement uncertainty to determine the threshold in Equation \ref{eq: rv criteria} by evaluating the completeness and purity (see Section \ref{sec: subsec 5.3 identify binary}). For the identified binaries, we fit their RV curves using \textit{The Joker} package \citep{price-whelanJokerCustomMonte2017,price-whelanBinaryCompanionsEvolved2018}, enabling us to obtain binary orbital properties and construct a catalog of well-characterized binaries (see Section \ref{subsec: section 5.4 fit binary}).

Specifically, a star is classified as an RV variable or a binary candidate if at least one pair of RV measurements satisfies both of the following criteria simultaneously:
\begin{equation}
\begin{aligned}
 \frac{\left|v_{i}-v_{j}\right|}{\sqrt{\sigma_{i}^{2}+\sigma_{j}^{2}}} &> A, \\
\left|v_{i}-v_{j}\right| &> C,
\end{aligned}
\label{eq: rv criteria}
\end{equation}where $v_{i}$ and $v_{j}$ are the RV measurements of the same star from two different exposures, and $\sigma_{i}$ and $\sigma_{j}$ are their uncertainties. 


The first criterion in Equation \ref{eq: rv criteria} evaluates the significance of RV variations, while the second criterion imposes a minimum amplitude threshold \(C\) to exclude small yet statistically significant variations. Here, we set thresholds of $A$ = 3 and $C$ = 5 $\mathrm{km\ s^{-1}}$ to ensure that the detected RV variations are statistically significant. These criteria help distinguish RV changes likely caused by binary motion or intrinsic stellar variability from those resulting from instrumental noise or measurement limitations. Following \citet{tianCatalogRVVariable2020}, we classify these stars as `RV variables’, which include potential binary systems as well as intrinsically variable stars. Applying the two criteria, we identified 41,382 RV variables in our sample.

Moreover, previous studies have shown that stellar activity can also produce significant RV variations. For instance, photospheric variations in supergiants can mimic RV variations with amplitudes of up to 20 $\mathrm{km\ s^{-1}}$ \citep{2009A&A...507.1585R}, and stellar wind effects can also contribute to such variations \citep{2016Natur.529..502L}. Therefore, the second criterion is particularly important when identifying binary stars, and the selection of threshold values $A$ and $C$ for binary identification will be further discussed in Section \ref{sec: subsec 5.3 identify binary}.

In addition to targets with observation coverage spanning multiple nights, 31,639 stars were observed on only a single night ($N_{\text{night}}$=1), accounting for approximately 30\% of the total sample. Some of these targets may display rapid RV variations over short periods, such as compact objects \citep{liuSampleCompactObject2024a}. To address this limitation, we introduced a new classification flag, `Short Time Scale RV Variable', specifically designed to capture significant RV variation within a short time span. We calculated the rate of RV change ($\mathrm{d}v/\mathrm{d}t$) between consecutive observations within a single night, along with the variance of these rates. Only nights with at least three exposures were considered, ensuring at least two $\mathrm{d}v/\mathrm{d}t$ measurements. 

Following \citet{liuSampleCompactObject2024a}, we select stars whose $\mathrm{d}v/\mathrm{d}t$ values satisfy the following conditions:

\begin{equation}
\label{eq: short time scale}
    \begin{aligned}
        &\left(\frac{\mathrm{d} v}{\mathrm{d} t}\right)_{\max} > 10 \mathrm{~km} \cdot \mathrm{s}^{-1} \cdot \mathrm{h}^{-1}, \\
        &\frac{\operatorname{Var}\left(\mathrm{dv/dt}\right)}{\left[\operatorname{Mean}\left(\mathrm{dv/dt}\right)\right]^{2}} < 0.1
    \end{aligned}
\end{equation}
   
Applying these criteria, we identified 160 stars in our sample as `Short Time Scale RV Variables', increasing the total number of RV variables in the LAMOST-Gaia-TESS catalog to 41,445.

\subsection{Identifying the Binaries}\label{sec: subsec 5.3 identify binary}
In order to further confirm the values for $A$ and $C$ (Equation \ref{eq: rv criteria}) for binaries, we employ a Monte Carlo approach to quantify their impacts on the detection purity (i.e, the proportion of identified binary candidates that are indeed true binary systems).

In the simulation, we generate a synthetic stellar population of $10^5$ systems, consisting of 45\% single stars and 55\% binaries, based on the median binary fraction from LAMOST \citep{tianCatalogRVVariable2020}. For the binary population, we assume the following distributions for key orbital parameters:
\begin{enumerate}
    \item \textbf{Mass Ratio ($q$) :} Follows a power-law distribution $p(q)\propto q^\gamma$, where the index $\gamma$ is $1.16 \pm 0.16$ for short period binaries with $\log P < 5.5$ and $-0.01 \pm 0.03$ for binaries with larger periods, as suggested by \cite{ducheneStellarMultiplicity2013}.
  
    \item \textbf{Orbital Period ($P$) :} Adopts a log-normal distribution characterized by a mean value of $\mu_{\log P}$ = 5.03 and a dispersion of $\sigma_{\log P}$ = 2.28 \citep{raghavanSurveyStellarFamilies2010}, with P expressed in days.
    \item \textbf{Eccentricity ($e$) :} Assumed to follow a uniform distribution in the range [0, 1], alternative distributions were tested and yielded similar results. 
\end{enumerate}

The synthetic populations are generated as follows:
\begin{enumerate}
    \item Each star in the synthetic population is assigned an observation list that includes the number of observation nights ($N_{\text{night}}$), the number of exposures per night ($n_i$), and the corresponding RV measurement errors from the observed dataset ($\sigma_i$).
    \item For the single stars, each is assigned an RV measurement along with the uncertainty from the multiple epoch catalog described in Section \ref{sec: subsec 5.1 crossmatch with LAMOST multiple epoch}. The RVs at the corresponding observation times are drawn from Gaussian distributions centered on their assigned RV measurements.
    
    \item For binary stars, the three-dimensional orbital orientation and the time of periastron passage are drawn randomly from 0 to 2$\pi$. The orbital period ($P$), mass ratio ($q$), and eccentricity ($e$) are also drawn randomly according to the distributions specified earlier.
    The RVs of the primary stars are then computed at the each observation epoch based on the orbital parameters using
    \begin{equation}
\begin{gathered}
    RV_0 = \sqrt[3]{\frac{2\pi G}{P(1-e^2)^{\frac{2}{3}}} \frac{(M \sin i)^3}{(M_* + M)^2} \cos\left(\omega + \frac{2\pi t}{P}\right)} \\
    RV \sim \mathcal{N}(RV_0, err_{RV})
\end{gathered}
\end{equation}
\end{enumerate}


We further evaluated the performance of our binary detection criteria on the mock sample by varying the thresholds (A, C) in Equation \ref{eq: rv criteria}. Specifically, we explored different cutoffs for $\Delta \text{RV} / \sigma_{\text{RV}}$, selecting values of 2, 3, and 4, as well as various values for the constant $C$, ranging from 5 to 20 $\mathrm{km\ s^{-1}}$. For each combination of (A, C), synthetic stars, regardless of whether single or binary, were flagged as binaries if they satisfied the criterion in Equation \ref{eq: rv criteria}. To minimize the effect of random sampling fluctuations, we repeated the above process for 10 times and adopted the average value of detection purity. The results, including the detection rate (i.e., the fraction of binary stars detected from the 55,000 true binary systems) and the corresponding purity, are summarized in Table \ref{tab: simulation result}.

Based on our simulation results, we adopt thresholds of $A = 4, C = 20$ $\mathrm{km\ s^{-1}}$ in Equation \ref{eq: rv criteria} to identify binary stars with higher purity while minimizing contamination from intrinsic stellar variability, as discussed in the previous section. These thresholds yield a 100\% purity in eliminating single star contamination (Table \ref{tab: simulation result}). However, it is important to note that this purity represents an upper bound, as we have not accounted for the effects of intrinsic stellar variability (e.g., eruptive stars, pulsating stars) that may affect the identification process. 

Additionally, we require that each star be observed on at least two separate nights ($N_{\text{night}} \geq 2$) to ensure that the RV measurements span enough of the orbital phases, enabling effective RV curve fitting (discussed later in Section \ref{subsec: section 5.4 fit binary}). Applying these stricter criteria, we classified 7,846 stars as binaries among the RV variables in our catalog.

\begin{figure}[!t]
    \centering
    \includegraphics[width=\linewidth]{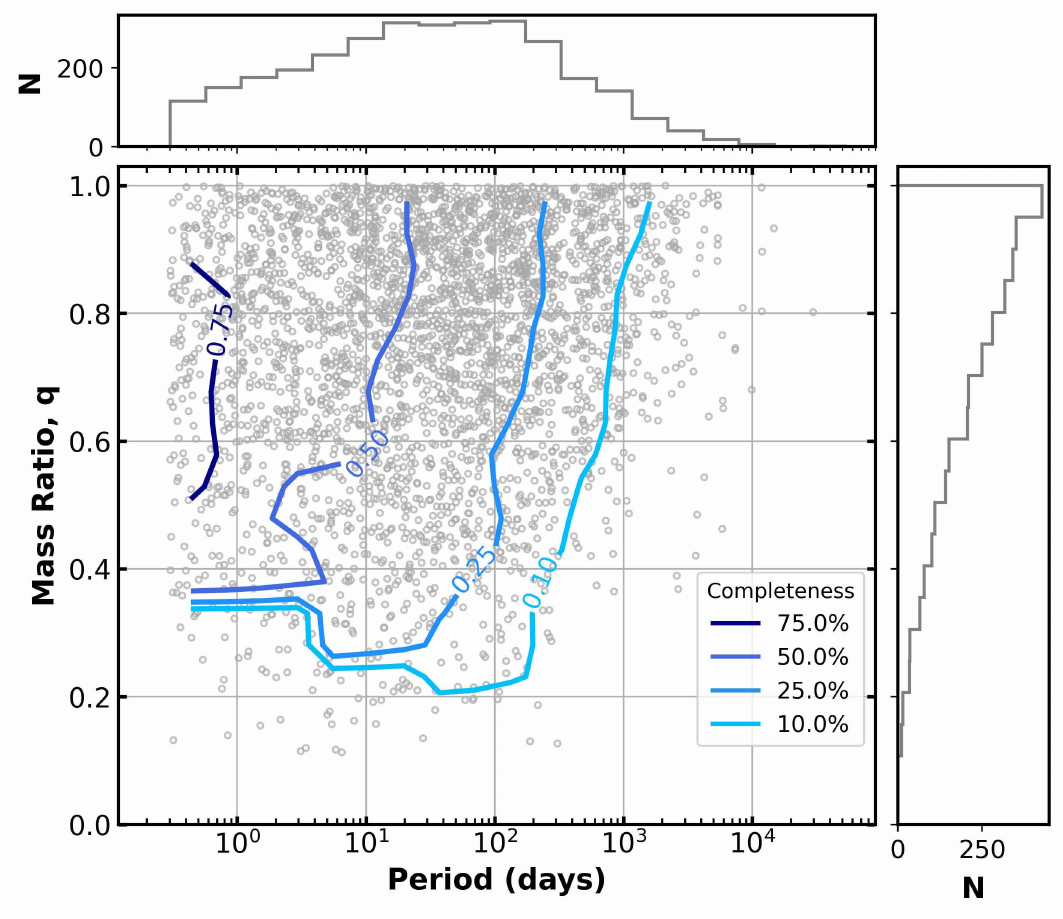}
    \caption{Distribution of binary mass ratio ($q$) and orbital period ($P$) for the simulated binaries which were retrieved by the detection method. Different detective completeness contour lines were plotted on the period-mass ratio plane. Histograms of $q$ and period were shown on the top and right side of the figure.}
    \label{fig: simulated q_p}
\end{figure}

Figure \ref{fig: simulated q_p} illustrates the detection efficiency and the distribution of mass ratio and orbital period for the detected binaries in the simulated sample, based on the adopted thresholds of A = 4, C = 20 $\mathrm{km\ s^{-1}}$. As shown, our detection method is primarily sensitive to binaries with orbital periods shorter than $10^3$ days, a limitation arising from the observational strategy and the time baseline of the LAMOST survey. The results highlight the survey’s inherent sensitivity toward detecting short-period binaries, which tend to exhibit more pronounced and periodic RV variations within the time span.

{\renewcommand{\arraystretch}{0.9}
\startlongtable
\begin{deluxetable*}{ccccl}
\tablecaption{The catalog of the LAMOST multiple epoch sample \label{tab: sub catalog}}
\tablehead{
\colhead{Column} & \colhead{Name} & \colhead{Format} & \colhead{Units} & \colhead{Description}
}
\tabletypesize{\footnotesize}
\startdata
\hline
\multicolumn{5}{c}{\small{\textbf{Parameters obtained from Gaia, LAMOST and TESS (41,445 rows)}}} \\
\hline
1 & Gaia DR3 ID & Long & & Unique Gaia source identifier \\
2 & LAMOST ID & String & & LAMOST unique spectral ID \\
3 & TIC & Integer & & TESS Input Catalog (TIC) ID \\
4-37& \vdots  &  & & Columns 4-37 from Table \ref{tab: MRS single catalog} \\
38 & $N_{\mathrm{p}}$ & Integer & & Planet (candidate) multiplicity \\
39 & Full TOI ID & String & & Unique TESS Object of Interest identifier \\
40 & LAMOST obsid list & String & & Spectrum ID for each observation of the target from LAMOST \\
41 & Exposure number & Float &  & Total exposure number for the target from LAMOST \\
42 & RV list & String & km s$^{-1}$ & Radial velocity list (`rv\_b1\_list' from LAMOST) \\
43 & e\_RV list & String & km s$^{-1}$ & Corresponding uncertainty of radial velocity \\
44 & SNR list & String &  & S/N for each exposure from LAMOST  \\
\hline
\multicolumn{5}{c}{\small\textbf{Parameters derived from multiple epoch data (41,445 rows)}} \\
\hline
45 & $N_{night}$ & Short & days & Number of effective observation nights \\
46 & $N_i$ & String & days & Number of exposures for each observation night \\
47& Delta exposure & Double & days & Time interval between exposures \\
48 & $T_{max}$ & Double & minutes & Time span for the target \\
49 & Delta\_RV\_max & Double & km s$^{-1}$ & Maximum RV difference for each target \\
50 & RV Variable & Boolean &  & True if satisfies the criteria for variable flag in Section \ref{sec: subsec 5.2 Identifying the RV variables}\\
51 & Short time scale RV variable & Boolean &  & True if satisfies the criteria for short time scale flag in Section \ref{sec: subsec 5.2 Identifying the RV variables} \\
52 & Binary & Boolean &  & True if satisfies the criteria for binary flag in Section \ref{sec: subsec 5.2 Identifying the RV variables} \\
\hline
\multicolumn{5}{c}{\small\textbf{Cross-match flag and results (41,445 rows)}} \\
\hline
53 & Gaia\_VCR\_flag & Boolean & & True if included in Gaia DR3 variabe catalog \\
54 & Class\_Gaia & String & & Gaia classification \\
55 & VSX\_flag & Boolean & & True if included in VSX \\
56 & Period\_VSX & Double & days &  Variable star period from VSX catalog \\
57 & Type\_VSX & String & & Variability type from VSX catalog \\
58 & Name\_VSX & String & & Variable star identifier from VSX catalog \\
59 & GCVS\_flag & Boolean & & True if included in GCVS \\
60 & Type\_GCVS & String & & Variability type from GCVS \\
61 & Name\_GCVS & String & & Variable star identifier from GCVS \\
62 & Period\_GCVS & Double & days & Variable period from GCVS \\
63 & ASAN-SN\_variable\_flag & Boolean & & True if included in ASAN-SN\_variable \\
64 & Type\_ASAN-SN\_variable & String & & Variability type from ASAN-SN\_variable catalog \\
65 & Period\_ASAN-SN\_variable & Double & days & Variable period from ASAN-SN\_variable catalog \\
66 & ASAN-SN\_binary\_flag & Boolean & & True if included in ASAN-SN\_binary catalog \\
67 & Period\_ASAN-SN\_binary & Double & days & Binary period from ASAN-SN binary catalog \\
68 & TESS\ EB\_flag & Boolean & & True if included in TESS EB catalog\\
69 & NSS\_SB1\_flag & Boolean & & True if included in Gaia-NSS SB1 catalog \\
70 & Period\_NSS\_SB1 & Double & days & Binary period from Gaia-NSS SB1 catalog \\
71 & NSS\_EB\_flag & Boolean & & True if included in Gaia-NSS EB catalog \\
72 & Period\_NSS\_EB & Double & days & Binary period from Gaia-NSS EB catalog \\
73 & NSS\_astrometry\_flag & Boolean & & True if included in Gaia-NSS Astrometry catalog \\
74 & Period\_NSS\_astrometry & Double & days & Binary period from Gaia-NSS Astrometry catalog \\
\hline
\multicolumn{5}{c}{\small\textbf{Parameters derived from fitting the radial velocity curve (7,846 rows)}} \\
\hline
75 & P\_unimodal & Boolean & & True if period is well constrained \\
76 & P\_bimodal & Boolean & & True if period is bimodal \\
77 & Period\ joker & String & days & Period calculated from \textit{The Joker}, two values if period is bimodal \\
78 & eccentricity & Double & & Eccentricity from \textit{The Joker} \\
79 & K & Double & km s$^{-1}$ & Semi-amplitude from \textit{The Joker} \\
80 &  $v_{0}$ & Double & km s$^{-1}$ & System barycentric velocity from \textit{The Joker} \\
81 &  $\omega$ & Double & rad & Argument of pericenter from \textit{The Joker} \\
82 &  $M_{0}$ & Double & rad & Mean anomaly from \textit{The Joker} \\
83 & $f(m)$ & Double & $M_\odot$ & Derived mass function \\
84 & $M_{comp}$ & Double & $M_\odot$ & Lower limit of companion mass \\
\hline
\multicolumn{5}{c}{\small \textbf{Kinematic properties for binaries (3,836 rows)}} \\
\hline
85 & $R$ & Double & kpc & Galactocentric Cylindrical radial distance \\
86 & $\theta$ & Double & deg & Galactocentric Cylindrical azimuth angle \\
87 & $Z$ & Double & kpc & Galactocentric Cylindrical vertical height \\
88 & $U_{\mathrm{LSR}}$ & Double & km s$^{-1}$ & Cartesian Galactocentric x-velocity to the LSR\\
89 & $V_{\mathrm{LSR}}$ & Double & km s$^{-1}$ & Cartesian Galactocentric y-velocity to the LSR \\
90 & $W_{\mathrm{LSR}}$ & Double & km s$^{-1}$ & Cartesian Galactocentric z-velocity to the LSR\\
91 & $L_X$ & Double & kpc km s$^{-1}$ & Galactocentric x-angular momentum \\
92 & $L_Y$ & Double & kpc km s$^{-1}$ & Galactocentric y-angular momentum \\
93 & $L_Z$ & Double & kpc km s$^{-1}$ & Galactocentric z-angular momentum \\
94 & $J_R$ & Double & kpc km s$^{-1}$ & Radial orbital action \\
95 & $J_Z$ & Double & kpc km s$^{-1}$ & Vertical orbital action \\
96 & $t_{J_Z}$ & Double & Gyr & kinematic age derived from $J_Z$ using \textit{zoomies} \\
97 & $t^{Up}_{J_Z}$ & Double & Gyr & 1-$\sigma$ upper limit of kinematic age derived from $J_Z$ using \textit{zoomies}\\
98 & $t^{Lower}_{J_Z}$ & Double & Gyr & 1-$\sigma$ lower limit of kinematic age derived from $J_Z$ using \textit{zoomies}\\
99 & $TD / D$ & Double & & Thick disc to thin disc membership probability \\
100 & $TD / H$ & Double & & Thick disc to halo membership probability \\
101 & $Herc / D$ & Double & & Hercules stream to thin disc membership probability \\
102 & $Herc / TD$ & Double & & Hercules stream to thick disc membership probability \\
103 & Component & String & & Classification of Galactic components \\
\hline
\hline
\enddata
\tablecomments{Table \ref{tab: sub catalog} is published in its entirety in the machine-readable format. A column description is shown here for guidance regarding its form and content.}
\end{deluxetable*}
}


\begin{table*}[!ht]

\caption{Detection Rate and Purity for different $\Delta RV$ thresholds and $C$ values.}

\label{tab: simulation result}

\begin{tabular}{|c|c|c|c|c|c|c|}
\hline
\multirow{2}{*}{C ($\mathrm{km\ s^{-1}}$)} 
& \multicolumn{2}{c|}{$\Delta RV > 2 \sigma$} 
& \multicolumn{2}{c|}{$\Delta RV > 3 \sigma $} 
& \multicolumn{2}{c|}{$\Delta RV > 4 \sigma$} \\ \cline{2-7} 
                
& Detection Rate & Purity & Detection Rate & Purity & Detection Rate & Purity \\ \hline
5                      & 0.3373       & 0.6024 & 0.2045       & 0.6946 & 0.1131       & 0.9460 \\ \hline
10                     & 0.1040       & 0.8969 & 0.0999       & 0.9269 & 0.09216       & 0.9828 \\ \hline
15                     & 0.0708       & 0.9898 & 0.0706       & 0.9923 & 0.0704       & 0.9974 \\ \hline
20                     & 0.0557       & 0.9977 & 0.0556       & 0.9983 & 0.0556       & 1.0000 \\ \hline
\end{tabular}

\end{table*}

\subsection{Binary Orbital Parameters from Radial Velocity Curve Fits}\label{subsec: section 5.4 fit binary}
In this subsection, we present the radial velocity curve fitting procedure for the 7,846 binary candidates. Given the sparse and non-uniform RV data, we employed the Python package \textit{The Joker} to derive orbital parameters (e.g. period $P$, eccentricity $e$, semi-amplitude $K$) for our identified binary candidates from the multiple-epoch RV data. \textit{The Joker} is a custom Markov Chain Monte Carlo (MCMC) sampler specifically designed to handle non-uniform datasets and can search over orbital parameter space for likely orbits \citep{price-whelanJokerCustomMonte2017}. It has been widely used in large sample surveys for statistical analysis of binary orbital posteriors from RV curves, such as APOGEE \citep{price-whelanBinaryCompanionsEvolved2018}, MUSE \citep{2019A&A...632A...3G}, as well as in exoplanet detection through RV data \citep{2021ApJS..255....8R}.


The RV fitting procedure is described as follows: we first fitted the RV curve with a Keplerian orbit in the form of six parameters, that is:
\begin{equation}
    v(t ; \boldsymbol{\theta})=v_{0}+K[\cos (\omega + f)+e \cos \omega],
\end{equation}
where $v_{0}$ is the barycentric velocity (center-of-mass RV), $K$ is the semi-amplitude of the RV curve, $f$ is the true anomaly which can be derived from mean anomaly $M$, $\omega$ is the argument of periastron, and $e$ is the eccentricity. For each of the 7,846 binary candidates, we ran \textit{The Joker} to generate posterior samplings for the Keplerian orbital parameters. We began by generating a cache of 20,000,000 prior samples for the nonlinear parameters, which were drawn from the prior probability density functions (PDFs) summarized in Table \ref{tab:summary_prior}, where $\sigma_K$ is defined by Equation \ref{eq: sigma_k} using $\sigma_K$ = $15\ \mathrm{km\ s^{-1}}$, $P_0$ = 365 days, following \citet{price-whelanBinaryCompanionsEvolved2018}:

\begin{equation}
\sigma_{K}=\sigma_{K, 0}\left(\frac{P}{P_{0}}\right)^{-1 / 3}\left(1-e^{2}\right)^{-1 / 2}
\label{eq: sigma_k}
\end{equation}

\begin{table*}[ht]
\centering
\caption{Summary and Description of Priors}
\begin{tabular}{c c c}
\hline 
Name & Prior & Description \\
\hline 
$P$ & $p(P)\propto \frac{1}{P};P\in (0.2,10000)$ day & Period \\
$e$ & $e \sim \operatorname{Beta}(0.867,3.03)$ & Eccentricity \\
$M_{0}$ & $M_{0} \sim \mathcal{U}(0,2 \pi)$ rad & Mean anomaly at reference time \\
$\omega$ & $\omega \sim \mathcal{U}(0,2 \pi)$ rad & Argument of pericenter \\
$K$ & $K \sim \mathcal{N}\left(0, \sigma_{K}\right)$ km s$^{-1}$ & Velocity semi-amplitude \\
$v_{0}$ & $\mathcal{N}\left(0, 100^{2}\right)$ km s$^{-1}$ & System barycentric velocity \\
\hline 
\end{tabular}
\label{tab:summary_prior}
\end{table*}


After sampling with \textit{The Joker}, we identified 3,836 binary systems exhibiting a unimodal orbital solution with determined parameters (hereinafter referred to as the unimodal sample). For these binaries, we derived the binary mass function $f(M)$ following Equation \ref{eq: f(m)} using the posterior samples from our RV modeling:
\begin{equation}
\label{eq: f(m)}
    f(M)=\frac{M_{1} q^3 \sin ^{3} i}{(1+q)^{2}}=\frac{P K^{3}\left(1-e^{2}\right)^{3 / 2}}{2 \pi G},
\end{equation}
where $M_2$ is the mass of the invisible star, $q = M_2/M_1$ is the mass ratio, $i$ is the system inclination angle, $K$ is the RV semi-amplitude of the visible star and $P$ is the orbital period. Using this mass function, we estimated the lower limit of the companion’s mass under the assumption of $i = 90^\circ$ (See Column $M_\text{comp}$ in Table \ref{tab: sub catalog}). 

The remaining 4,010 systems failed to yield well-determined periods due to insufficient sampling. Among them, 561 systems exhibited two distinct period modes, indicating that one or a few additional radial velocity measurements would be sufficient to uniquely determine their orbits, as noted by \cite{price-whelanBinaryCompanionsEvolved2018}.

For systems with a unique orbit solution, we generated additional prior samples to mitigate sampling limitations and utilized the posterior samples returned by \textit{The Joker} to initiate an MCMC process for deriving the posterior distribution. We visually inspected the inferred orbital solutions and the resulting RV curves for all systems in this sample, flagging those with questionable or invalid fits (an example RV curve is shown in Figure \ref{fig: mcmc_example}). After filtering out poorly fitted RV curves (e.g., those lacking significant periodic variations), we retained a subset of 297 binary stars with well-folded RV curves (hereinafter referred to as the well-fitted sample), which is presented in Table \ref{tab: 299 gold sample}. 

For the fitted binary sample, we calculated spatial velocities and angular momenta using the fitted systemic velocities, and determined their Galactic components following the procedure described in Section \ref{sec: section 3. classification}. We found that 91.3$\pm$1.6\% of the binaries belong to the thin disk, a significantly higher fraction than in the single star sample (83.0$\pm$0.1\%), while only small fractions are in the thick disk (1.7\%) and stellar streams (1.3\%). This distribution suggests the binary population may be preferentially found in the younger thin disk component.


\begin{figure*}
    \centering
    \includegraphics[width=0.9\linewidth]{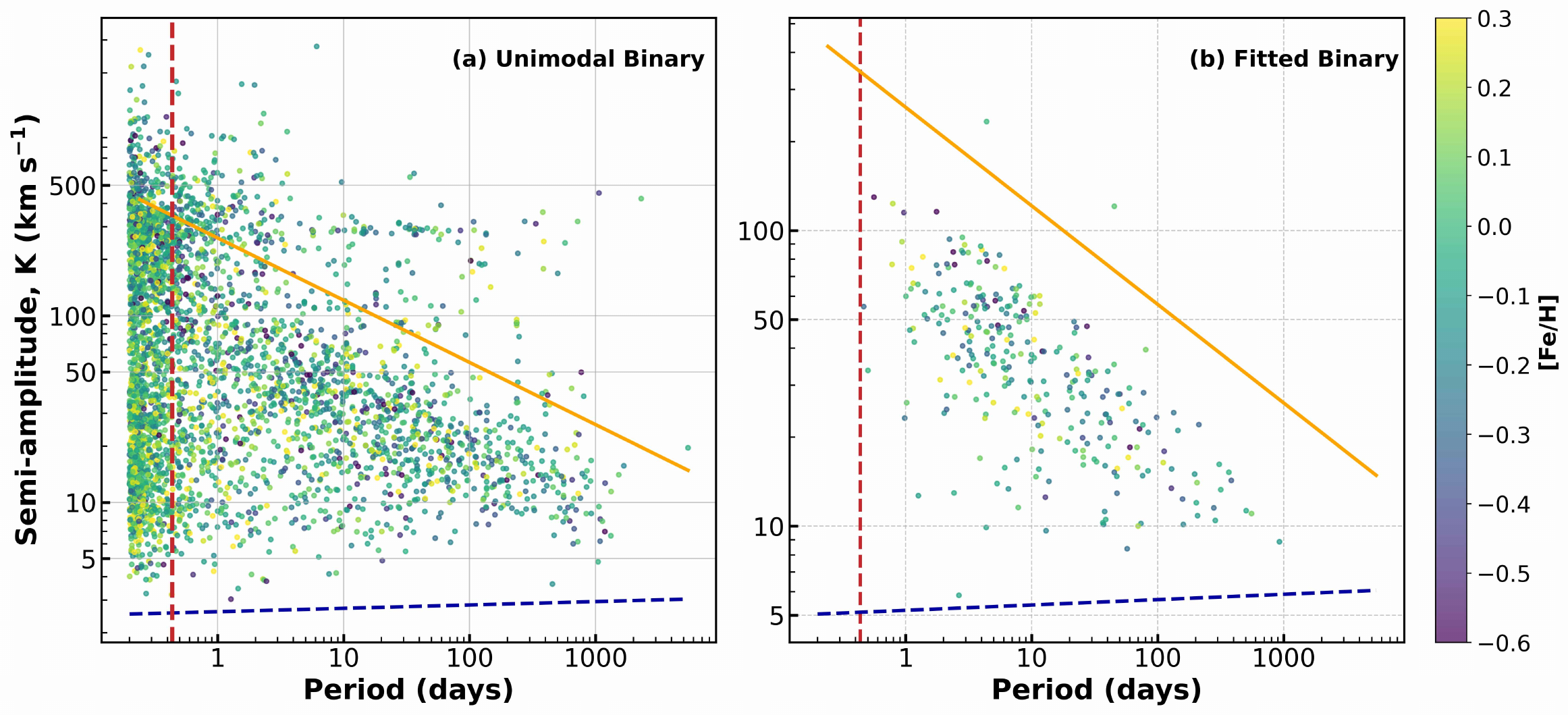}
    \caption{Left: Inferred semi-amplitude $K$ vs. orbital periods for the binary samples, color-coded by [Fe/H]. The left panel shows the 3,836 binaries with unimodal samplings, while the right panel shows the 297 well-fitted samples. The three colored reference lines, from top to bottom, represent the following: The bottom dashed blue line corresponds to the $\Delta RV > 4\sigma$ threshold, used to distinguish significant RV variations. The left red dashed line indicates the lower limit of the period calculated from the Roche limit. The top orange line represents a rough estimation of upper limit for the semi-amplitude K.}
    \label{fig: K-P}
\end{figure*}

Figure \ref{fig: K-P} illustrates the relationship between orbital periods and semi-amplitudes for stars in the unimodal sample (left panel) and the well-fitted sample (right panel). We found that the majority of our binary stars are short-period binaries, commonly referred to as close binaries \citep{moeMindYourPs2017}, with a typical orbital period of 7.5 days. 

Notably, a significant fraction of stars in the unimodal sample exhibit short orbital periods of less than 0.5 days. Among these, some display large mass ratios, suggesting potential inaccuracies in the derived parameters. Such discrepancies may result from sparse RV sampling or low-quality data and could also in turn affect the reliability of the inferred kinematics. To assist in evaluating the reliability and physical plausibility of the data, we have added three reference lines in Figure \ref{fig: K-P}, which serve as benchmarks for further scrutiny.



The bottom line represents the 4$\sigma$ threshold for RV variations, as determined by our selection criteria outlined in Section \ref{sec: subsec 5.2 Identifying the RV variables}. This threshold reflects the minimum RV amplitude required for a star to be classified as a binary in our sample. 

The vertical red line marks the lower limit for orbital periods as the binary will fill its Roche lobe below a critical distance. Using 
\begin{equation}
    a_R = (\frac{R_p}{0.462}) (\frac{M_{*}}{M_p})^{\frac{1}{3}} \ \  \mathrm{\citep{2005Icar..175..248F}},
\end{equation}
we derive a reference period of 0.436 days by adopting $\mathrm{R_{comp}} = 0.8\,\mathrm{R_{median}}$ and $\mathrm{M_{comp}} = 0.8\,\mathrm{M_{median}}$, where $\mathrm{R_{comp}}$ and $\mathrm{M_{comp}}$ are the radius and mass of the unseen companion, while $\mathrm{R_{median}}$ and $\mathrm{M_{median}}$ represent the typical values of the primary star.

The top line indicates an upper limit for the RV semi-amplitude, derived from the mass function $f(M)$ (Equation \ref{eq: f(m)}) under the assumption of a circular orbit ($e\ =\ 0$). The constraint for $K$ is given by:
\begin{equation}
    K \leq \frac{f(M)}{1.0361\times10^7} \left(\frac{P_{\mathrm{orb}}}{\text { days }}\right)^{-1} \left({\mathrm{~km} \mathrm{~s}^{-1}}\right)
\end{equation}


As shown in the right panel of Figure \ref{fig: K-P}, the majority of stars in the well-fitted sample reside within the region defined by these reference lines. This alignment indicates that our fitting method is robust and effective in identifying binaries with realistic orbital parameters, ensuring reliability and consistency in the derived results. 

\begin{figure*}
    \centering
    \includegraphics[width=0.9\linewidth]{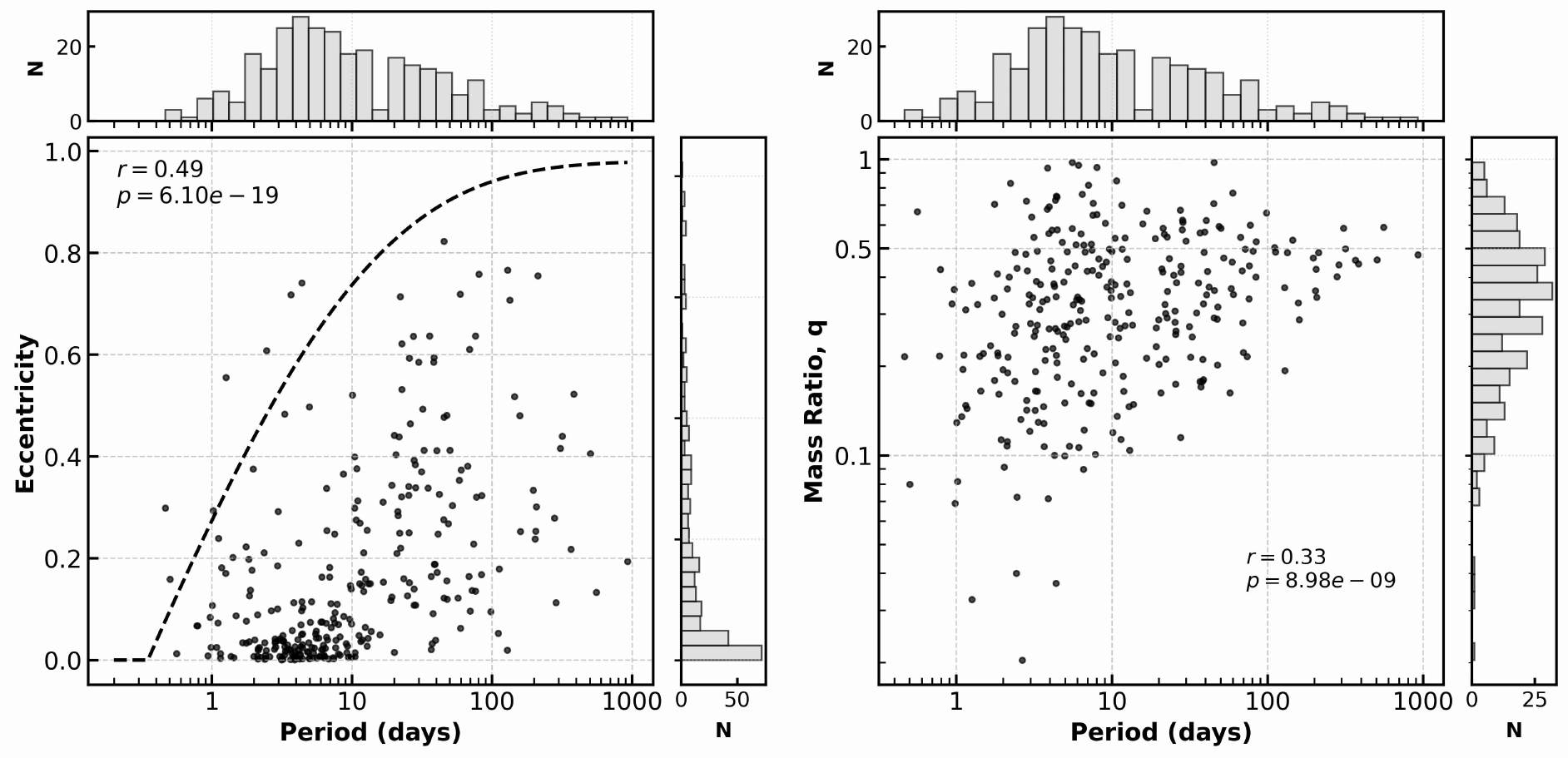}
    \caption{Orbital eccentricity (left panel) and binary mass ratio (right panel) as functions of orbital period for the well-fitted binary samples. Both parameters show positive correlations with orbital period, with Pearson correlation coefficients of 0.48 and 0.33, respectively, and p-values $<$ 0.05. Dotted lines in the left panel is the upper envelope from \citet{2008EAS....29....1M}. On the eccentricity-period plane, the well-fitted sample shows a concentrated distribution of eccentricities across all periods, with a tendency for lower eccentricities at shorter periods indicating tidal circularization effort.}
    \label{fig: binary period-eccentricity}
\end{figure*}

Figure \ref{fig: binary period-eccentricity} displays the orbital period as a function of eccentricity (left panel) and mass ratio (right panel) for the well-fitted sample. As illustrated, the $q$-$P$ distribution reveals a peak in $q$ around 0.3-0.5, while also displaying sparsely populated region in both the upper-left and upper-right corners of the right panel. In addition to the selection bias caused by the LAMOST observational cadence and time baseline, limiting our ability to detect binaries with either very short or long periods ($P$ $>$ 1000 days). Systems with low mass ratios are also more likely to be missed due to their smaller RV amplitudes being lost in the noise. Moreover, in cases where an SB2 system is misclassified as a single-lined spectroscopic binary, the companion’s spectral contribution may blend with that of the primary, potentially biasing the observed RV amplitudes toward lower values and resulting in underestimated mass ratios \citep{2010A&A...521A..24J}, thereby reducing the number of detected high-$q$ systems. Beyond selection effects, the lack of systems in the upper-left (short-period, high-$q$ systems) and upper-right (long-period, high-$q$ systems) regions of the $q$-$P$ distribution may also reflect physical mechanisms. For instance, short-period, high-$q$ binaries may undergo orbital evolution due to tidal dissipation  \citep{1975A&A....41..329Z, zahnTidalDissipationBinary2008,2008EAS....29....1M,2023ASPC..534..275O}. The observed deficit of high-$q$ binaries with intermediate periods (20–500 days) may be linked to the intrinsic properties of early-type binaries, as suggested by \citet{moeMindYourPs2017}.

The positive correlation observed between $q$ and period may be partially influenced by selection effects, as long-period systems with low $q$ are more difficult to detect (see Figure \ref{fig: simulated q_p}). Meanwhile, the positive correlation between $e$ and period in the left panel demonstrates the impact of tidal circularization \citep{meibomRobustMeasureTidal2005,2008EAS....29....1M,zahnTidalDissipationBinary2008}, with eccentricities of systems having $P < 10$ days being more concentrated near $e = 0$, compared to those with longer orbital periods, this trend has been observed across various samples \citep{raghavanSurveyStellarFamilies2010,price-whelanBinaryCompanionsEvolved2018,2019MNRAS.489.1644W,price-whelanCloseBinaryCompanions2020,2021AJ....162..184K,2023ApJS..269...41C,2025ApJS..278...46G}. The dotted line in Figure \ref{fig: binary period-eccentricity} represents the upper envelope adopted from \citet{2008EAS....29....1M}, which was derived from 2,751 binary systems in the 9th Catalog of Spectroscopic Binaries \citep{2004A&A...424..727P}. As shown, the majority of our sample resides within the envelope while only a few stars exceed above the envelope, similar trend has also been reported in LAMOST LRS sample \citep{2023ApJS..269...41C}. For these outliers above the envelope, the elevated eccentricities may be linked to their relatively high effective temperatures (typically exceeding 6,250 K), or they may reflect long-term RV trends suggestive of a third body. Alternatively, these deviations could also be attributed to high primordial eccentricity. Nevertheless, the precise origin of these outliers remains uncertain.

Despite being widely observed, the underlying mechanisms of the positive $e$-$p$ correlation still remains an open question. \citet{meibomRobustMeasureTidal2005} introduced the concept of `circularization period' ($P_{circ}$), defined as the period below which most binaries appear to be circular. Subsequent studies found that $P_{circ}$ increases with stellar age \citep[e.g., ]{meibomRobustMeasureTidal2005,2006ApJ...653..621M,2012AJ....144...54G,2014AJ....148...38M,2015AJ....150...10L,2020AJ....160..169N,2021AJ....161..190G},  indicating active tidal circularization throughout the main-sequence phase and providing constraints the tidal theories \citep{1975A&A....41..329Z,1989A&A...220..112Z,1989A&A...223..112Z,2007ApJ...661.1180O}. Nevertheless,  \citet{2022ApJ...929L..27Z} suggested the apparent age dependence of $P_{circ}$ may be influenced by a `cold core', and highlighted the role of resonance locking in addition to equilibrium and dynamical tides. In contrast, other studies using large samples of eclipsing binaries have found that the eccentricity distribution also varies with stellar temperature: the circularization period decreases with increasing temperature, with hotter binaries tend to be more eccentric at shorter periods \citep{2016ApJ...824...15V,2017A&A...608A.129T,2021ApJ...912..123J,2024A&A...691A.242I}. By analyzing spectroscopic binaries, \citet{2023MNRAS.522.1184B} found the dependence of cutoff period on age is not significant comparing with the dependence on temperature. These observations indicate that equilibrium tides may dominate tidal circularization in the pre-main-sequence, while additional effective mechanisms likely contribute to tidal evolution at later stages \citep{2011MNRAS.411.2804K,2022ApJ...927L..36B}.

To uncover the process of tidal evolution in binary systems, it is essential to disentangle how properties such as circularization period vary with stellar age, metallicity, and other parameters individually. However, this task is complicated by the strong correlations among stellar properties (e.g., age, temperature, metallicity). In this work, we construct a large catalog of binaries with orbital parameters, stellar parameters, as well as kinematic age estimates, offering a solid foundation for such investigations. Based on this catalog, by controlling for confounding parameters, one can examine the temporal evolution of binary characteristics (e.g., orbital eccentricity, mass ratio), thereby shedding light on the mechanisms driving tidal circularization of binary systems (to be discussed in a follow-up paper, Wu et al. in prep.).



\begin{table*}[ht]\centering\rotatebox{90}{ 
\begin{minipage}{\textheight} 
\centering
\caption{Catalog of 297 Well Fitted Binary}
\label{tab: 299 gold sample}
\begin{tabular}{cccccccccccccc}
\hline\hline
TIC & Gaia DR3 ID & Period & $e$ & $K$ & $v_0$ & $\omega$ & $M_0$ & $M_{\text{comp}}$ & $P_{\text{mcmc}}$ & $e_{\text{mcmc}}$ & $K_{\text{mcmc}}$ & $v_{\text{0 mcmc}}$  & $\bar{\hat{R}}$ \\
- & - & (days) & - & ($\mathrm{km\ s^{-1}}$) & ($\mathrm{km\ s^{-1}}$) & (rad) & (rad) & ($M_\odot$) & (days) & - & ($\mathrm{km\ s^{-1}}$) & ($\mathrm{km\ s^{-1}}$)  & - \\
1 & 2 & 3 & 4 & 5 & 6 & 7 & 8 & 9 & 10 & 11 & 12 & 13 & 14  \\
\hline
60288033 & 609809150428649472 & 307.90 & 0.42 & 16.19 & 21.76 & 0.62 & 1.08 & 0.77 & $307.92^{+0.10}_{-0.10}$ & $0.41^{+0.01}_{-0.01}$ & $16.29^{+0.28}_{-0.25}$ & $21.63^{+0.20}_{-0.19}$ & 1.00 \\
63209649 & 2126356983051726720 & 4.43 & 0.05 & 78.88 & -17.71 & 0.57 & 0.30 & 0.87 & $4.71^{+0.43}_{-0.87}$ & $0.54^{+0.23}_{-0.18}$ & $83.53^{+31.71}_{-9.29}$ & $-8.74^{+12.00}_{-7.20}$ & 6.67 \\
1001066189 & 1505618931851239552 & 12.13 & 0.21 & 22.26 & 3.31 & 1.91 & 0.50 & 0.27 & $12.13^{+0.02}_{-1.60}$ & $0.23^{+0.52}_{-0.02}$ & $22.17^{+16.89}_{-0.32}$ & $3.20^{+0.31}_{-6.61}$ & 1.58 \\
276662038 & 3217291698571122816 & 11.58 & 0.02 & 28.87 & 19.78 & 3.86 & -0.51 & 0.36 & $11.94^{+0.39}_{-1.18}$ & $0.40^{+0.33}_{-0.09}$ & $27.19^{+27.59}_{-3.75}$ & $14.22^{+6.85}_{-8.62}$ & 3.35 \\
99236062 & 1009391604610898176 & 5.40 & 0.02 & 47.80 & 73.90 & -2.76 & 1.03 & 0.28 & $5.62^{+0.41}_{-0.39}$ & $0.83^{+0.13}_{-0.22}$ & $47.13^{+41.82}_{-9.45}$ & $73.34^{+5.48}_{-7.61}$ & 6.07 \\
46361693 & 608190188275369728 & 13.71 & 0.06 & 14.18 & 92.77 & 2.63 & 2.74 & 0.15 & $13.96^{+0.46}_{-0.77}$ & $0.40^{+0.38}_{-0.10}$ & $13.89^{+3.27}_{-2.25}$ & $94.13^{+2.88}_{-2.97}$ & 3.35 \\
46305454 & 604943639676920448 & 207.22 & 0.30 & 10.85 & -3.27 & 2.72 & -1.06 & 0.34 & $207.22^{+0.10}_{-0.10}$ & $0.30^{+0.03}_{-0.03}$ & $10.59^{+0.35}_{-0.35}$ & $-3.26^{+0.24}_{-0.24}$ & 1.00 \\
60392646 & 3419723643353175808 & 6.69 & 0.04 & 55.58 & -49.76 & 1.96 & 0.64 & 0.67 & $6.49^{+0.94}_{-0.42}$ & $0.66^{+0.18}_{-0.19}$ & $67.32^{+16.96}_{-10.46}$ & $-35.12^{+21.24}_{-19.75}$ & 5.84 \\
46361691 & 608189290627289856 & 0.79 & 0.07 & 123.33 & 61.19 & 2.44 & -1.70 & 0.43 & $1.00^{+0.50}_{-0.46}$ & $0.74^{+0.16}_{-0.15}$ & $104.34^{+26.79}_{-20.28}$ & $33.68^{+9.44}_{-12.08}$ & 7.13 \\
165624130 & 1612920305965260032 & 8.20 & 0.01 & 50.91 & 4.94 & 4.66 & 2.54 & 0.41 & $8.56^{+0.32}_{-1.08}$ & $0.70^{+0.15}_{-0.30}$ & $50.13^{+27.48}_{-13.58}$ & $-5.30^{+10.66}_{-22.55}$ & 5.39 \\
196973956 & 397635910581888512 & 3.19 & 0.04 & 34.94 & 3.74 & 3.14 & 1.68 & 0.21 & $2.95^{+1.02}_{-0.23}$ & $0.70^{+0.10}_{-0.19}$ & $32.57^{+4.24}_{-4.89}$ & $-6.49^{+8.59}_{-5.33}$ & 5.59 \\
198152789 & 1505620405023410944 & 12.82 & 0.25 & 24.46 & -15.35 & -0.33 & 2.35 & 0.30 & $12.84^{+0.69}_{-1.00}$ & $0.45^{+0.26}_{-0.20}$ & $25.44^{+3.98}_{-1.52}$ & $-13.78^{+4.13}_{-7.85}$ & 3.48 \\
458424681 & 4031670377528216320 & 31.98 & 0.49 & 33.42 & -13.20 & 3.02 & 0.72 & 0.20 & $31.98^{+0.02}_{-0.02}$ & $0.34^{+0.08}_{-0.06}$ & $13.12^{+2.22}_{-1.79}$ & $-10.18^{+0.69}_{-0.66}$ & 1.22 \\
46307919 & 611686051136877440 & 3.58 & 0.02 & 54.21 & 11.58 & 2.59 & -0.14 & 0.35 & $3.44^{+0.42}_{-0.55}$ & $0.72^{+0.13}_{-0.18}$ & $56.04^{+78.14}_{-8.62}$ & $24.58^{+11.42}_{-13.47}$ & 5.94 \\
429554271 & 3424894333005909504 & 1.02 & 0.29 & 26.15 & 8.38 & 1.47 & 2.55 & 0.09 & $1.11^{+0.44}_{-0.68}$ & $0.73^{+0.25}_{-0.25}$ & $26.72^{+29.57}_{-4.81}$ & $11.59^{+3.65}_{-2.76}$ & 7.25 \\
26690153 & 1893702345536080256 & 2.36 & 0.21 & 53.30 & -42.62 & 6.28 & 1.96 & 0.34 & $2.34^{+0.79}_{-0.76}$ & $0.73^{+0.08}_{-0.25}$ & $58.33^{+4.96}_{-3.97}$ & $-61.65^{+15.22}_{-8.72}$ & 6.75 \\
20237221 & 1391723889583920768 & 4.32 & 0.07 & 23.89 & -3.23 & 0.32 & 1.68 & 0.24 & $4.50^{+0.65}_{-0.59}$ & $0.50^{+0.17}_{-0.24}$ & $31.91^{+17.03}_{-6.51}$ & $-1.47^{+4.99}_{-6.25}$ & 6.06 \\
20216313 & 1392056491851280640 & 212.34 & 0.76 & 22.05 & -0.76 & 2.19 & 2.32 & 0.50 & $212.35^{+0.10}_{-0.10}$ & $0.79^{+0.04}_{-0.04}$ & $21.89^{+2.20}_{-1.56}$ & $-1.76^{+0.84}_{-0.78}$ & 1.00 \\
219037973 & 1560561149934463104 & 171.56 & 0.32 & 17.42 & -15.75 & 2.04 & -0.98 & 0.61 & $171.55^{+0.10}_{-0.10}$ & $0.32^{+0.01}_{-0.01}$ & $17.30^{+0.26}_{-0.26}$ & $-15.75^{+0.16}_{-0.16}$ & 1.00 \\
417546685 & 1384466288207118592 & 1.85 & 0.13 & 61.13 & -32.11 & -0.82 & -1.54 & 0.28 & $2.07^{+0.50}_{-0.46}$ & $0.58^{+0.22}_{-0.28}$ & $42.91^{+46.37}_{-8.41}$ & $-23.30^{+13.15}_{-7.27}$ & 6.03 \\
99252577 & 1009377860715543936 & 202.63 & 0.24 & 12.00 & 10.86 & 5.41 & -0.04 & 0.48 & $202.62^{+0.08}_{-0.08}$ & $0.22^{+0.01}_{-0.01}$ & $12.03^{+0.14}_{-0.14}$ & $10.71^{+0.11}_{-0.11}$ & 1.00 \\
81638394 & 3426751098906874624 & 37.21 & 0.03 & 10.14 & 16.68 & 0.97 & 0.41 & 0.22 & $37.21^{+0.71}_{-0.93}$ & $0.17^{+0.65}_{-0.15}$ & $10.96^{+11.16}_{-1.03}$ & $16.61^{+1.81}_{-2.11}$ & 1.83 \\

... &  ... &... &... &... &... &... &... &... &... &... &... &... &... \\
\hline
\end{tabular}
\vspace*{0.15cm} 

\tablecomments{Columns 1-2 provide the identifiers for the target stars. Columns 4-9 list the orbital parameters for the binary systems, which were calculated using \textit{The Joker}, along with the derived companion masses. The corresponding MCMC results are presented in columns 10-13. The final column shows the mean value of the MCMC Gelman–Rubin convergence statistic for all parameters. For detailed usage and caveats related to this table, please refer to Section \ref{sec: section 6. caveats}. The complete table, which includes 297 stars, is available in its entirety in machine-readable format. A portion of the table is shown here to guide the reader regarding its form and content.}
\end{minipage}
\digitalasset
}
\end{table*}

\subsection{Characteristics of the RV Variable Catalog} \label{subsec: section 5.5 characteristics of the RV variable}

\begin{table*}[!ht]
\centering

\caption{Summary of RV Variable Catalog.}
\begin{tabular}{ccccccc}
\toprule
Sample & $N_s$ & $N_p$ & [Fe/H] (dex) & Exposure Number & $\Delta RV_{max} (\rm {km\ s^{-1}})$ & $T_{max}$ (hour) \\
\midrule
Whole Multiple-epoch & 113,792 & 114 & $-0.095_{-0.25}^{+0.19}$ & $6_{-3}^{+16}$ & $5.19_{-3.25}^{+6.16}$ & $1389.12_{-1388.33}^{+19531.28}$ \\
RV Variable & 41,382 & 37& $-0.106_{-0.24}^{+0.18}$ & $14_{-8}^{+30}$ & $10.16_{-3.32}^{+16.17}$ & $10268.92_{-9883.97}^{+17136.52}$ \\
Binary & 7,846 & 6 & $-0.054_{-0.22}^{+0.20}$ & $15_{-9}^{+29}$ & $46.16_{-21.95}^{+201.06}$ & $17090.95_{-16130.93}^{+10858.65}$ \\
RV Variable Without Binary & 33,536 & 31 & $-0.120_{-0.25}^{+0.18}$ & $14_{-8}^{+30}$ & $9.10_{-2.55}^{+4.31}$ & $9863.13_{-9693.85}^{+17396.96}$ \\
Short Time Scale & 160 & 0 & $-0.101_{-0.27}^{+0.22}$ & $3_{-0}^{+0}$ & $8.26_{-2.83}^{+6.85}$ & $0.73_{-0.28}^{+0.05}$ \\
\bottomrule
\end{tabular}
\label{tab: sub-catalog description}
\end{table*}

In the above Sections, we have identified 41,445 RV variables, including 7,846 binaries including a sample of 297 systems with well-constrained orbits (summarized as in Figure \ref{fig: flow chart} and Table \ref{tab: sub-catalog description}). However, as noted in Section \ref{sec: subsec 5.3 identify binary}, in addition to stellar activity such as photospheric variations that can produce significant RV shifts, the binary catalog may also be contaminated by intrinsically variable stars. Specifically, pulsating stars such as Cepheids, e.g., $\delta$ Cephei \citep[10–50 km s$^{-1}$,][]{2023A&A...674A..17R,2024A&A...686A.177A}, Mira variables \citep[5–20 km s$^{-1}$,][]{samusGeneralCatalogueVariable2017}, and RR Lyrae stars \citep[up to 50–100 km s$^{-1}$,][]{2020ApJ...896L..15B,2023A&A...674A..18C,2024ApJS..272...31W} can exhibit RV amplitudes comparable to those of binaries. Eruptive or active stars, including T Tauri and FU Orionis types, may also induce RV shifts of up to $\sim$ 5 km s$^{-1}$ due to accretion shocks, jets, or chromospheric activity \citep{2021ApJ...917...80S}. To further validate the identified binaries and characterize the remaining variables, we performed cross-match with the following variable star and binary star catalogs:



\begin{enumerate}
    \item \textit{\textbf{VSX Catalog}\footnote{\url{https://www.aavso.org}}}: The International Variable Star Index (VSX) is a comprehensive, continuously updated database maintained by the American Association of Variable Star Observers (AAVSO). 
    It includes detailed information such as their variability types, periods, and amplitudes \cite{watsonInternationalVariableStar2006} on approximately 2,280,000 variable stars.
    
    We cross-matched the VSX database with our sample and identified a total of 1,138 matched targets, which are composed of eclipsing binaries (32.6\%), pulsating variables (42.88\%), rotating variables (21.44\%) and eruptive and other variable types (0.73\%). 
    Among the 371 eclipsing binaries, 263 sources are identified as binary stars in our RV variable catalog.
    

    \item \textit{\textbf{Gaia DR3 Variable Catalog} \footnote{\url{https://gea.esac.esa.int/archive/}}}: The Gaia DR3 classifies 9,976,881 sources into 24 variability classes using supervised machine learning \citep{eyerGaiaDataRelease2023}. These variables are mainly composed of pulsation, rotation, and eclipsing systems, accounting for 34\%, 28\%, and 22\%, respectively.
    
    The cross-match with our radial velocity (RV) variable star catalog yielded 2,605 common sources, including 1,216 rotating variables (such as RS Canum Venaticorum, $\alpha^2$ CVn, and stars with solar-like variability), 1,085 pulsating variables (e.g., $\delta$ Scuti, $\gamma$ Doradus), 296 eclipsing binaries, 1 RR Lyrae star, 6 young stellar objects, and 2 exoplanetary transits. Among the 296 eclipsing binaries, 218 were identified as binary stars in our catalog.
    \item \textit{\textbf{GCVS} \footnote{\url{https://heasarc.gsfc.nasa.gov/W3Browse/all/gcvs.html}}}: The General Catalog of Variable Stars (GCVS) is a comprehensive database of variable stars, maintained by the Sternberg Astronomical Institute at Moscow University. The fifth edition (GCVS 5.1) contains 58,202 individually identified and named variable stars \citep{samusGeneralCatalogueVariable2017}.

    Our RV variable catalog overlaps with GCVS 5.1 for 107 sources, including 81 eclipsing binaries, 7 $\delta$ Scuti stars, 14 eruptive variables (e.g., UV Ceti), 4 rotational variables, and 1 RR Lyrae star. Of the 81 eclipsing binaries, 61 were classified as binary stars in our catalog.

    \item \textit{\textbf{ASAN-SN} \footnote{\url{https://www.astronomy.ohio-state.edu/asassn/index.shtml}}}: ASAS-SN (All-Sky Automated Survey for Supernovae) is the first optical survey to monitor the entire sky, with a cadence of about 24 hours down to $\mathit{g}$$\le$ 18.5 mag. As of 2024, ASAS-SN has discovered over 687,000 variable stars in the V-band \citep{shappeeMANCURTAINXRAYS2014,jayasingheASASSNCatalogueVariable2021}. We cross-matched the RV variable catalog with the ASAS-SN V-band catalog and identified 464 common sources. Among these, 343 were marked as variable stars in our catalog, including 222 eclipsing binaries, 71 spotted variables with rotational modulation, 16 $\delta$ Scuti-type variables (including 3 high-amplitude types), 4 young stellar objects, 3 RR Lyrae stars, 3 eruptive variables, and 24 variables with unspecified types.
    
    In addition, ASAS-SN provides a binary star database, which includes various physical parameters of binary systems derived from light curves \citep{rowanValueaddedCatalogueASASSN2023}. They modeled the light curves of more than 30,000 detached eclipsing binaries from ASAS-SN and fitted the V- and g-band light curves, producing a catalog of orbital periods, eccentricities, and inclinations. Upon cross-matching this catalog with our RV variable catalog, we found 69 common sources, of which 57 were identified as binary stars in our catalog. Eleven of these were included in our well-fitted binary catalog, and the periods obtained from the RV curve fitting agreed with those from ASAS-SN.
    
    \item \textit{\textbf{TESS Eclipsing Binary Catalog} \footnote{\url{https://tessebs.villanova.edu}}}: The TESS Eclipsing Binary (EB) catalog provides a comprehensive list of eclipsing binary stars identified by the TESS mission \citep{prsaTESSEclipsingBinary2022}. It includes detailed information on the eclipse depths, periods, and other properties of 4,584 systems. Using the TICs, we identified 36 matches between the TESS EB catalog and our RV variable catalog. Among these, 31 were labeled as binary stars in our catalog.

    \item \textit{\textbf{Gaia Non-Single Stars (NSS) Catalog} \footnote{\url{https://gea.esac.esa.int/archive/}}}:The Gaia Non-Single Star (NSS) catalogs, released as part of Gaia DR3, provide a valuable resource for studying binary and multiple star systems by combining astrometric, spectroscopic, and photometric data from the Gaia mission \citep{2023A&A...674A..34G}. These catalogs include orbital or trend solutions for approximately 800,000 systems covering various binary types. We cross-matched our RV-variable catalog with Gaia NSS sources classified under the eclipsing binary model, the single-lined spectroscopic binary (SB1) model, and the astrometric binary model. This yielded a total of 29 eclipsing binaries (EBs), 490 astrometric binaries, and 1,300 SB1s, with some sources appearing in multiple categories. Among these, 22 EBs, 102 astrometric binaries, and 711 SB1s are also labeled as binaries in our catalog.

\end{enumerate}

In total, there are 4,901 common sources between our RV variable catalog and the referred variable catalogs such as VSX, Gaia DR3 variables, Gaia-NSS catalog, GCVS, and ASAS-SN variable catalogs. These include both extrinsic variables (e.g., binary systems), and intrinsic variables (e.g., pulsating and eruptive stars). None of the 297 well-fitted binary systems in our sample are identified as known variables of other types, among these, 192 are newly discovered binary star systems. 

Among the matched sources with known binary classifications from referred catalogs, some systems are labeled as `RV variable' in our catalog but not classified as `binary’. This discrepancy can be attributed to two main reasons. Taking the cross-match with the VSX catalog as an example, out of 108 such binaries, 64.8\% (70 sources) exhibit RV amplitudes smaller than our threshold of 20 km s$^{-1}$, which we adopted to avoid potential contamination from spurious low-amplitude variability. Among the remaining 38 sources, the vast majority (35 sources out of 38) are very short-period binaries of periods less than 0.5 days. Due to the observing cadence and typical exposure intervals of the LAMOST survey, such short-period systems are inherently difficult to detect and accurately characterize in our catalog.

\section{Notable Caveats and Limitations} \label{sec: section 6. caveats}
Some of the caveats and assumptions of this study have been discussed in Section \ref{sec: section 3. classification}-\ref{sec: Section 5. catalog of RV variables}. These include 
the contamination of the binary catalog by intrinsic variable stars, and simplifications adopted when evaluating the selection criteria of binary candidates and deriving orbital parameters. During the RV curve fitting process, we found that RV measurements from certain LAMOST observing nights exhibit abnormally high or low values, likely due to instrumental or calibration issues. Moreover, incomplete RV phase coverage or poorly constrained RV fits may result in inaccurate orbital period estimates, such as overestimated values or periods that are integer multiples of the true value. Below, we summarize other key limitations and considerations:
\begin{enumerate}
    \item 
    Since our catalog includes stars with multiple observations, some of which are binary or variable stars, we incorporate different types of RV data. To accommodate the diversity of RV data, we introduced an `RV Source Flag' for each star in our catalog (Column 38 in Table \ref{tab: MRS single catalog}), indicating the type of RV measurement associated with each star: If the `RV Source Flag' is set to 0, it indicates that the star does not have multiple epoch RV measurements, and its RV value corresponds to a single measurement from the LAMOST DR10 dataset. If the `RV Source Flag' is set to 1, it indicates that the star has been identified as a binary system, and the RV value corresponds to the systemic velocity derived from fitting the binary star’s radial velocity curve. If the `RV Source Flag' is set to 2, it indicates that the star has multiple RV measurements, and the RV value is the average of those measurements. The RV data for stars with an `RV Source Flag' of 2 should be used with caution, since they may exhibit RV variability across multiple epochs. Especially for those labeled as binaries (`Binary' flag = 1). This combination suggests that the RV difference exceeds 20 $\mathrm{km\ s^{-1}}$, but the RV curve is poorly fitted, leading to an unreliable systemic velocity estimate. 
    \item While the kinematic ages with uncertainty $\sim $ 10\%-20\% were given for disk stars, the age-velocity dispersion relation is not suitable for halo and stream stars (as described in \cite{chenPlanetsSpaceTime2021a}). Moreover, the ages for each star calculated from $J_Z$ using \textit{zoomies} should only be used as a rough range, as the typical uncertainty is 3-4 Gyrs.
    \item The use of LAMOST multiple RV observations to identify binary systems has certain limitations due to selection effects and observational constraints. The time span of the LAMOST MRS (less than 5 years) is not sufficient to effectively detect long-period binaries with orbital periods exceeding 50 years. Consequently, the binaries identified in this work are predominantly close binary systems, with a typical orbital period of $\sim$10 days (See Section \ref{subsec: section 5.4 fit binary}). In addition, LAMOST's typical exposure time of $\sim$20 minutes and observation cadence pose challenges in detecting very short-period variables with rapidly varying RV.

    \item  For stars classified as `RV variables' but not identified as `binary' in Table \ref{tab: sub catalog}, the observed RV variability may be attributed to long-period binary companions, intrinsic stellar variability (as discussed in Section \ref{subsec: section 5.5 characteristics of the RV variable}), or photospheric activity. Stars classified as `short time scale RV variables’ may be compact object systems or binaries with very short orbital periods. For the identified binary systems, contamination from intrinsic variable stars cannot be fully excluded. To account for known variables identified in other catalogs, we include an additional flags in our catalog (Table \ref{tab: sub catalog}).
    
    \item For each of the binary candidates, we fitted their RV curve to derive orbital parameters (see Table \ref{tab: sub catalog}). For the well-fitted systems, we also performed Markov Chain Monte Carlo (MCMC) sampling to obtain posterior distributions (see Table \ref{tab: 299 gold sample}). However, we recommend using the orbital solutions returned by the sampling from $The\ Joker$ software, as some of the MCMC results did not converge properly. Both sets of results are provided in Table \ref{tab: 299 gold sample}, along with the Gelman-Rubin statistic to assess convergence.

    \item During the RV curve fitting process, we found that some stars exhibit abnormally high or low RV values at specific observing epochs, deviating significantly (up to $\sim$ 500 km s$^{-1}$) from the rest of their measurements, even in sources with more than 30 exposures. Upon further inspection, we noticed that many of these outlier measurements cluster around specific observing dates (e.g., MJD 58450, 58800, 59250), suggesting that they may be associated with instrumental or calibration issues during certain LAMOST observing nights. To mitigate the impact of such issues, we applied additional filtering stpdf. In particular, we removed exposures from nights with only a single observation when constructing our multi-epoch sample, and we excluded sources affected by these outliers during the selection of the well-fitted sample. Nonetheless, we acknowledge that some residual contamination may still remain. We therefore commend users of the LAMOST time-domain RV data, especially those analyzing individual sources, to be aware of possible instrumental artifacts and consider them as a potential source of contamination when doing a large sample analysis.
    
\end{enumerate}

\section{Summary}\label{sec: summary}

In this work, we constructed kinematic-chemical catalogs (Table \ref{tab: MRS single catalog} and a similar table for the LRS sample) by combining TESS, Gaia DR3, and LAMOST DR10 data (Section \ref{sec: section 2. Data collection}), we also identified RV variables in the sample and characterized kinematic and orbital properties of the identified binary candidates (Table \ref{tab: sub catalog} and \ref{tab: 299 gold sample}). Detailed caveats and recommended guidelines for using the released catalogs are provided in Section \ref{sec: section 6. caveats}. The kinematic catalogs include 207,690 stars with MRS spectral data and 452,803 stars with LRS spectral data (Section \ref{sec: section 4. MRS single catalog}, Table \ref{tab: MRS single catalog}). Using LAMOST spectroscopic and Gaia astrometric data, we calculated each star's space position and Galactic orbit parameters (i.e., velocity, angular momentum, actions; Section \ref{subsec: section 3.1 space velocity}). Applying the revised kinematic methods from PAST \uppercase\expandafter{\romannumeral1} \citep{chenPlanetsSpaceTime2021a}, we derived the Galactic component membership probabilities and then classified stars into different components (Section \ref{subsec: section 3.2 Classification of Galactic Components}). Additionally, we obtained the ages for individual stars based on vertical actions $J_Z$ using the \textit{zoomies} package, though the typical uncertainty is relatively large (60\%-80\%). 
Alternatively, the average kinematic age for a group of stars can be derived from velocity dispersion, with a much smaller typical uncertainty of 10\% – 20\% by adopting the revised AVR from the PAST series (Section \ref{subsec: section 3.3 Calculating Kinematic Age and uncertainty}).

Based on our constructed LAMOST-Gaia-TESS catalogs, we further investigated the kinematics and chemical abundances of different Galactic components (Table \ref{tab: properties of LAMOST-Gaia-TESS star sample.}). As expected, from the thin disk, thick disk to the halo, Galactic velocities and radial orbital action ($J_R$) increase, while vertical angular momentum ($L_Z$) decreases (Figure \ref{fig: MRS toomre}-\ref{fig: Lz disks MRS}). Additionally, stellar ages grow older, [Fe/H] decreases, and [$\alpha$/Fe] increases (Figures \ref{fig: Fe/H vs alpha/Fe disks MRS}–\ref{fig: MRS chemical_TDD}). By adopting the kinematic planes from \cite{2019A&A...631A..47K}, we also identified stars associated with 12 nearby streams (Figure \ref{fig: Streams MRS}), and found that stars within a given stream exhibit very similar kinematic and chemical properties. Our results show that most TESS-detected planets are hot Jupiters orbiting thin disk stars, while no hot planets (period $<$ 10 days) are detected around thick disk stars (Figure \ref{fig: planet P-R}).


Using multiple-epoch RV measurements from LAMOST, we identified 41,445 RV variables in our LAMOST-Gaia-TESS sample based on the selection criteria in Equations \ref{eq: rv criteria} and \ref{eq: short time scale} (Section \ref{sec: subsec 5.2 Identifying the RV variables}, Table \ref{tab: sub catalog}). Among them, 7,846 were classified as high-confidence binary candidates since they exhibit significant RV variations and have peak-to-peak RV variation exceeding 20 km s$^{-1}$. To evaluate the effectiveness of our selection method, we conducted a Monte Carlo simulation to assess the impact of the adopted parameters on purity, detection rate, and completeness of the binary star identification (Table \ref{tab: simulation result}, Figure \ref{fig: simulated q_p}).

For these binary candidates, we further derived orbital parameters by fitting their RV curves using \textit{The Joker}. Combing the fitted systemic velocities with astrometry data, we also determined their kinematic properties (see columns in Table \ref{tab: sub catalog}). After inspecting the RV curves, we obtained a sample of 297 binaries with well-constrained orbital solutions (Table \ref{tab: 299 gold sample}). We found the sample is dominated by close binaries with orbital periods $<$ 10 days and the mass ratio distributions peaking between 0.3 and 0.5 (Figure \ref{fig: binary period-eccentricity}). We also observed a significantly positive correlation between orbital period and eccentricity, indicating the effects of tidal circularization (Section \ref{subsec: section 5.4 fit binary}). The combination of orbital and kinematic parameters from our well-fitted binary catalog provides a data basis for exploring how these correlations vary with kinematic age and other stellar properties, such as effective temperature and mass ratio, which will be addressed in future work.


The LAMOST-Gaia-TESS kinematic catalog constructed in this work is useful for future research on exoplanets in different Galactic environments (e.g., position, components, age, history, and chemistry). Additionally, applying our RV variable identification method to future LAMOST releases will yield more binaries, including planet-hosting stars. With the characterized kinematics and orbital properties, the RV variable and binary sample we identified will facilitate statistical analyses of the evolution and interactions of close binaries, as well as their impact on planetary systems. Together, these studies based on our catalogs will deepen the understanding of stellar and planetary formation and evolution within the context of the Milky Way.

\begin{acknowledgments}
This work is supported by the National Key R\&D Program of China (2024YFA1611803) and the National Natural Science Foundation of China (NSFC; grant No. 12273011, 12150009, 12403071).
We also acknowledge the science research grants from the China Manned Space Project with NO.CMS-CSST-2021-B12. 
J.-W.X. also acknowledges the support from the National Youth Talent Support Program.
H.F.W. is supported by the Department of Physics and Astronomy of Padova University through the 2022 ARPE grant: Rediscovering our Galaxy with Machines. 
W.Z. acknowledges the support from the Central Guidance for Local Science and Technology Development Fund, No. ZYYD2025QY27. 
S.D. is supported by the National Natural Science Foundation of China (Grant No. 12133005) and the China Manned Space Program with grant no. CMS-CSST-2025-A16. S.D. acknowledges the New Cornerstone Science Foundation through the XPLORER PRIZE.

This paper includes data collected by the TESS mission. Funding for the TESS mission is provided by the NASA's Science Mission Directorate. This work has included data from Guoshoujing Telescope (the Large Sky AreaMulti-Object Fiber Spectroscopic Telescope LAMOST), which is a National Major Scientific Project built by the Chinese Academy of Sciences. Funding for the project has been provided by the National Development and Reform Commission. LAMOST is operated and managed by the National Astronomical Observatories, Chinese Academy of Sciences. This work has made use of data from the European Space Agency (ESA) mission
{\it Gaia} (\url{https://www.cosmos.esa.int/gaia}), processed by the {\it Gaia}
Data Processing and Analysis Consortium (DPAC,
\url{https://www.cosmos.esa.int/web/gaia/dpac/consortium}). Funding for the DPAC has been provided by national institutions, in particular the institutions participating in the {\it Gaia} Multilateral Agreement.
\end{acknowledgments}

\bibliography{Project2023}{}
\bibliographystyle{aasjournal}

\newpage
\appendix
\renewcommand\thefigure{\Alph{section}\arabic{figure}}
\renewcommand\thetable{\Alph{section}\arabic{table}}
\setcounter{figure}{0} 
\setcounter{table}{0}
\section{Results for LRS} \label{appedndix:LRS}

\begin{figure}[ht!]
    \centering
    \begin{minipage}[t]{0.4\linewidth}
        \centering
        \includegraphics[width=\linewidth,height=\linewidth,keepaspectratio=false]{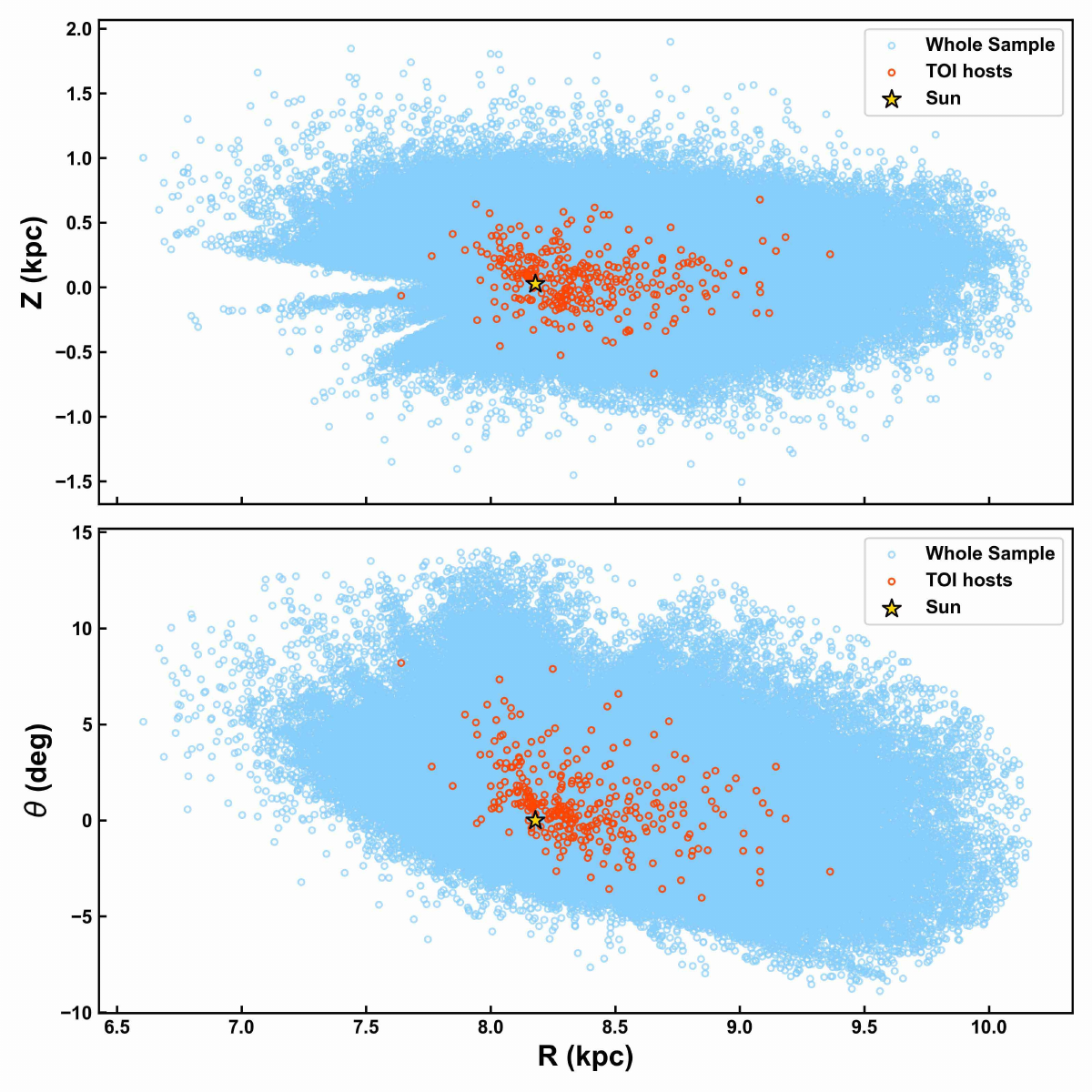}
        \caption{Same as Figure \ref{fig: fig 4. distribution of mrs sample} for the LRS sample.}
        \label{figappendix: lrs distribution}
    \end{minipage}
    \hfill
    \begin{minipage}[t]{0.4\linewidth}
        \centering
        \includegraphics[width=\linewidth]{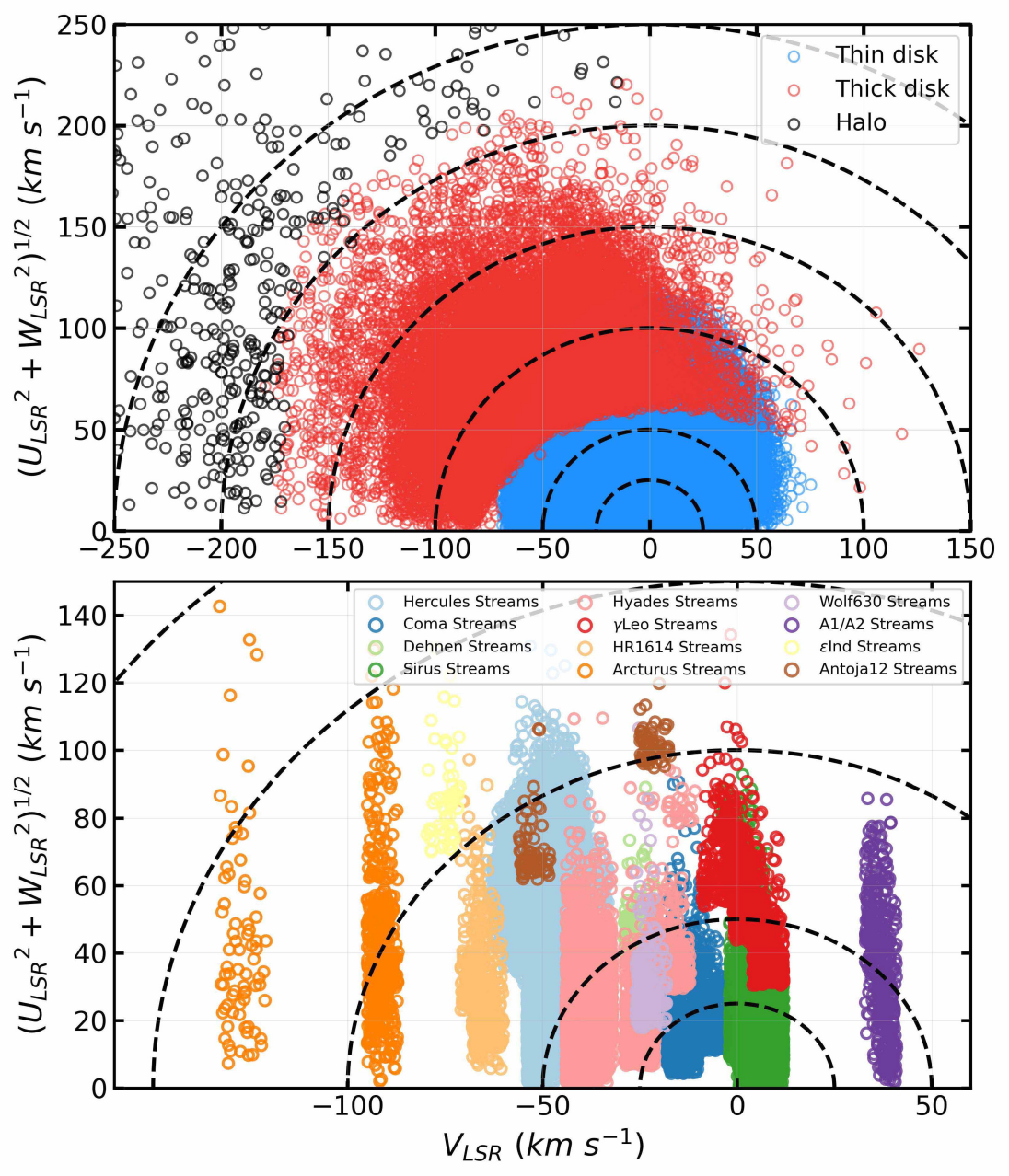}
        \caption{Same as Figure \ref{fig: MRS toomre} for the LRS sample.}
        \label{figappendix: lrs toomre}
    \end{minipage}
\end{figure}

\begin{figure}
    \begin{minipage}[t]{0.48\linewidth}
        \centering
        \includegraphics[width=\linewidth]{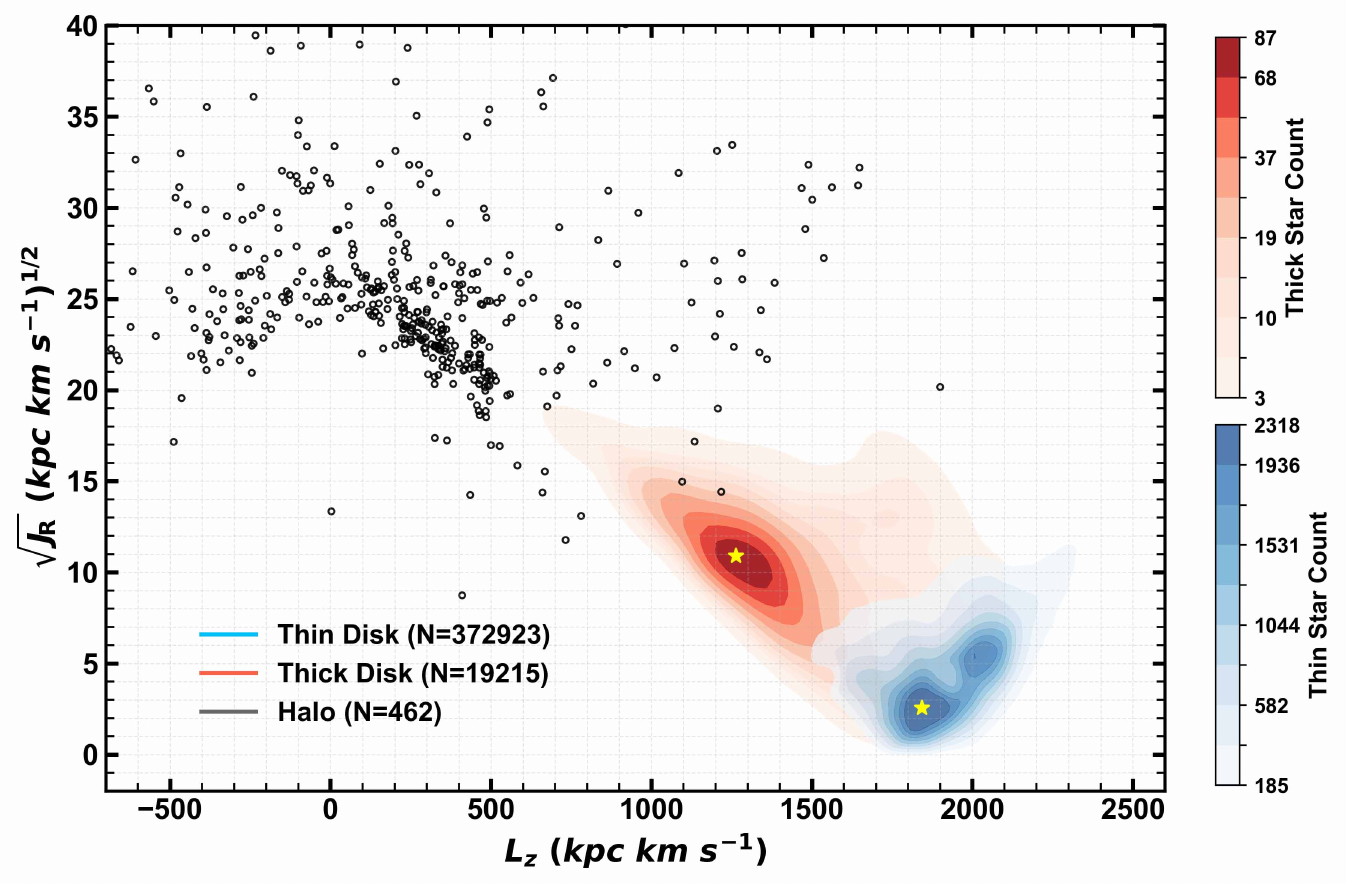}
        \caption{Same as Figure \ref{fig: Lz disks MRS} for the LRS sample.}
        \label{figappendix: lrs LZ}
    \end{minipage}
    \hfill
    \begin{minipage}[t]{0.48\linewidth}
        \centering
        \includegraphics[width=\linewidth]{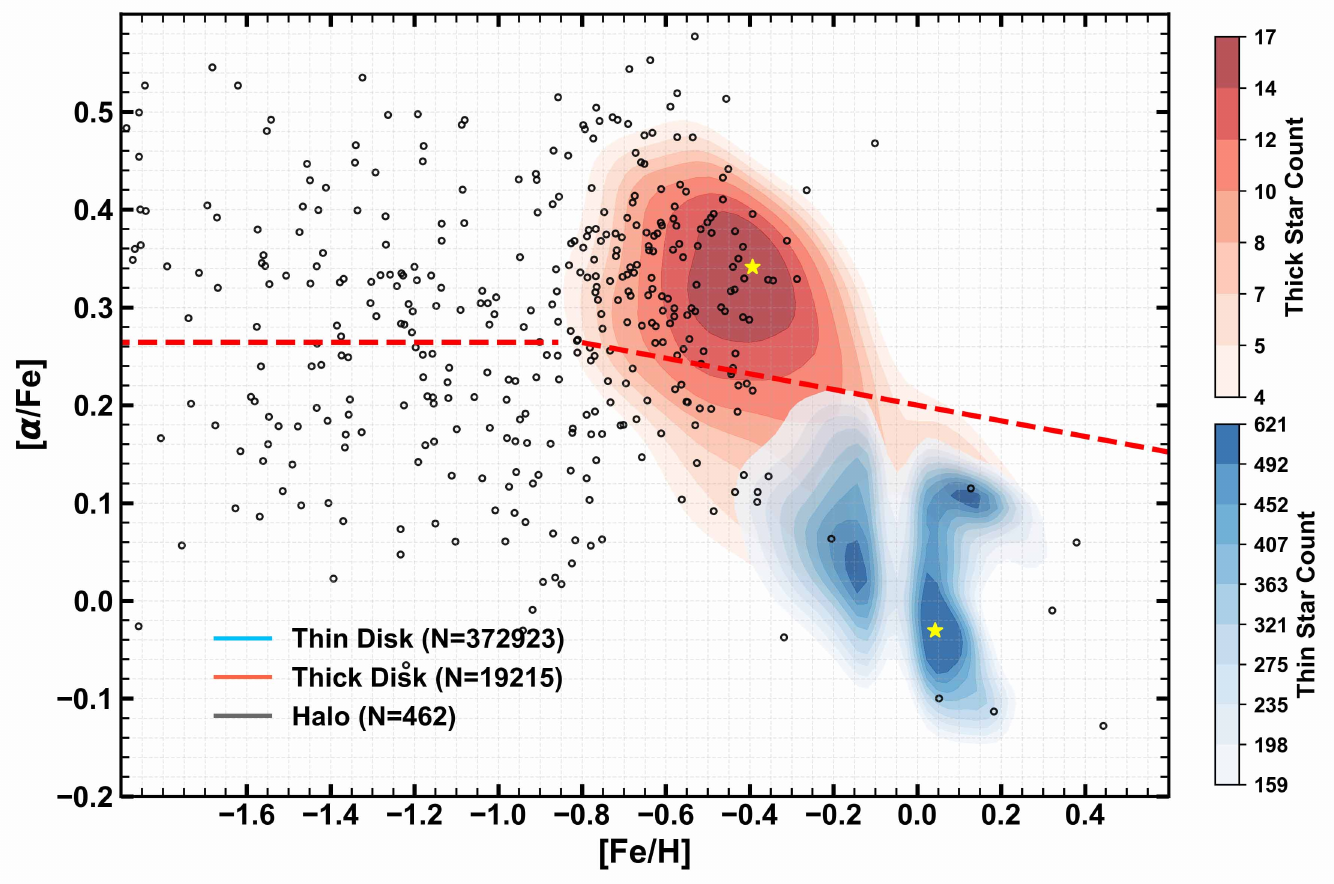}
        \caption{Same as Figure \ref{fig: Fe/H vs alpha/Fe disks MRS} for the LRS sample.}
        \label{figappendix: lrs alpha fe}
    \end{minipage}
\end{figure}

\begin{figure}[t]
  \centering
  \includegraphics[width=0.7\linewidth]{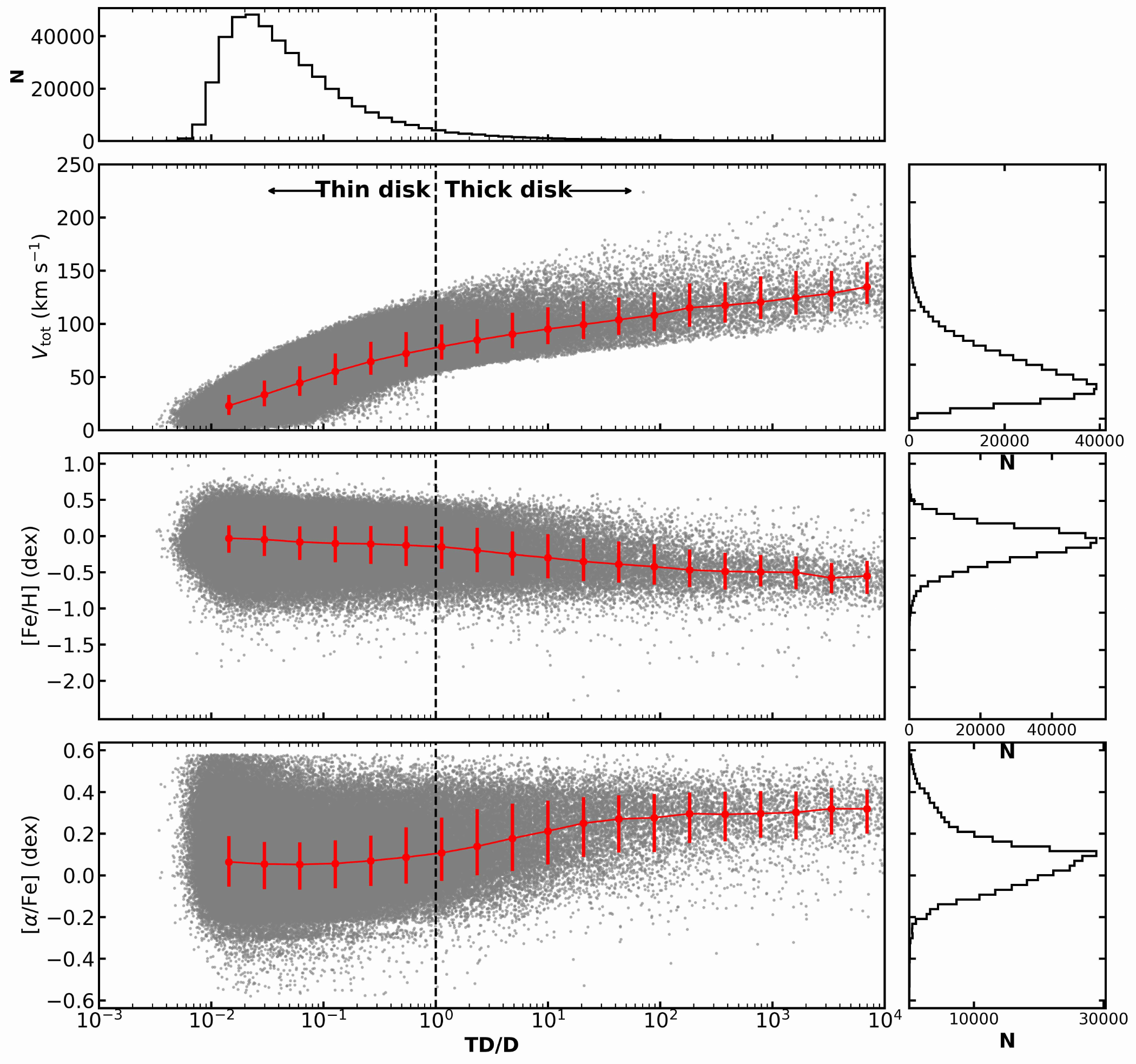}
  \caption{Same as Figure \ref{fig: MRS chemical_TDD} for the LRS sample.}
  \label{figappendix: lrs chemical}
\end{figure}

\begin{figure}[t]
  \centering
  \includegraphics[width=0.7\linewidth]{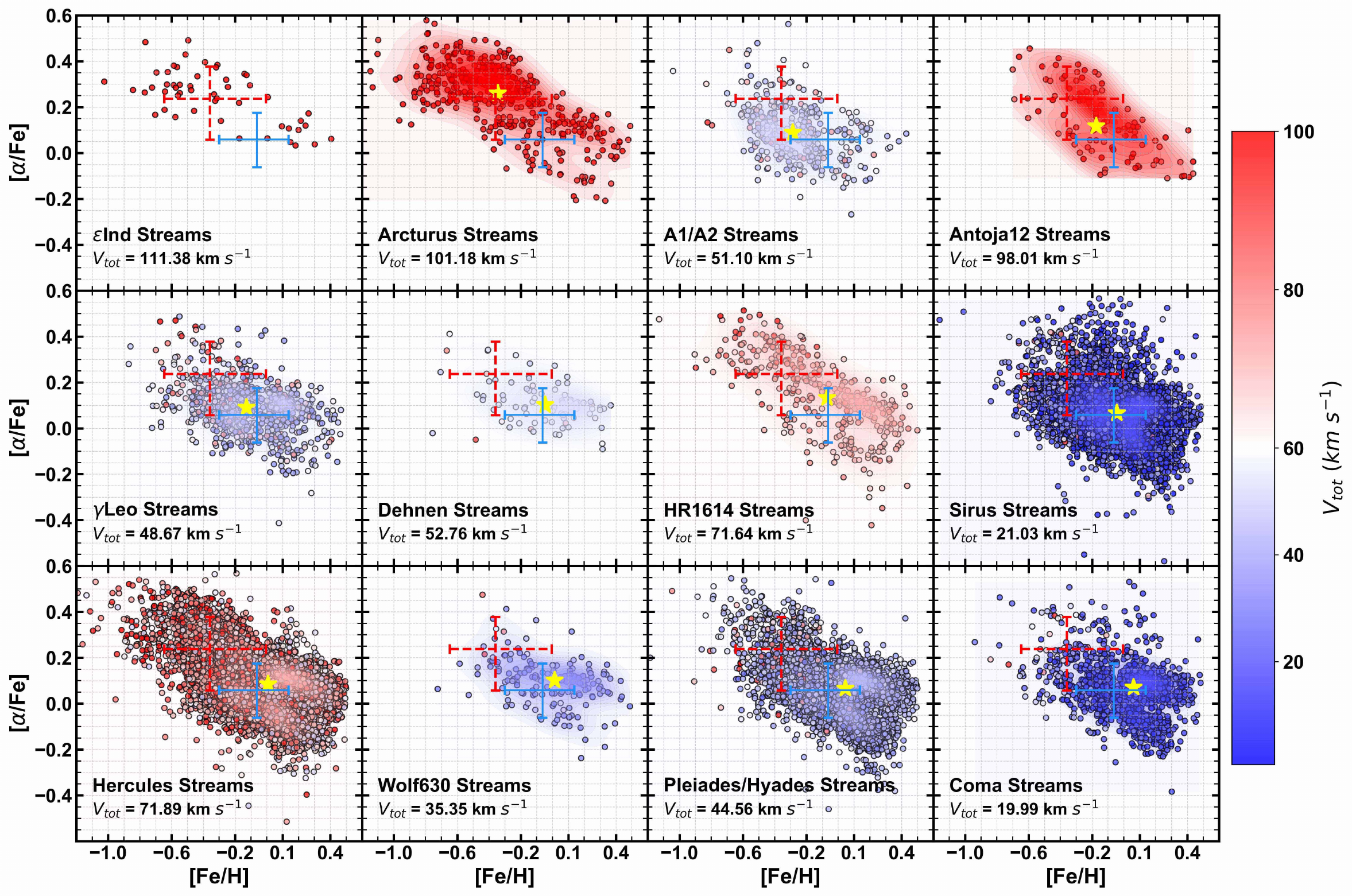}
  \caption{Same as Figure \ref{fig: Streams MRS} for the LRS sample.}
  \label{figappendix: lrs streams}
\end{figure}

The kinematic and chemical properties of the LRS sample, in the same format as Table \ref{tab: properties of LAMOST-Gaia-TESS star sample.}, are presented here (Table \ref{tab: properties of LAMOST_LRS-Gaia-TESS star sample.}). Corresponding figures for the LRS sample are also shown (Figure \ref{figappendix: lrs distribution}-\ref{figappendix: lrs streams}), analogous to Figures \ref{fig: fig 4. distribution of mrs sample}–\ref{fig: Streams MRS} for the MRS sample.
\begin{table*}[ht!]
    \caption{Kinematic and chemical properties of different Galactic components for the LAMOST\_LRS-Gaia-TESS sample.}
    \centering
    \begin{tabular}{c|cccccccccc}
    \hline Components & $N_\mathrm{s}$ & $N_\mathrm{p}$ & $V_\mathrm{tot}$ & $[\mathrm{Fe}/\mathrm{H}]$ & $[\alpha/\mathrm{Fe}]$ & $L_Z$ & $\sqrt{J_R}$ & $\ln(J_Z)$ & {$\rm Age_{\rm kin}$} & $\rm Age_{\rm zoomies}$ \\
    \hline
    \multicolumn{11}{c}{LRS} \\ \hline
    Thin disk & 372,923 & 270 & $35.4_{-15.6}^{+22.1}$ & $-0.06_{-0.24}^{+0.20}$ & $0.06_{-0.12}^{+0.11}$ & $1889^{+180}_{-239}$  & $4.6^{+3.2}_{-2.4}$ & $1.3^{+1.1}_{-1.5}$ & $2.52^{+0.31}_{-0.24}$ & $1.18^{+1.86}_{-1.18}$ \\
    Thick disk & 19,215 & 5 & $100.1_{-19.2}^{+31.0}$ & $-0.36_{-0.29}^{+0.36}$ & $0.19_{-0.19}^{+0.17}$ & $1389^{+375}_{-212}$ & $10.6^{+3.7}_{-3.9}$ & $3.1^{+0.9}_{-1.4}$ & $10.94_{-1.26}^{+1.62}$  & $8.92^{+5.08}_{-8.91}$ \\
    Halo & 462 & 0 & $241.81_{-36.89}^{+80.89}$ & $-0.87_{-0.63}^{+0.32}$ & $0.27_{-0.27}^{+0.13}$ & $259^{+300}_{-493}$ & $24.5^{+5.6}_{-3.3}$ & $3.6^{+1.6}_{-1.6}$ & NA & $10.14^{+3.86}_{-8.12}$ \\
    Sirus & 10,036 & 8 & $21.24_{-8.61}^{+10.74}$ & $-0.05_{-0.21}^{+0.18}$ & $0.01_{-0.03}^{+0.12}$ & $1925^{+25}_{-25}$ & $4.0^{+0.7}_{-0.6}$  & $1.3^{+1.1}_{-1.2}$ & NA & $1.65^{+2.68}_{-1.65}$ \\
    Hercules & 8,170 & 8 & $72.09_{-14.95}^{+15.10}$ & $-0.03_{-0.29}^{+0.24}$ & $0.03_{-0.06}^{+0.16}$ & $1510^{+76}_{-54}$ & $7.8^{+1.9}_{-1.6}$ & $1.5^{+1.0}_{-1.0}$  & NA & $2.05^{+3.19}_{-2.05}$ \\
    Pleiades/Hyades  & 5,668 & 4 & $44.69_{-6.40}^{+7.30}$ & $0.00_{-0.27}^{+0.21}$ & $0.01_{-0.05}^{+0.13}$ & $1573^{+88}_{-32}$ & $4.5^{+0.6}_{_0.7}$ & $1.5^{+0.9}_{-1.2}$ & NA  & $2.06^{+2.86}_{-2.06}$ \\
    Coma  & 1,823 & 2 & $20.00_{-3.58}^{+9.30}$ & $0.03_{-0.25}^{+0.18}$ & $0.01_{-0.02}^{+0.12}$ & $1800^{+50}_{-41}$ & $1.90^{+0.6}_{-0.7}$ & $1.0^{+1.3}_{-1.2}$ & NA  & $1.62^{+2.56}_{-1.62}$ \\
    $\gamma$Leo & 1,202 & 0 & $49.10_{-11.79}^{+18.55}$ & $-0.10_{-0.25}^{+0.21}$ & $0.04_{-0.04}^{+0.12}$ & $1929^{+39}_{-57}$ & $6.7^{+1.8}_{-0.9}$ & $1.8^{+1.0}_{-1.3}$ & NA & $3.57^{+5.05}_{-3.57}$ \\
    Arcturus & 523 & 2 & $101.18_{-6.59}^{+28.60}$ & $-0.33_{-0.33}^{+0.37}$ & $0.23_{-0.23}^{+0.14}$ & $1137^{+19}_{-37}$ & $11.5^{+1.5}_{-0.8}$  & $2.2^{+1.5}_{-1.1}$ &  NA  & $5.39^{+8.61}_{-5.39}$ \\
    HR1614  & 469 & 1 & $71.64_{-5.52}^{+7.87}$ & $-0.08_{-0.37}^{+0.30}$ & $0.09_{-0.09}^{+0.21}$ & $1355^{+26}_{-36}$ & $8.1^{+0.8}_{-0.6}$ & $2.0^{+1.0}_{-1.1}$ & NA & $4.17^{+6.68}_{-4.17}$ \\
    A1/A2  & 432 & 0 & $51.15_{-10.09}^{+15.22}$ & $-0.27_{-0.22}^{+0.24}$ & $0.06_{-0.06}^{+0.12}$ & $2186^{+22}_{-35}$ & $10.8^{+1.2}_{-0.5}$ & $2.0^{+0.9}_{-1.1}$ & NA & $3.84^{+5.18}_{-3.17}$ \\
    Wolf630  & 273 & 3 & $35.43_{-3.34}^{+8.42}$ & $-0.02_{-0.26}^{+0.21}$ & $0.04_{-0.04}^{+0.11}$ & $1709^{+20}_{-32}$ & $2.9^{+0.4}_{_0.2}$  & $1.5^{+1.1}_{-1.1}$ & NA & $2.51^{+6.50}_{-2.51}$ \\
    Dehnen  & 118 & 0 & $52.67_{-3.36}^{+7.66}$ & $-0.09_{-0.28}^{+0.24}$ & $0.05_{-0.05}^{+0.12}$ & $1688^{+21}_{-30}$ & $5.5^{+0.4}_{-0.3}$ & $1.8^{+1.2}_{-1.1}$ & NA  & $3.83^{+6.11}_{-3.83}$ \\
    Antoja12  & 128 & 0 & $98.26_{-13.87}^{+7.62}$ & $-0.14_{-0.21}^{+0.22}$ & $0.08_{-0.08}^{+0.19}$ & $1495^{+243}_{-39}$ & $9.7^{+1.9}_{-1.3}$ & $1.9^{+1.0}_{-1.0}$ & NA & $3.50^{+5.36}_{-3.50}$ \\
    $\epsilon$Ind  & 61 & 0 & $111.38_{-6.02}^{+12.31}$ &  $-0.33_{-0.30}^{+0.33}$ & $0.25_{-0.25}^{+0.09}$ & $1268^{+22}_{-37}$ & $12.5^{+0.5}_{-1.0}$ & $2.3^{+1.4}_{-0.8}$ & NA & $6.00^{+8.00}_{-6.00}$ \\
    \hline
    
    \end{tabular}
    \label{tab: properties of LAMOST_LRS-Gaia-TESS star sample.}
    {\raggedright \tablecomments{Units are the same as Table \ref{tab: properties of LAMOST-Gaia-TESS star sample.}.}\par}
    
\end{table*}


  


    \vspace{0.1cm}

\section{MCMC results from \textit{The Joker}}
\setcounter{figure}{0}  
For systems with a unique orbital solution returned by the rejection sampling from $The\ Joker$, we first generated 80,000,000 prior samples to mitigate sampling incompleteness.
If a unique solution was still obtained after this extensive sampling, the resulting posterior samples from The Joker were used to initialize a standard MCMC process to further refine the posterior distributions of the orbital parameters (i.e., $P$, $e$, $\omega$, $M_0$, $K$, and $v_0$). The MCMC sampling was performed using the PyMC3 package \citep{abril-plaPyMCModernComprehensive2023}, with 5,000 draws and 20 chains. The 1$\sigma$ uncertainties were defined by the 15.9th and 84.1st percentiles of the marginalized posterior distributions. As an illustrative example, Figure \ref{fig: mcmc_example} shows the corner plot for TIC 168704855, which demonstrates good convergence. In some cases, the MCMC chains may exhibit poor convergence. To ensure reliability,  we recommend utilizing posterior distribution results only when the Gelman-Rubin statistic is less than 1.1 \citep{gelmanInferenceIterativeSimulation1992}, indicating well-converged chains.

\begin{figure}[htb]
	\centering
	\includegraphics[width=0.5\linewidth]{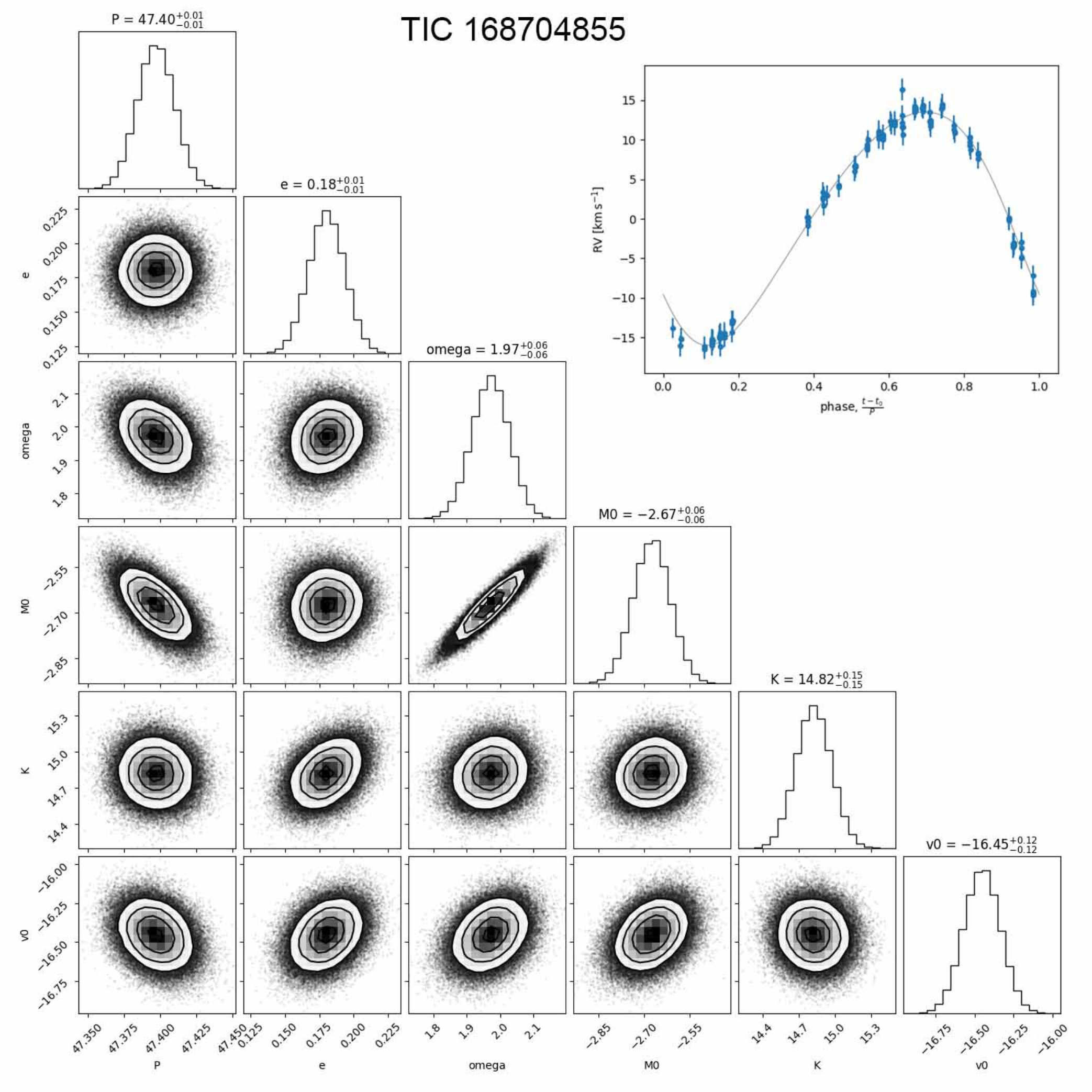}
	\caption{Corner plot for TIC 168704855 on the bottom left showing the posterior distributions of orbital parameters. The phase-folded RV curve and the corresponding fitting result are displayed in the top right panel.}
    \label{fig: mcmc_example}
\end{figure}

\end{document}